\documentclass[preprint,showpacs,preprintnumbers,amsmath,amssymb]{revtex4-1}
\renewcommand\thesection{\arabic{section}}
\renewcommand\thesubsection{\thesection.\arabic{subsection} }
\usepackage[utf8x]{inputenc}
\usepackage{graphicx}
\usepackage{blindtext}
\usepackage{float}
\usepackage{dcolumn}% Align table columns on decimal point
\usepackage{bm}% bold math
\usepackage{sidecap} 
\usepackage{amssymb}
\usepackage{epsfig}
\usepackage{psfrag}
\usepackage{gensymb}
\usepackage{amsmath}
\usepackage{amsfonts}
\usepackage{braket}

\begin{document}

\title{Insights into vibrational and electronic properties of MoS$_2$ using Raman, photoluminescence and transport studies}

\author{Achintya Bera}
\author{A K Sood\footnote{electronic mail: asood@physics.iisc.ernet.in}}

\affiliation{Department of Physics, Indian Institute of Science, Bangalore 560 012, India}
\date{\today\\}
\pacs{} 
\begin{abstract}

We review vibrational and electronic properties of single and a few layer MoS$_2$ relevant to understand their resonant and non-resonant Raman scattering results. In particular, the optical modes and low frequency shear and layer breathing modes show significant dependence on the number of MoS$_2$ layers. Further, the electron doping of the MoS$_2$ single layer achieved using top-gating in a field effect transistor renormalizes the two optical modes A$_{1g}$ and E$^1_{2g}$ differently due to symmetry-dependent electron-phonon coupling. The issues related to carrier mobility, the Schottky barrier at the MoS$_2$-metal contact pads and the modifications of the dielectric environment are addressed. The direct optical transitions for single layer-MoS$_2$ involve two excitons at K-point in the Brillouin zone and their stability with temperature and pressure has been reviewed. Finally, the Fermi-level dependence of spectral shift for a quasiparticle, called trion, has been discussed. \\

\smallskip
\noindent \textbf{Keywords.} MoS$_2$, Raman, electron-phonon coupling, field effect transistor, mobility, photoluminescence
\end{abstract}

\maketitle

\clearpage
%\tableofcontents
\clearpage

\section{Introduction}

Scientific challenges and potential applications of single and a few layer graphene in the last few years \cite{rise3,rmp3,raman3,jpcl,bisurev1,bisurev2} have given imputes to look for similar opportunities in other layered materials like BN, MoS$_2$, MoSe$_2$ and WS$_2$. The field-effect transistors made of two-dimensional semi-conducting materials are of particular interest for technological applications. In past few-years, the enormously studied single-layer graphene has made its importance in sensors and high-frequency devices \cite{novo1,roger1} because of the high carrier mobility ($\sim$ 10$^5$ for suspended graphene \cite{bolo}) and the response to the surrounding charges; although, in logical devices it is not very useful due to the vanishing electronic band gap at the K-points of the Brillouin zone (BZ). Recently, a single layer transition dichalcogenide material Molybdenum disulfide (MoS$_2$) field effect transistor has been demonstrated at room temperature to have a high on-off ratio $\sim$ 10$^8$ \cite{radi}, whereas $\sim$ 10 nm thick MoS$_2$ device gives a phonon limited-mobility value as high as $\sim$ 700 cm$^2$/V-sec \cite{das}. Production of single layer MoS$_2$ is not limited to the mechanical exfoliation method pioneered by Geim et al for graphene, it is being produced by the chemical vapor deposition (CVD) \cite{zhangw}, high temperature annealing method \cite{liun1}, solution method \cite{zhou1,lin2,colu1}, intercalation method \cite{eda,raon2,raonm} and so many other techniques \cite{liq,bale,peng1,lauri}. CVD based MoS$_2$ is used with graphene to form heterojunction which produces photoresponsitivity value $\sim$ 10$^7$ A/W \cite{zhangw}. MoS$_2$ nano-particles deposited on reduced graphene oxide act as an advanced catalyst for the hydrogen evolution reaction \cite{liy}. The diamagnetic single layer MoS$_2$ shows a magnetic ordering at room temperature upon exposure of proton of 2 Mev energy which is attributed to the formation of various defects and edge states \cite{malau,mathews}.    

The photoluminescence (PL) spectra confirms the gradual transition from an indirect band gap of bulk-MoS$_2$ to the direct band gap of single layer-MoS$_2$ \cite{mak}. The presence of the spin-filtered valence band at the K-point valleys of BZ and the time reversal symmetry guarantees the strong couplings of the spin and the valley index of the charge carriers and hence consequent optical selection rules \cite{yao1}. It has been shown with the circularly polarized light that the observed chirality of the PL emission has almost the same degree of polarization as the incident one at low temperatures \cite{mak2}. The voltage dependent transfer of the spectral weight \cite{mak3} of the exciton complexes $\sim$ 1.9 eV in PL spectra is associated with the observed metal-insulator transition \cite{kis} in the two-dimensional MoS$_2$ at low temperature. The measured ratio of the Coulomb potential energy to the kinetic energy is $\approx$ 60 at a charge density of $\sim$ 1 $\times$ 10$^{11}$ cm$^{-2}$ for the electron gas \cite{mak3}. Such high electron-electron interactions set the single layer MoS$_2$ as a platform to observe Wigner crystallization in the insulating regime, yet to be experimentally verified. Moreover, the existence of non-zero Berry curvature predicts the Hall effect in presence of circularly polarized light without applying magnetic field \cite{cao}. The single layer MoS$_2$ shows an electric-field induced superconductivity in the metallic phase with transition temperature of T$_c$ $\sim$ 9.4K \cite{tan1}. This observed superconductivity is attributed to the electron-electron interactions and the weak electron-phonon interactions \cite{rold}. In this review, we will discuss various aspects of phonon-assisted phenomena and their probing by Raman spectroscopy as a function of temperature, pressure, carrier doping and the layer numbers; and the dependence of carrier's mobility on various factors. Finally, the PL spectra at different gate voltages has been discussed leading to the discovery of negatively charged excitons or trions in single layer MoS$_2$.

\section{Crystal lattice and electronic band structure of MoS$_2$ : indirect to direct band gap}
\begin{figure}[h!]
 \centering
%\leavevmode
%\includegraphics[trim=0 0 20 25, scale=0.6]{fig_01.pdf}
\includegraphics[trim=0.5cm 10cm 0.5cm 2cm, width=1.00\textwidth]{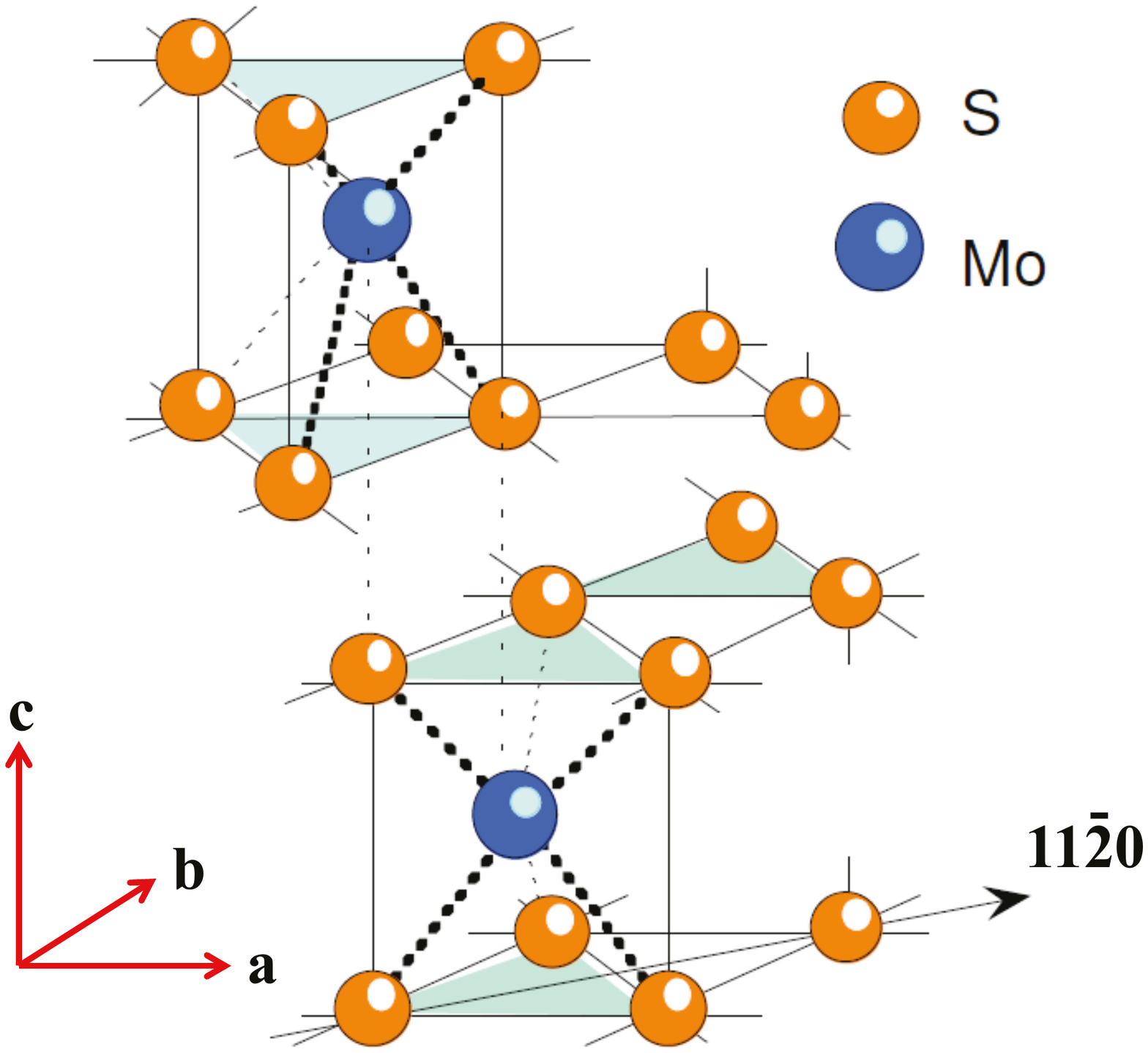}
  \caption{(color online)--The trigonal prismatic configurations of Mo atoms are shown for the structure of 2H-MoS$_2$ type. The (a,b) axis is in the x-y plane and the c-axis is along the z-direction. Along c-axis S-Mo-S units are stacked via van der Waals interactions. Taken from ref. \cite{ramana}.}
 \label{ramana}
\end{figure}

Bulk Molybdenum disulphide (MoS$_2$) is a layered semiconducting material, each layer having hexagonal symmetry as graphene. Unlike graphene which consists of only carbon atoms, each MoS$_2$ layer unit is formed by stacking of three hexagonal planes of S-Mo-S atoms in order along the c-axis \cite{ramana} as shown in Fig.~\ref{ramana}. Within  stable unit (S-Mo-S), atoms are bonded in a prismatic configuration through ionic-covalent interactions and each such stable units are connected via van der Waals interactions along the c-axis to form the bulk MoS$_2$ crystal. Two units, forming the unit cell of bulk MoS$_2$, are stacked vertically in such a way that the Mo atom of one layer sits on top of the S atoms of other layer as shown in Fig.~\ref{ramana}. There are two polytypes \cite{bromley} of MoS$_2$: (i) 2H-MoS$_2$ [shown in Fig.~\ref{ramana}] is more commonly available in nature as molybdenite salt and it has two layers per unit cell. Our discussions will be related to this type of structure; and (ii) 3R-MoS$_2$ has three layers per unit cell, which are stacked in rhombohedral symmetry.

\begin{figure}[h!]
 \centering
%\leavevmode
%\includegraphics[trim=0 0 20 25, scale=0.6]{fig_02.pdf}
\includegraphics[trim=0.5cm 15cm 0.5cm 2cm, width=1.00\textwidth]{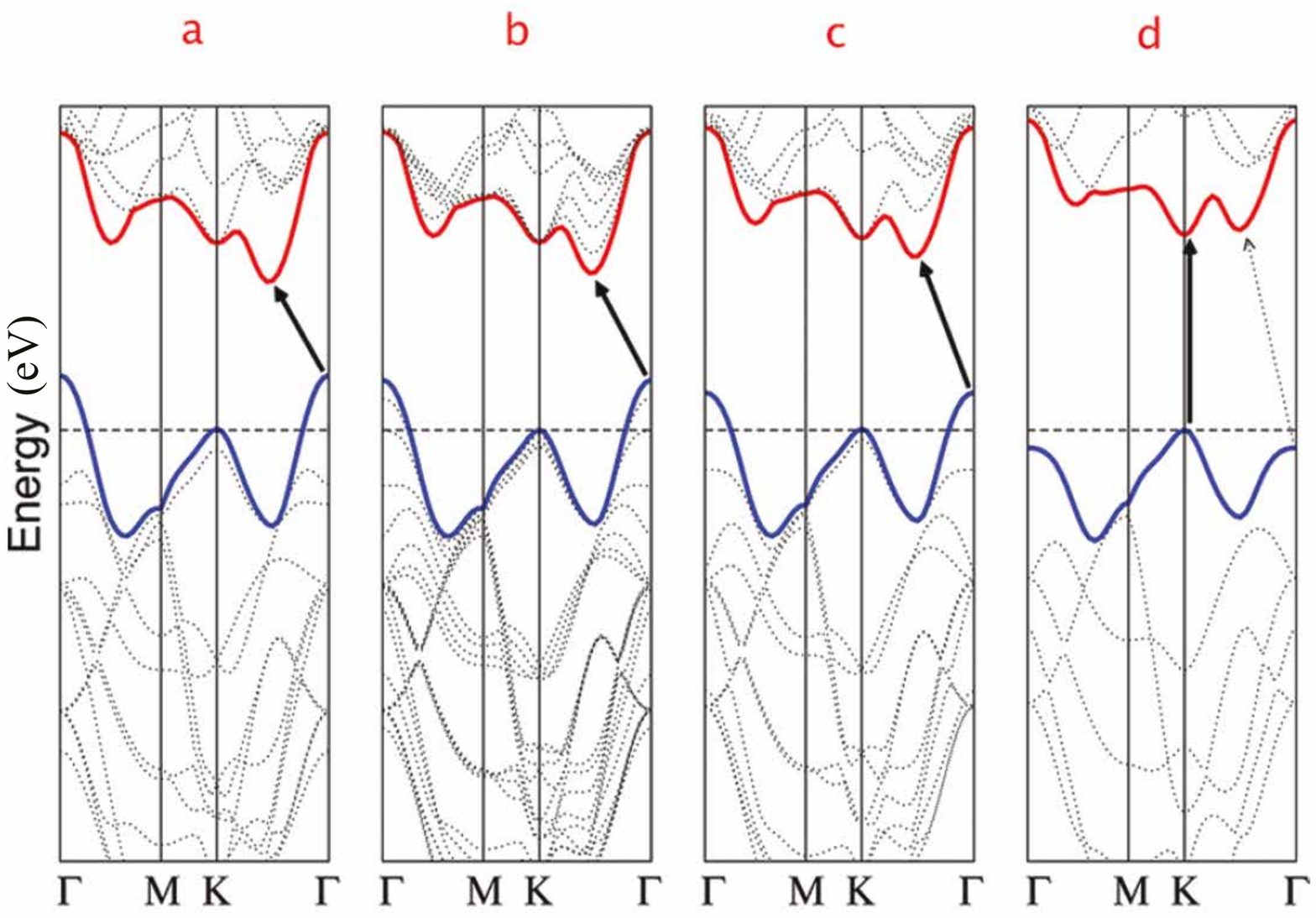}
  \caption{(color online)--Calculated electronic energy band-diagrams are shown for (a) bulk-MoS$_2$, (b) 4-layer MoS$_2$, (c) 2-layer MoS$_2$ and (d) single-layer MoS$_2$. The solid arrow lines shown in Figs.(a), (b) and (c) indicate the indirect transitions from valence band (at $\Gamma$ point) to the conduction band minimum (along $\Gamma$-K direction), whereas in (d) it indicates the direct electronic transition at the K-point. Taken from ref. \cite{splend}.}
 \label{splend}
\end{figure}

\begin{figure}[h]
 \centering
%\leavevmode
\includegraphics[trim=0.5cm 13cm 0.5cm 2cm, width=1.00\textwidth]{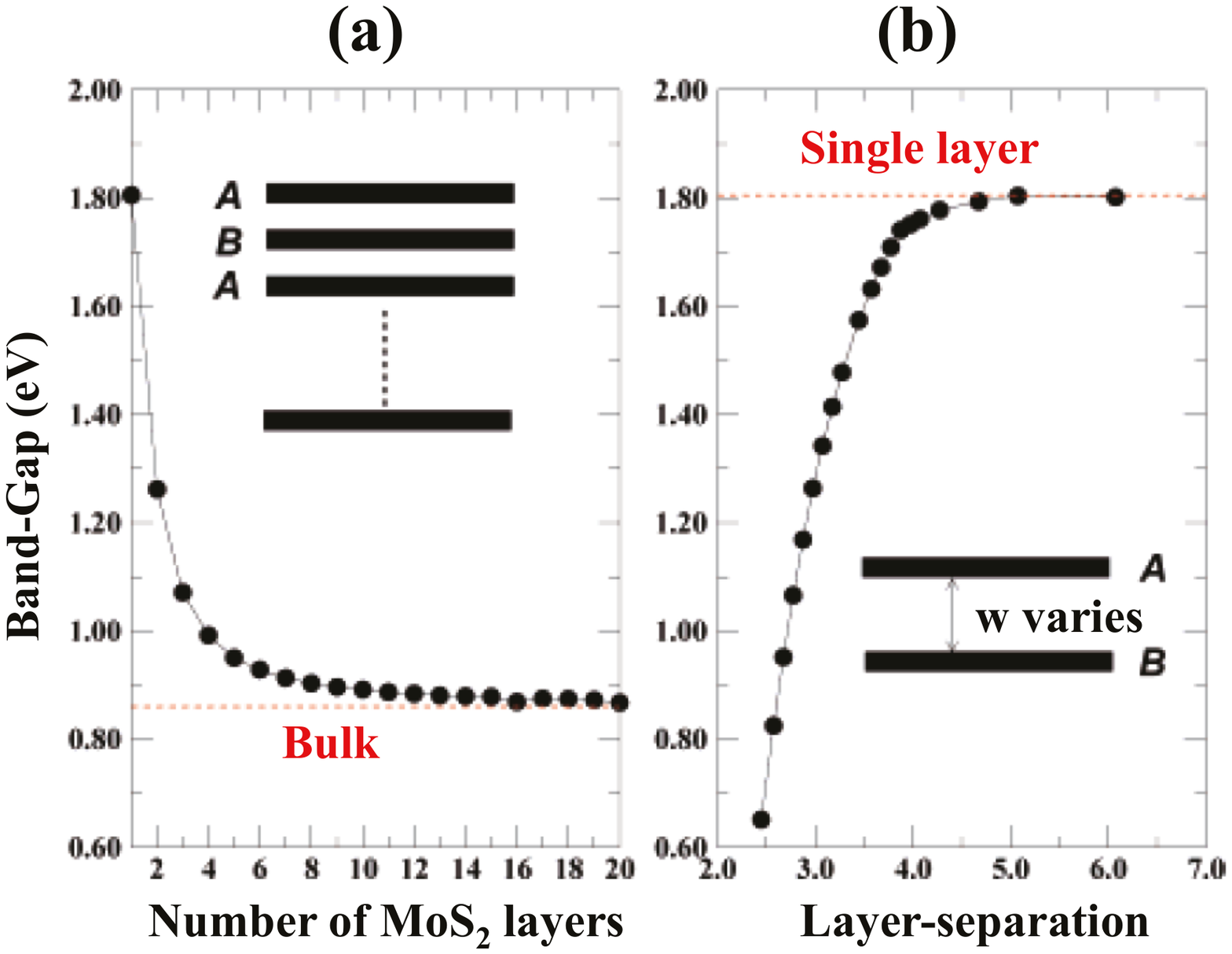}
  \caption{(color online)--Calculated changes in the band gap for MoS$_2$ flakes as a function of (a) number (N) of layers and (b) the separation (w) between two-layers. Here A and B denote S-Mo-S unit. Taken from Ref. \cite{li}.}
 \label{li}
\end{figure}

Like Si, bulk MoS$_2$ has an indirect band gap of $\sim$ 1.2 eV \cite{matt,leb,mak} with valence band maximum at $\Gamma$ point and conduction band minimum in between the $\Gamma-K$ direction as shown in Fig.~\ref{splend}(a). Now if we cleave the bulk MoS$_2$ one by one stable unit (S-Mo-S) and examine the electronic band structure, we will see that there is a transition from indirect band gap to a direct band gap $\sim$ 1.9 eV at $K$ point for single layer (S-Mo-S unit) MoS$_2$ (SL-MoS$_2$). The calculated electronic band structure with the number of layers \cite{splend} is shown in Fig.~\ref{li}. The photoluminescence spectra \cite{mak,splend} confirms the direct band gap of SL-MoS$_2$, which will be discussed later. For the bulk or a few layers of MoS$_2$, it has been shown \cite{li} that the states near the indirect band gap consist of linear combinations of $d$ orbitals on $Mo$ atoms and $p_z$ orbitals on S atoms. The states near the conduction band at $K$ point consist of only localized $d$ orbitals on $Mo$ atoms. Since Mo atoms are sandwiched between two planes of S atoms in single S-Mo-S unit, the direct band gap has less dependence than the indirect band gap on number of layers (or interlayer coupling). The quantum confinement effect along the c-axis increases the indirect band gap, whereas direct band gap does not vary much. \\

\section{Phonons in bulk and single layer MoS$_2$}

\begin{figure}[h!]
 \centering
%\leavevmode
%\includegraphics[width=0.4\textwidth]{fig_04.pdf}
\includegraphics[trim=0.5cm 20cm 0.5cm 2cm, width=1.00\textwidth]{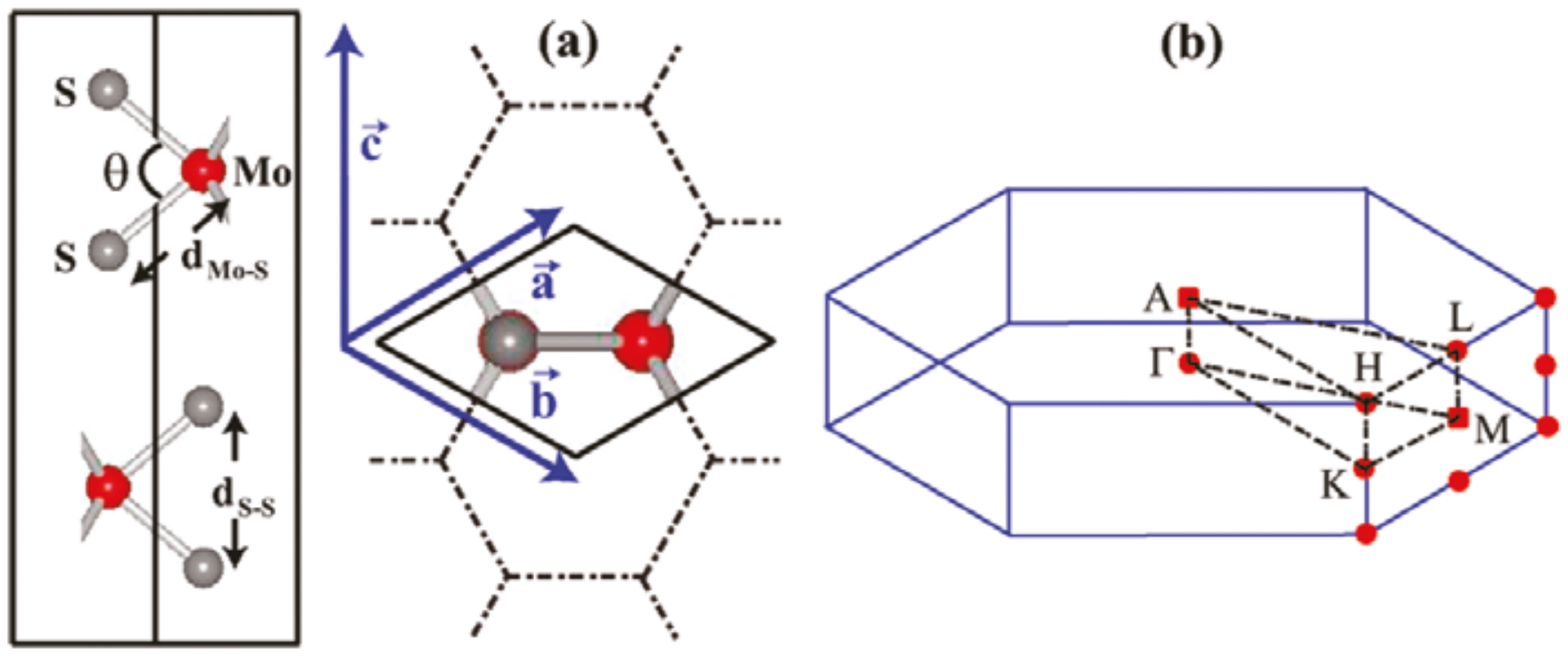}
  \caption{(color online)--Hexagonal lattice structure of 2H-MoS$_2$ is shown in (a) for side and top views. The unit cell is marked by solid black lines, which contains one Mo atom (shown by red balls) and one S atom (shown by grey balls). The corresponding Brillouin zone is shown in (b), where red dots mark the high-symmetry points and  dashed lines the corresponding directions. Taken from Ref. \cite{ataca}.}
 \label{ataca1}
\end{figure}

\begin{figure}[h!]
 \centering
%\leavevmode
%\includegraphics[trim=0 0 20 25, scale=0.6]{fig_05.pdf}
\includegraphics[trim=0.5cm 4cm 0.5cm 5cm, width=1.00\textwidth]{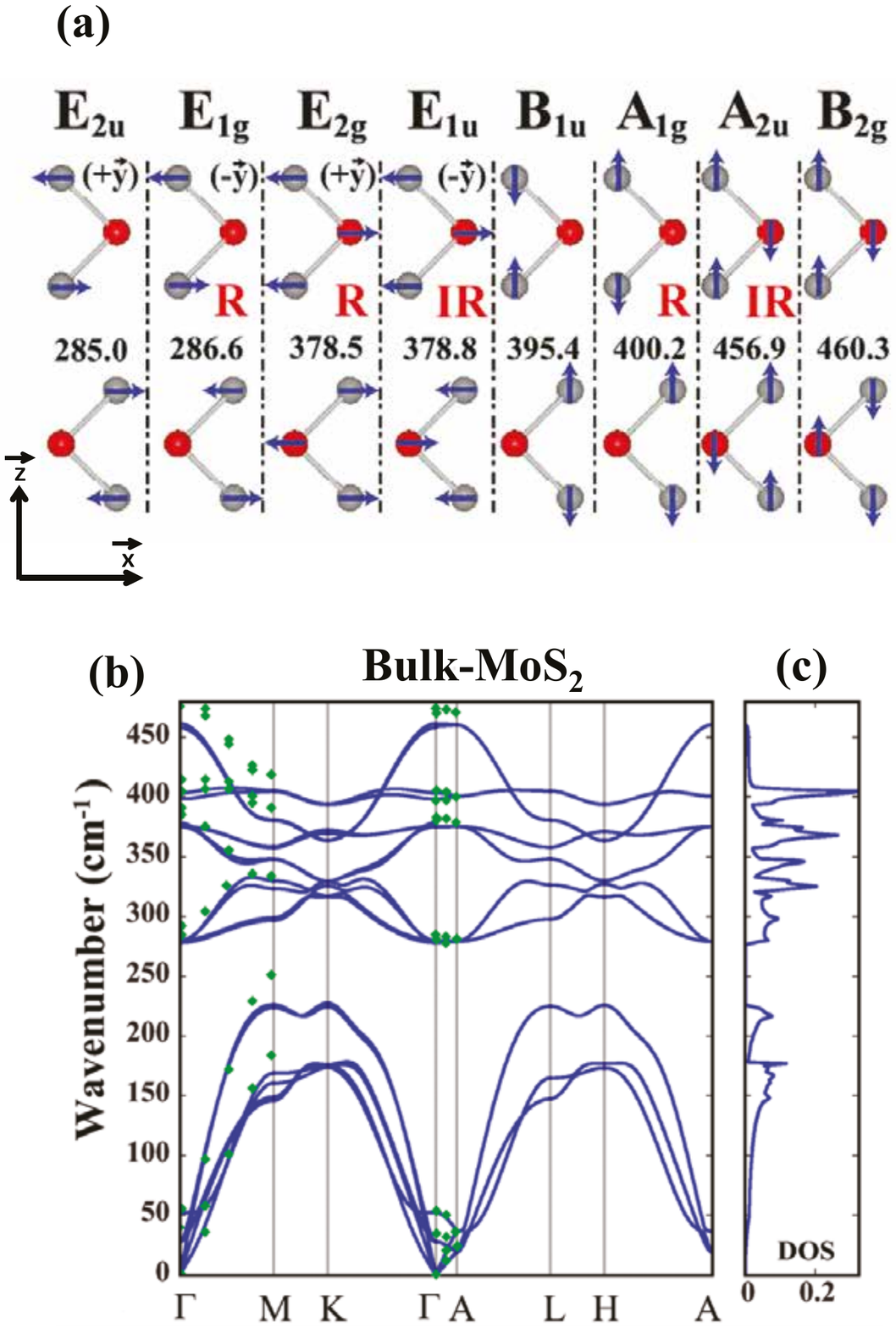}
  \caption{(color online)--(a) Different optical phonon modes at the $\Gamma$ point with the corresponding symmetries and vibrational configurations for bulk-MoS$_2$. The numbers are the corresponding vibrational frequencies in cm$^{-1}$. The phonon dispersion curves along different symmetry directions and the corresponding density of states are shown in (b) and (c), respectively. Taken from Ref. \cite{ataca}.}
 \label{ataca2}
\end{figure}

Bulk MoS$_2$ belongs to the point group $D_{6h}$ (space group $P3m1$) \cite{moli}. Each unit cell with lattice parameters $\it{a}$=3.12 $\AA{}$ and $\it{c}$=12.3 $\AA{}$   \cite{moli} contains two molecular units shown in Fig.~\ref{ataca1} \cite{ataca} and hence, it has total 18 normal modes of vibrations corresponding to 2A$_{2u}$+2E$_{1u}$+2B$_{2g}$+2E$_{2g}$+A$_{1g}$+E$_{1g}$+B$_{1u}$+E$_{2u}$  irreducible representations \cite{verbel,wiet,zhang}. The calculated values of the frequencies and optical activity of all the phonon modes are shown in Table 1 \cite{moli}. A$^1_{2u}$ and E$^1_{1u}$ phonon modes belong to acoustic branches. Since bulk MoS$_2$ has center of inversion symmetry, the infra-red (IR) active phonon modes (A$^2_{2u}$ and E$^2_{1u}$) and Raman active phonon modes are mutually exclusive. Excluding the inactive phonon modes (2B$_{2g}$, B$_{1u}$ and E$_{2u}$), the total four Raman-active optical phonon modes are A$_{1g}$, E$_{1g}$ and 2E$_{2g}$. The vibrational configurations for all the phonon modes are shown in Fig.~\ref{ataca2}(a) and the phonon dispersion curves for bulk MoS$_2$ are shown in Fig.~\ref{ataca2}(b) \cite{ataca}.

%Fig-5

\begin{figure}[h!]
 \centering
%\leavevmode
%\includegraphics[trim=0 0 20 25, scale=0.5]{fig_06.pdf}
\includegraphics[trim=0.5cm 3cm 0.5cm 3cm, scale=0.6]{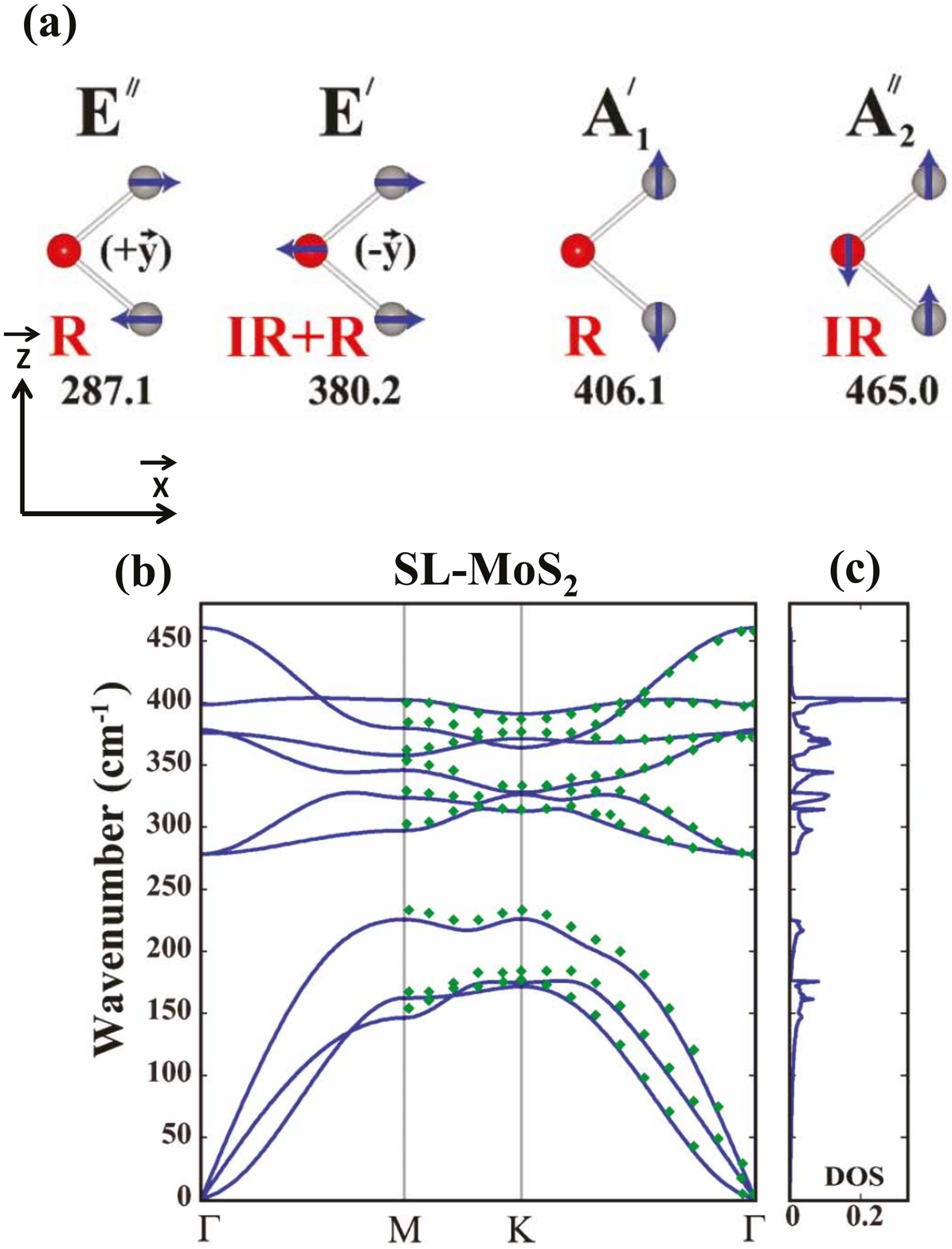}
  \caption{(color online)--(a) Different optical phonon modes at the $\Gamma$ point with the corresponding symmetries and vibrational configurations for single layer-MoS$_2$. The numbers are the corresponding vibrational frequencies in cm$^{-1}$. (b) Phonon dispersion curves along different symmetry directions and (c) the density of states. Taken from Ref. \cite{ataca}.}
 \label{ataca3}
\end{figure}

Single layer MoS$_2$ has a point group symmetry of $D_{3h}$ (space group $P6m2$) \cite{moli}. Since SL-MoS$_2$ has no inversion symmetry, the labellings for the phonon modes gets changed. The total number of phonon branches is reduced to nine and the modes at $\Gamma$-point are given by 2A$^{''}_{2}$+2E$^{'}$+A$^{'}_{1}$+E$^{''}$ \cite{zhang}. Here A$^{''}_{2}$ and E$^{'}$ are acoustic phonons; another E$^{'}$ mode is both IR and Raman active. Second A$^{''}_{2}$ mode is IR active. The total Raman active phonon modes are A$^{'}_{1}$, E$^{'}$ and E$^{''}$ corresponding to bulk Raman modes A$_{1g}$, E$_{2g}$ and E$_{1g}$ respectively. The eigenvectors of the phonon modes and the calculated dispersions are shown in Fig.~\ref{ataca3}(a) and Fig.~\ref{ataca3}(b), respectively \cite{ataca}. Even number of S-Mo-S units belong to $D_{6h}$ (space group $P3m1$) point group symmetry having inversion center and odd numbers belong to $D_{3h}$ (space group $P6m2$) point group symmetry without inversion center \cite{moli}.

\begin{table}
\caption{\label{tab:table1} List of high-frequency optical phonon modes for bulk and SL-MoS2.}
\begin{ruledtabular}
\begin{tabular}{llllcrr}
%\\

SL-MoS$_2$  & Bulk            & Character            & Direction        & Atoms        & $\omega (cm^{-1})$      & $\omega (cm^{-1})$  \\
D$_{3h}$(S) & D$_{6h}$(B)     &                      &                  & Involved     & calculated$^a$          & exp.       \\                             

\hline
%\hline
%\\

A$_2^{''}$ & $A_{2u}$       & Acoustic              & Out of plane      & Mo+S        & 0.0 (S)   0.0 (B)         &\\
           & $B^2_{2g}$     & Inactive              & Out of plane      & Mo+S        & 55.7 (B)                  &\\  
           & $E^2_{2g}$     & Raman                 & In plane          & Mo+S        & 35.2 (B)                  & 22 (Bi)  33 (B)$^{c,d}$\\
A$_1^{'}$  & $A_{1g}$       & Raman                 & Out of plane      & S           & 410.3 (S) 412.0 (B)       & 402 (S) 408(B)$^b$\\
           & $B_{1u}$       & Inactive              & Out of plane      & S           & 407.8 (B)                 &\\
A$_2^{''}$ & $A_{2u}$       & Inactive (E $||$ c)   & Out of plane      & Mo+S        & 476.0 (S) 469.4 (B)       & \\
           & $B^1_{2g}$     & Inactive              & Out of plane      & Mo+S        & 473.2 (B)                 &\\
E$_2^{'}$  & $E^1_{2g}$     & Raman                 & In plane          & Mo+S        & 391.7 (S) 387.8 (B)       & 382 (S) 380(B)$^b$\\
           & $E_{1u}$       & Infrared (E $\bot$ c) & In plane          & Mo+S        & 391.2 (B)                 &\\
E$_2^{''}$ & $E_{1g}$       & Raman                 & In plane          & S           & 289.2 (S) 288.7 (B)       &\\          
           & $E_{2u}$       & Inactive              & In plane          & S           & 287.1 (B)                 &\\
                                      
\end{tabular}
{$^a$ Ref. \cite{moli}. $^b$ Ref. \cite{lee}.  $^c$ Ref. \cite{zeng}. $^d$ Ref. \cite{ple}. Here 'Bi' indicates Bi-layer}
\end{ruledtabular}
\end{table}

\section{Layer dependence of optical Raman modes: off-resonance}

With the ability of making single and bi-layer samples by mechanical exfoliation technique, the layered materials have potential technological applications in nano devices and logic circuits. The non-destructive Raman spectroscopy has been widely used to measure the number of layers precisely for single, bi-layer and multi-layer graphene systems via the signatures of the 2D phonon band near $\sim$ 2700 cm$^{-1}$ (two-phonon double resonance feature) \cite{d2g}. Similarly, different number of  MoS$_2$ layers have been identified through the mode frequencies. In this section, we will discuss the off-resonance Raman spectra (laser excitation energy 2.41 eV and 2.33 eV as compared to the direct band gap of 1.9 eV).

\begin{figure}[h!]
 \centering
%\leavevmode
%\includegraphics[width=0.5\textwidth]{fig_07.pdf}
%\includegraphics[trim=0 0 20 25, scale=0.4]{fig_07.pdf}
\includegraphics[trim=0.5cm 18cm 0.5cm 2cm, width=1.00\textwidth]{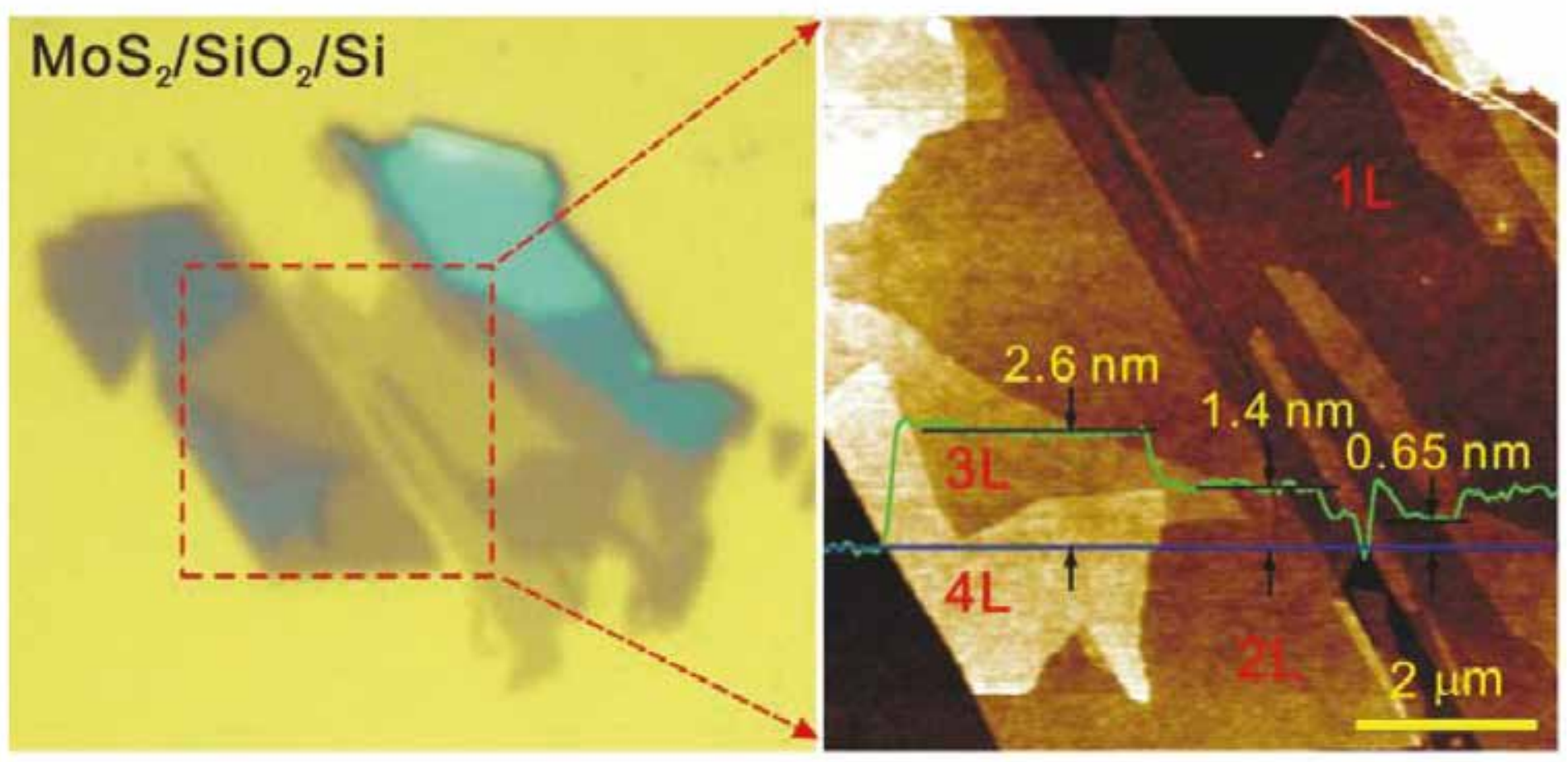}
  \caption{(color online)--(a) Optical image of the MoS$_2$ flake deposited on SiO$_2$/Si wafer. The dashed box indicates the area of 8 $\times$ 8 $\mu$ m$^2$ for which the AFM image has been taken and shown in (b). The different layers and the corresponding height profiles are shown in AFM image. Taken from Ref. \cite{lee}.}
 \label{lee1}
\end{figure}

\subsection{High frequency optical modes (around 400 cm$^{-1}$): opposite trends of A$_{1g}$ and E$^1_{2g}$ modes as a function of number of layers}

\begin{figure}[h!]
 \centering
%\leavevmode
%\includegraphics[trim=0 0 20 25, scale=0.5]{fig_08.pdf}
\includegraphics[trim=0.5cm 10cm 0.5cm 2cm, width=1.00\textwidth]{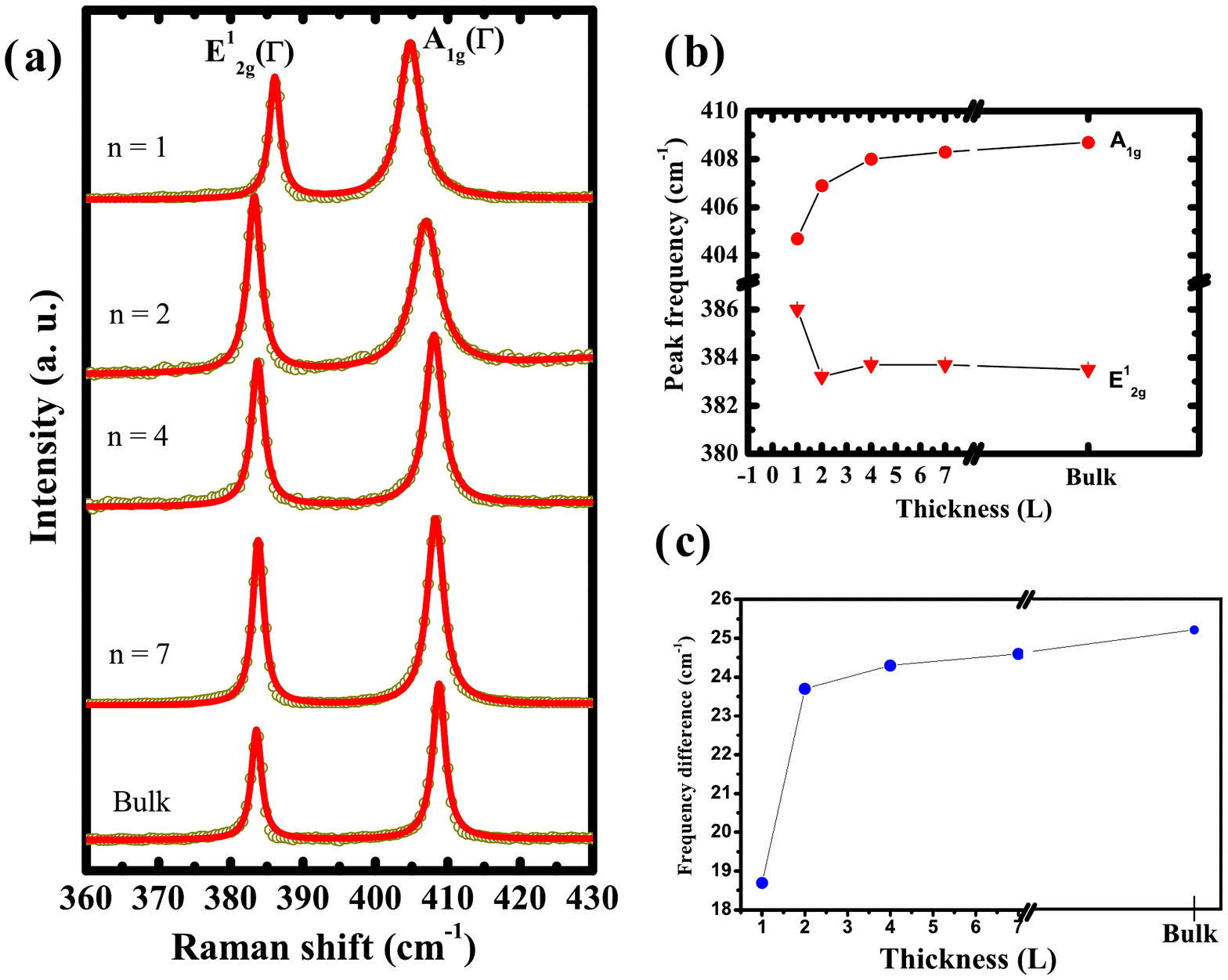}
  \caption{(color online)--(a) Raman spectra of a few layers and bulk MoS$_2$. The laser excitation energy used is 2.41 eV (514 nm) \cite{bisu}. Solid lines represent the Lorentzian fits to the experimental data (circles). (b) Raman shifts of these two modes and (c) the frequency difference ($\omega_{A_{1g}}$-$\omega_{E^1_{2g}}$) as a function of thickness of the sample. Taken from Ref. \cite{bisu}.}
 \label{lee2}
\end{figure}

\begin{figure}[h!]
 \centering
%\leavevmode
%\includegraphics[trim=0 0 20 25, scale=0.5]{fig_09.pdf}
\includegraphics[trim=0.5cm 5cm 0.5cm 2cm, width=1.00\textwidth]{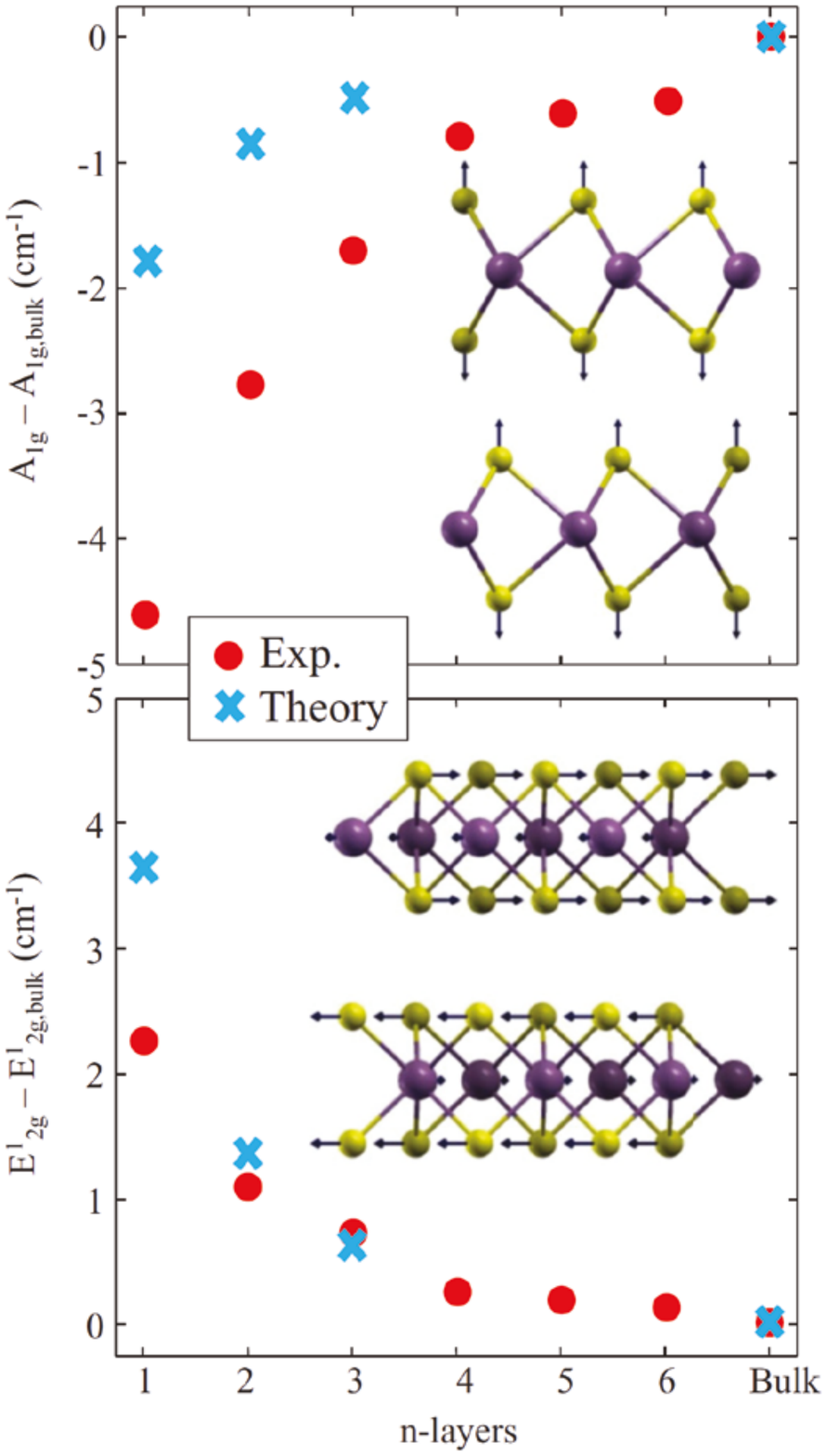}
  \caption{(color online)-- The phonon frequency differences between n-layers and bulk for A$_{1g}$ (top panel)   and E$^1_{2g}$ (bottom panel). The calculated data from Moli et al \cite{moli} are shown with blue cross symbol and the experimental data from Lee et al \cite{lee} are shown with red circles. Taken from Ref. \cite{moli}.} 
 \label{moli}
\end{figure}

The single and few layers of MoS$_2$ deposited on SiO$_2$/Si wafer are shown in Fig.~\ref{lee1}(a) taken by optical microscope and the thickness of the corresponding flakes measured by atomic force microscopy (AFM) are shown in Fig.~\ref{lee1}(b). The measured thickness of a single layer MoS$_2$ is about 0.6-0.7 nm. The Raman spectra for bulk and few layers MoS$_2$ are shown in Fig.~\ref{lee2}(a) and the observed frequencies of the two modes are shown in Fig.~\ref{lee2}(b) \cite{bisu}. The A$_{1g}$ Raman mode hardens with increasing the number of layers by $\sim$ 6 cm$^{-1}$ , whereas E$^1_{2g}$ softens by $\sim$ 2 cm$^{-1}$. Above four layers, both modes show saturation values with the Bulk values $\sim$ 408 and 382 cm$^{-1}$, respectively. The opposite trend of the two Raman modes leads to increase in frequency difference [shown in Fig.~\ref{lee2}(c)] which can be used to probe the number of layers of MoS$_2$ system. The hardening behavior of out-of-plane mode (A$_{1g}$) can be understood in terms of increasing the restoring force constant perpendicular to the basal plane through van der Waals interactions as we increase the number of layers. In comparison, the in-plane vibrational mode (E$^1_{2g}$) shows anomaly. The difference between these two vibrational modes is that E$^1_{2g}$ involves vibration of Mo and S atoms in the basal plane; whereas in A$_{1g}$ vibrational configuration, Mo atoms remain fixed and only S atoms vibrate along c-axis. It has been shown \cite{moli} that the long range coulomb interaction part of the self interaction remains negligible for S atoms, whereas for Mo atoms it decreases considerably as we increase the number of stacking layers. Because of increased screening and consequent increase in dielectric tensor with the number of layers, the in-plane mode E$^1_{2g}$ softens. The qualitative agreement between the theory \cite{moli} and the experiment \cite{lee} for the observed two Raman modes is shown in Fig.~\ref{moli}.

\begin{figure}[h!]
 \centering
%\leavevmode
%\includegraphics[trim=0 0 20 25, scale=0.5]{fig_10.pdf}
\includegraphics[trim=0.5cm 18cm 0.5cm 2cm, width=1.00\textwidth]{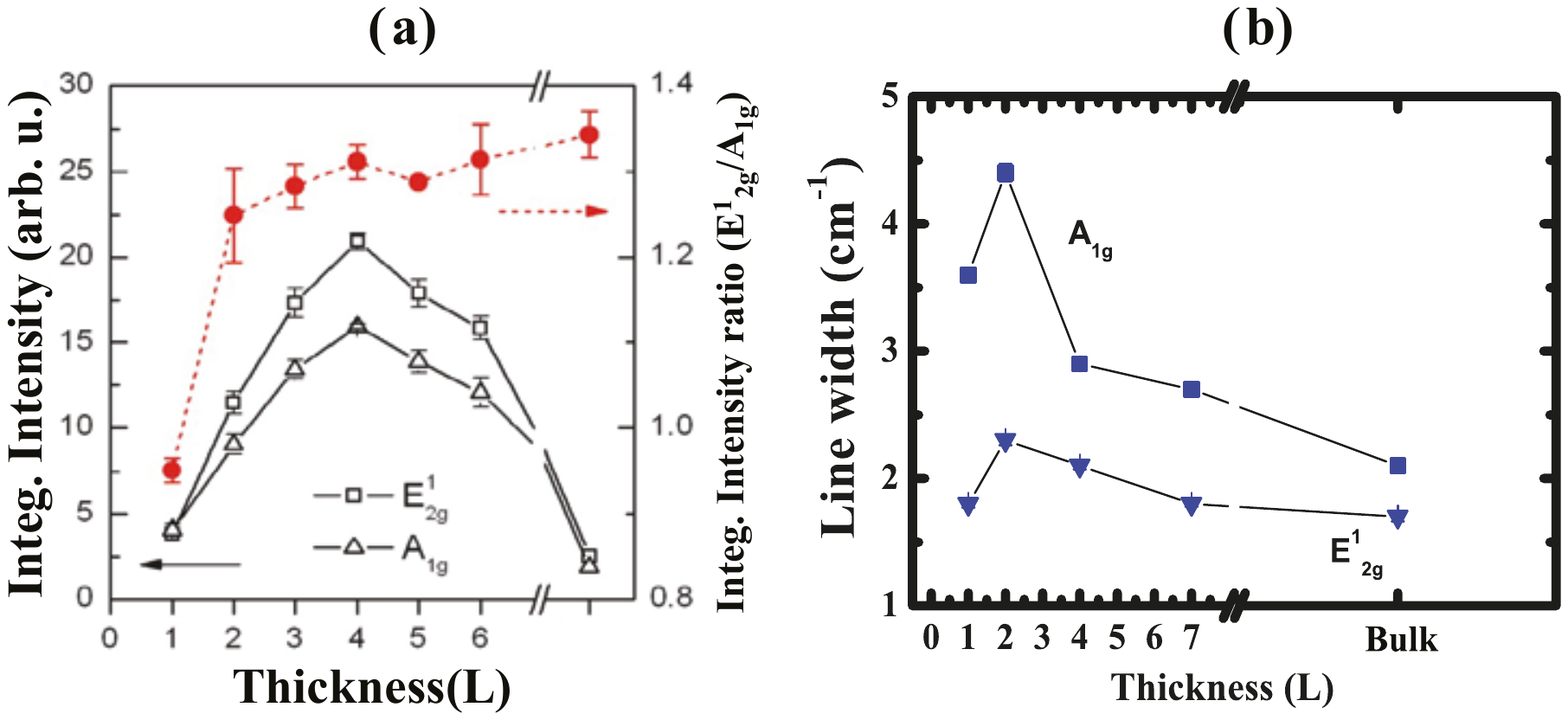}
  \caption{(color online)--(a) The integrated intensity of A$_{1g}$ and E$^1_{2g}$ modes (corresponding to left vertical scale) and their ratio (corresponding to right vertical scale). Taken from Ref. \cite{lee}. (b) The line width (FWHM) of these two modes. Taken from Ref. \cite{bisu}.}
 \label{lee3}
\end{figure}
%FIG-10

The integrated intensity variation of the A$_{1g}$ and E$^1_{2g}$ modes with the number of layers is shown in Fig.~\ref{lee3}(a). The Raman intensity increases linearly upto four-layers and then decreases to the bulk value. The increase in intensity for both the Raman modes can be attributed to the optical field enhancement due to multiple reflections of Raman field within the sample layers, and consequent interference effects between these reflected lights and the those coming from the thin layer of SiO$_2$ (300nm) deposited on Si substrate. The optical field enhancement and interference phenomena also leads to increase in intensity of Raman modes for graphene systems and shows a maximum upto 9 layers \cite{wang}. It has been observed \cite{lee} that SL-MoS$_2$ deposited on quartz substrate has Raman intensity of $\sim$ 20 $\%$ of that of the SL-MoS$_2$ deposited on SiO$_2$/Si substrate. The integrated intensity ratio of the two modes, I(E$^1_{2g}$)/I(A$_{1g}$), is also shown in Fig.~\ref{lee3}(a). Since the interference enhancement phenomena affects both the modes equally, the abrupt increase of the integrated intensity ratio from single to bi-layer is not clear and, it may be due to the differences of underlying crystal symmetries between them. The full width at half maximum (FWHM) of A$_{1g}$ and E$^1_{2g}$ modes are shown in Fig.~\ref{lee3}(b) \cite{bisu}. The FWHM of A$_{1g}$ Raman mode increases from 5 to 7 cm$^{-1}$ as layer number decreases from 7 to 2 layers. The E$^1_{2g}$ mode shows no thickness dependent variations ($\sim$ 1 cm$^{-1}$) of the FWHM. The FWHM dependence of both the modes A$_{1g}$ and E$^1_{2g}$ needs further understanding. Lee at el \cite{lee} showed that the broadening of the A$_{1g}$ phonon is also observed for the suspended MoS$_2$ layers and hence cannot be attributed to the interaction with the substrate.

\subsection{Low frequency optical modes (below 50 cm$^{-1}$): shear (E$^2_{2g}$) and compressional modes}

\begin{figure}[h!]
 \centering
%\leavevmode
%\includegraphics[trim=0 0 20 25, scale=0.5]{fig_11.pdf}
\includegraphics[trim=0.5cm 0cm 0.5cm 2cm, width=0.65\textwidth]{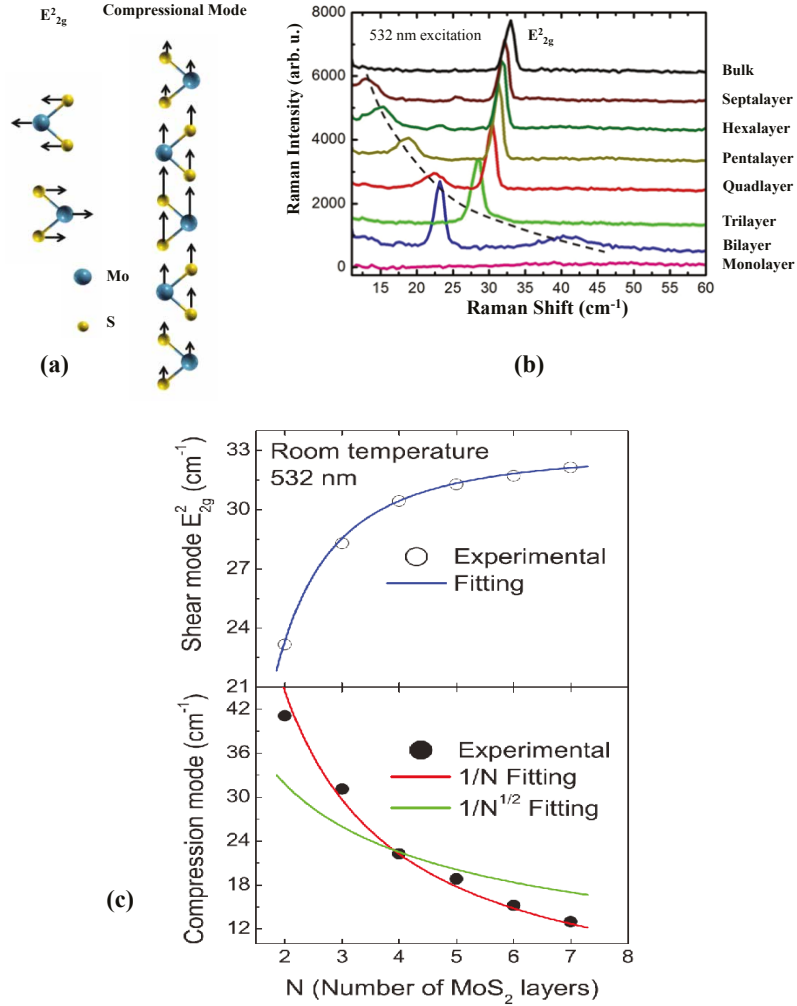}
  \caption{(color online)--(a) The vibrational configurations of the E$^2_{2g}$ and the compressional modes. (b) Raman spectra of two modes for different layers and bulk MoS$_2$ deposited on SiO$_2$/Si substrate. The dashed line is guide to the eye. (c) The peak positions as a function of thickness of the samples. The interlayer force constant per unit area ($\alpha$) and the unit layer mass per unit area ($\mu$) used for the least square fitting of E$^2_{2g}$ mode are 2.9 $\times$ 10$^{19}$ N m$^{−3}$ and 30.3 Kg $\AA^{-2}$, respectively. Taken from Ref. \cite{zeng}.}
 \label{zeng1}
\end{figure}

In the E$^2_{2g}$ optical phonon, which is a in-plane rigid layer vibration (Fig.~\ref{zeng1}a), the atoms in each layer vibrates in phase such that each layer rigidly oscillates against the adjacent layers and the only restoring force is the weak van der Waals interactions acting between the layers. This mode, also known as a shear mode, exists in all layered compound materials with low frequency ($\sim$ 30-40 cm$^{-1}$). Raman spectra \cite{zeng} of a few layers and bulk  MoS$_2$  for the low frequency optical phonons are shown in Fig.~\ref{zeng1}(b). The E$^2_{2g}$ mode shows stiffening behavior with the thickness of the sample from $\sim$ 22 cm$^{-1}$ (2 layers) to $\sim$ 33 cm$^{-1}$ (bulk)[Fig.~\ref{zeng1}c]. Another low frequency mode with broad feature attributed to the compressional mode (C-mode) shows softening behavior from $\sim$ 42 cm$^{-1}$ (2 layers) to $\sim$ 15 cm$^{-1}$ (7-layers). The experimentally observed and theoretically calculated \cite{zeng} frequencies for both the modes are shown in Fig.~\ref{zeng1}(c). For the shear E$^2_{2g}$ mode, the fitted frequency (Fig.~\ref{zeng1}c) follows the expression (obtained from the linear-chain model) $\omega=\frac{1}{\sqrt{2} \pi c}\sqrt{\frac{\alpha}{\mu}}\sqrt{1+cos(\frac{\pi}{N})}$ \cite{tan}. Here $\omega$ denotes the peak position of the shear mode; N, the number of layers; c, the speed of light in cm/s; $\alpha$, the interlayer force constant per unit area and $\mu$, the unit layer mass per unit area. For multi-layer graphene \cite{tan}, the low frequency shear modes also follow the same trends i.e. with the increasing layer numbers, the restoring force constant acting between the layers increases. \\

The thickness dependence of the C-mode shows $\frac{1}{N}$ dependence instead of $\frac{1}{\sqrt{N}}$. The $\frac{1}{N}$ behavior also observed for the longitudinal acoustic phonons on Na films [deposited on Cu (001)] \cite{has}, is due to open standing waves in multilayer (or thin films) systems \cite{luo}. A strong coupling between the substrate and the atomic layer of the sample does not allow the atoms close to the substrate vibrate in the C-mode and creates a node there, while there is no restrictions for the atoms in the top layer; thus creating open standing waves (kind of organ-pipe modes) and hence $\frac{1}{N}$ dependence. 

\subsection{Thickness monitoring of MoS$_2$ flakes via Raman imaging: two pairs of Raman modes}
\label{sec:introduction}

\begin{figure}[h!]
 \centering
%\leavevmode
%\includegraphics[trim=0 0 20 25, scale=0.5]{fig_12.pdf}
\includegraphics[trim=0.5cm 4cm 0.5cm 2cm, width=0.65\textwidth]{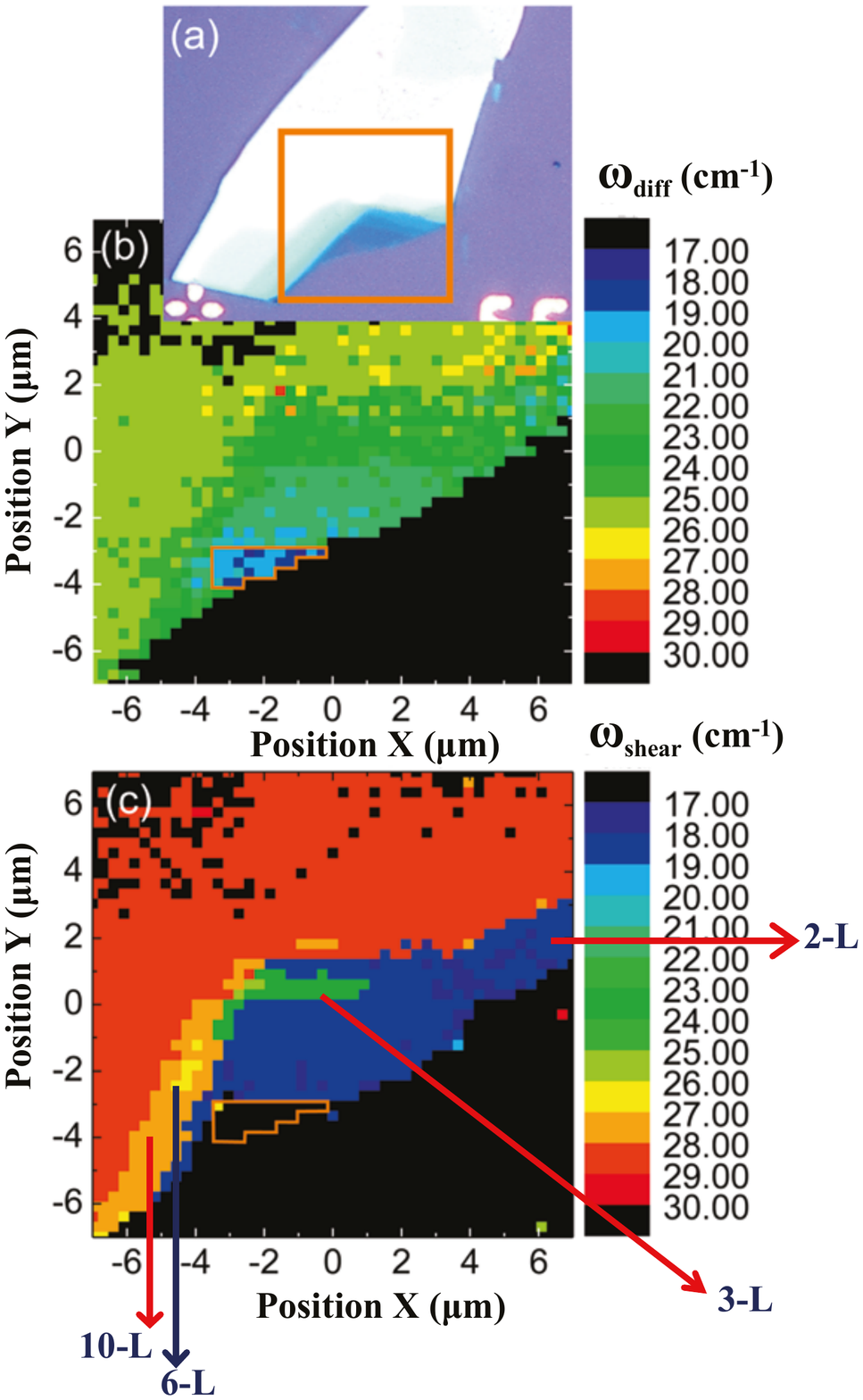}
  \caption{(color online)--(a) The optical image of different layers of MoS$_2$. The boxed area used for Raman imaging. (b) The Raman image taken by locking the $\omega_{diff}$= $\omega_{A_{1g}}$-$\omega_{E^1_{2g}}$. Single layer is marked by a boxed area. (c) Raman image recorded by locking the $\omega_{shear}$=$\omega_{E^2_{2g}}$. Here the color scales used for $\omega_{diff}$ and $\omega_{shear}$ are arbitrary. Taken from Ref. \cite{ple}.}
 \label{ple}
\end{figure}

From the above two subsections, we have seen that the frequencies of the four Raman active modes (2 -pair) change differently by sensing only the number of MoS$_2$ layers. The 2-pairs of optical phonons are $\{$A$_{1g}$ and E$^1_{2g}$ $\}$ and $\{$E$^2_{2g}$ and C-mode$\}$. Each member of the first pair shows opposite trend with the result that the difference $\omega_{A_{1g}}$ - $\omega_{ E^1_{2g}}$ shows a sizable change, which helps in Raman imaging. Since, the C-mode shows softening behavior and it has low frequency (15 cm$^{-1}$ for 7-layers) than the E$^2_{2g}$ mode (32 cm$^{-1}$ for 7-layers), the shear E$^2_{2g}$ mode is the feasible one from experimental point of view. Moreover, the shear mode shows a $\sim$ 50 $\%$ changes of peak positions going form bi-layer (22 cm$^{-1}$ ) to bulk value (33 cm$^{-1}$). Plechinger et al \cite{ple} recorded (shown in Fig.~\ref{ple}) Raman images of a few layers of MoS$_2$ by locking the frequency difference of two high frequency Raman modes ($\omega_{A_{1g}}$ - $\omega_{ E^1_{2g}}$) and a low frequency shear mode (E$^2_{2g}$). The Raman imaging of MoS$_2$ flakes (rectangular area marked in Fig.~\ref{ple}a) is shown in Fig.~\ref{ple}(b) for the frequency difference of $\omega_{A_{1g}}$-$\omega_{ E^1_{2g}}$. The single layer MoS$_2$ for which, $\omega_{A_{1g}}$ - $\omega_{ E^1_{2g}}$=18 cm$^{-1}$ (from Fig.~\ref{lee2}b), is shown by marked area with false color in Fig.~\ref{ple}(b). The signature of E$^2_{2g}$ mode on single layer is zero and is shown in Fig.~\ref{ple}(c) by a closed marked region. The bi-layer (2L), tri-layer (3L), hexa-layer (6L) and 10-layers (10L) and their boundaries are clearly seen by tracking the shear mode (in Fig.~\ref{ple}c), whereas the counterpart for the frequency difference is not clear as shown in Fig.~\ref{ple}(b).

\subsection{Shear and layer breathing modes of N-layers of MoS$_2$ $\{$ N = 1 $\rightarrow$ 19 $\}$}
\begin{figure}[h!]
 \centering
%\leavevmode
%\includegraphics[trim=0 0 20 25, scale=0.5]{fig_13.pdf}
\includegraphics[trim=0.5cm 12cm 0.5cm 2cm, width=0.8\textwidth]{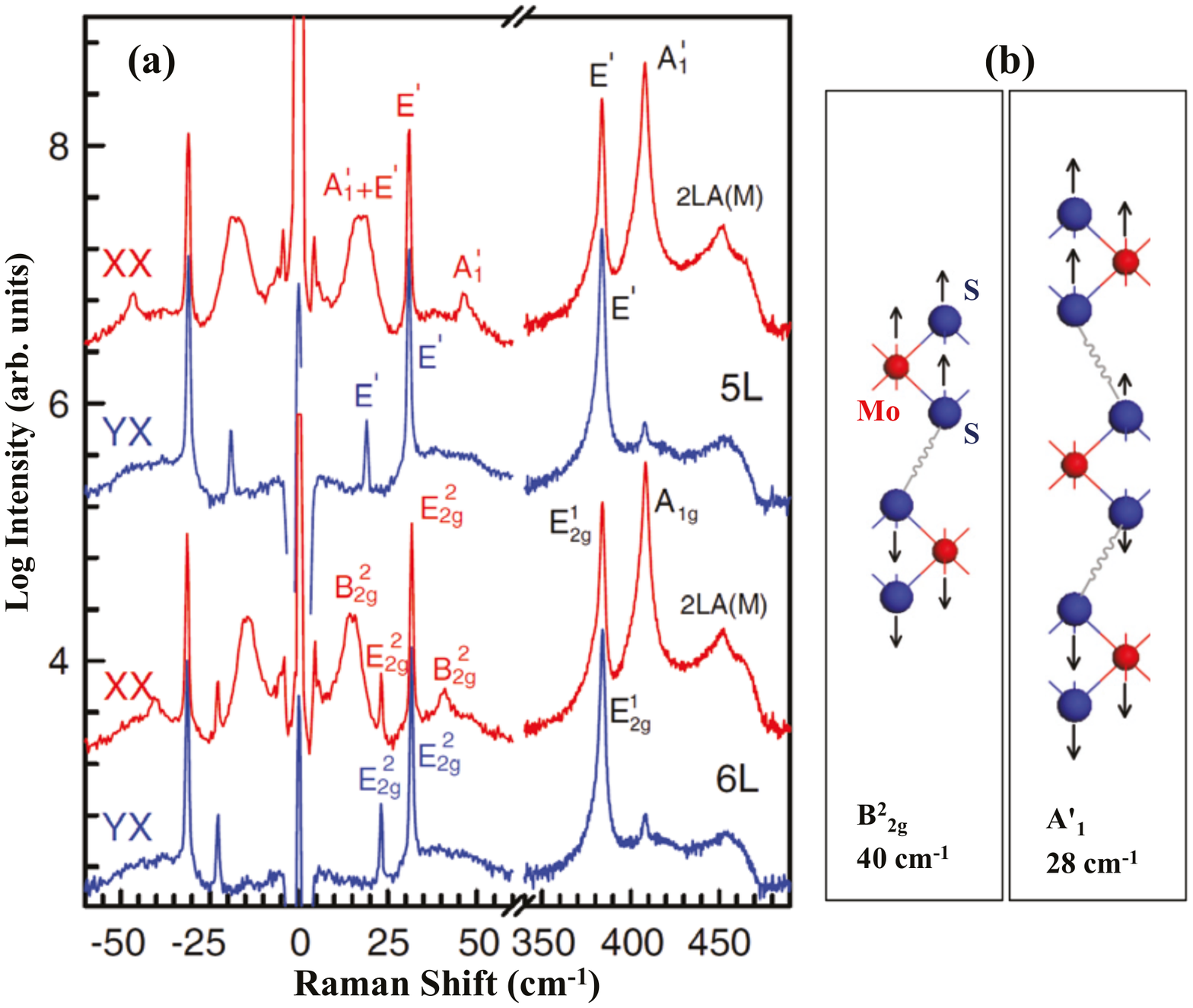}
  \caption{(color online)--(a) Polarized Raman spectra for 5L and 6L of MoS$_2$. YX represents the direction of incident laser polarization (Y) and the direction of analyzer's polarization (X); both lie in the basal plane. (b) The vibrational configurations of the atoms in A$_1^{'}$ and B$^2_{2g}$ modes. Taken from Ref. \cite{zhang}.}
 \label{zhang1}
\end{figure}

For the rigid layer vibrations of N-layer MoS$_2$, there are N-1 shear modes parallel to the basal plane (perpendicular to the c-axis) and N-1 layer breathing modes (LBMs) along the c-axis. For even N, LMBs are not Raman active and $\frac{N}{2}$ shear modes are Raman active (doubly degenerate). For odd, $\frac{N-1}{2}$ LBMs and N-1 shear modes are Raman active \cite{zhang}. Fig.~\ref{zhang1}(a) shows the newly observed low frequency shear and LBMs Raman modes for 5L and 6L with different polarizations \cite{zhang}. As stated earlier, the Raman active modes A$_1^{'}$ and E$^{'}$ for odd-layers (ONL-MoS$_2$) are the counterpart of the A$_{1g}$ and E$^2_{2g}$ for even-layers (ENL-MoS$_2$), respectively. The eigenmodes of A$_1^{'}$ and B$^2_{2g}$ are shown in Fig.~\ref{zhang1}(b). The B$^2_{2g}$ mode is Raman inactive and the reason of observation is discussed later. All the observed low frequency modes are listed in Table II.\\ 

\begin{table}
\caption{\label{tab:table2} List of low-frequency optical phonon modes for 5-layer and 6-layers of MoS$_2$ \cite{zhang}.}
\begin{ruledtabular}
\begin{tabular}{llllcrrrr}

5-Layers (5L)     &                  &                &                  &\vline    &6-Layers (6L)     &                       &                & \\ \hline\hline
 
Shear modes       & Exp.            &LBMs            & Exp.              &\vline    &Shear modes       &Exp.                  &LBMs             &Exp. \\ \hline

2E$^{'}$ (R, IR)  &19 $\&$ 33$^a$   &2A$_1^{'}$ (R) &17$^a$ $\&$ 47$^b$  &\vline    &3E$^2_{2g}$ (R)   &23$^a$ $\&$ 32 $^b$   &3B$^2_{2g}$ $^c$  &15 $\&$ 41$^a$ \\
$\{XX, YX\}$      &(in cm$^{-1}$)   &$\{XX\}$       &(in cm$^{-1}$)      &\vline    &$\{XX, YX\}$      &(in cm$^{-1}$)        &$\{XX\}$         &(in cm$^{-1}$) \\                                                                                                                                                    
2E$^{''}$ (R)     &   --            &2A$_2^{''}$ (IR) &  --              &\vline    &2E$_{1u}$ (IR)    &  --                   &2A$^2_{2u}$(IR)     & --\\     
$\{XZ, YZ\}$      &   --            & --             &  --               &\vline    & --               &  --                   & --             & --\\

\end{tabular}
{$^a$ Ref. \cite{zhang}. $^b$ Ref. \cite{zeng}. $^c$ This mode is a silent mode; it is neither Raman nor infrared (IR) active.}
\end{ruledtabular}
\end{table}

\begin{figure}[h!]
 \centering
\leavevmode
\includegraphics[trim=0.5cm 12cm 0.5cm 2cm, width=1.0\textwidth]{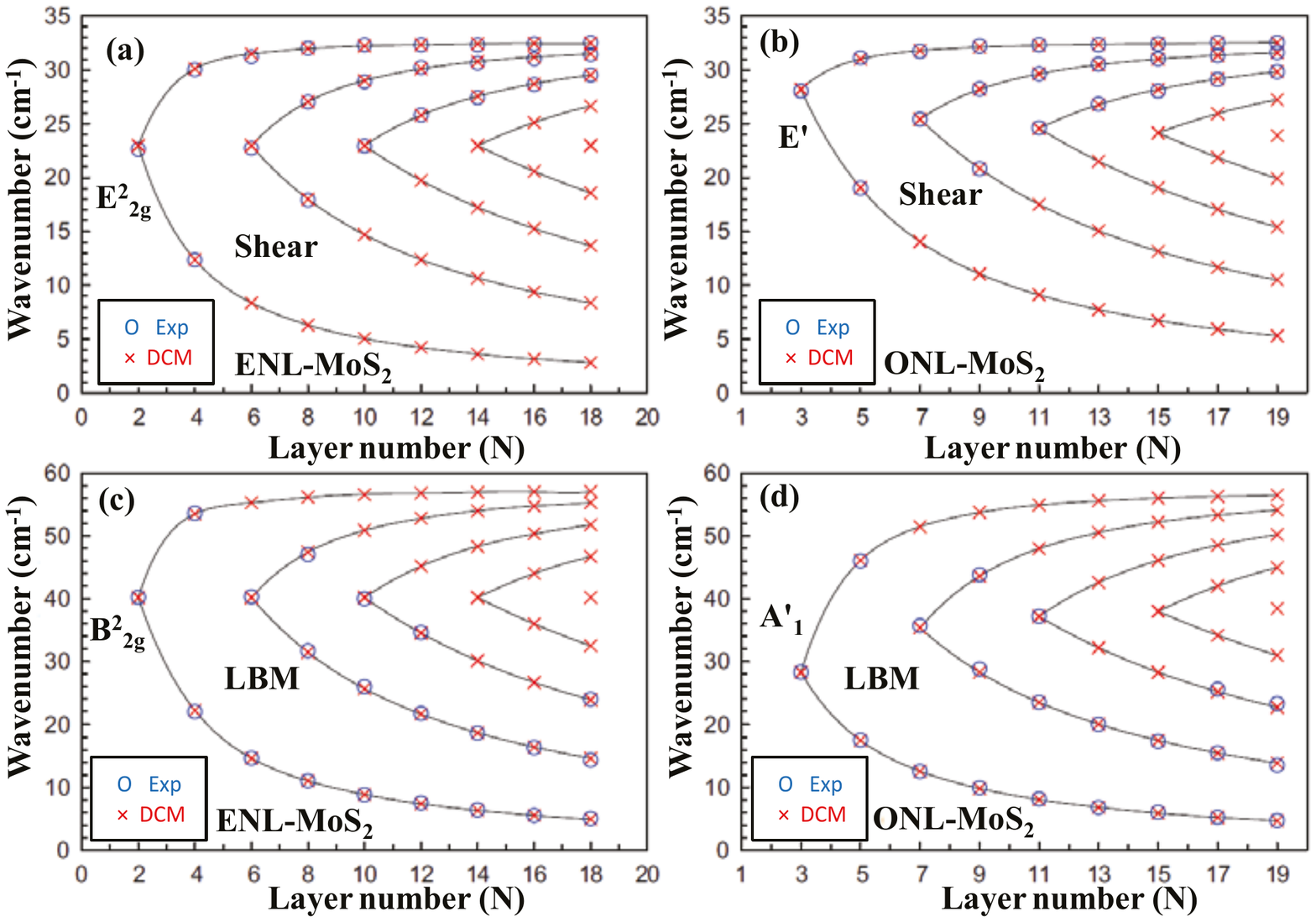}
  \caption{(color online)--Calculated (red cross) according to the diatomic chain model (DCM) and the experimentally observed frequencies or wavenumbers (blue circle) (a) with the even (ENL) and (b) odd number (ONL) of layers for the shear modes, and for the layer breathing modes (LBM) in (c) and (d). Here the connected solid lines are guide to the eye. Taken from Ref. \cite{zhang}.}
 \label{zhang4}
\end{figure}

To understand the low frequency rigid layer modes, the calculated \cite{zhang} frequencies with the diatomic chain model (DCM) and the experimentally observed ones for the ENL-MoS$_2$ and ONL-MoS$_2$ are shown in Fig.~\ref{zhang4}, separately. For the ONL-MoS$_2$, the shear mode E$^{'}$ (28 cm$^{-1}$ for 3L) stiffens and the one LBM A$_1^{'}$ (29 cm$^{-1}$for 3L) softens with increasing the thickness of the sample from 3L to 19L (see Figs.~\ref{zhang4} b and d). According to DCM calculations, a new mode appears as N reaches a value of 4N+3, where N=1,2,3,..... And that new mode splits into two branches such that one branch softens and other stiffens until N reaches a value of 4N+3 and so on. The similar softening and stiffening behavior of a shear mode E$^2_{2g}$ (23 cm$^{-1}$ for 2L) and a LBM mode B$^2_{2g}$ (40 cm$^{-1}$ for 2L) for ENL-MoS$_2$ are shown in Figs.~\ref{zhang4} (a) and (b), respectively. For ENL-MoS$_2$, the appearance and splitting of a new mode into two branches follows according to 4N+2. The DCM calculated value matches well with the experimentally observed Raman inactive B$^2_{2g}$ mode and this is the reason, along with the polarization behavior that Zhang et al \cite{zhang} named it as a B$^2_{2g}$ mode. Although, the B$^2_{2g}$ for ENL-MoS$_2$ follows exactly the same polarization behavior as A$_1^{'}$ for ONL-MoS$_2$ (see Fig.~\ref{zhang1}a), the deeper understanding of its appearance, polarization behavior and matching with the DCM calculated value needs further work. The most striking features from Figs.~\ref{zhang4} is that the experimentally 
observed shear modes are coming from the upper branch of cone like curve with stiffening character and LBMs are from the lower branch with softening feature for both the ENL-MoS$_2$ and ONL-MoS$_2$. The ratio of the shear E$^2_{2g}$ mode for bulk to 2L is given by $\omega_{Bulk} / \omega_{2L}$=32.7/22.6=1.447 \cite{zhang}, which is very close to the value of graphene ($\omega_{Bulk} / \omega_{2L}=\sqrt 2$) \cite{tan}.

%FIG-16

\begin{figure}[h!]
 \centering
%\leavevmode
%\includegraphics[trim=0 0 20 25, scale=0.5]{fig_15.pdf}
\includegraphics[trim=0.5cm 22cm 0.5cm 2cm, width=1.0\textwidth]{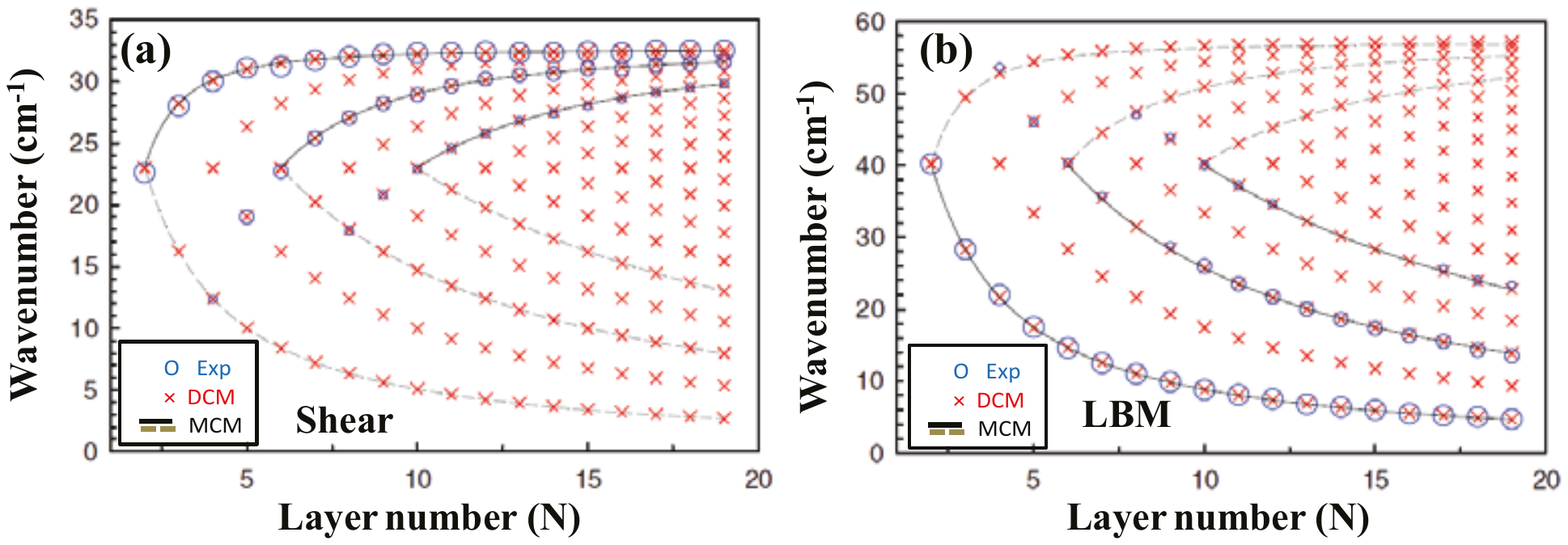}
  \caption{(color online)--Calculated (red cross) according to the diatomic chain model (DCM) and the experimentally observed frequencies or wavenumbers (blue circle) of all the layers together for the shear modes in (a) and the LBM modes in (b). Size of the circle represents the observed Raman intensity. The black solid lines (in a and b) are fitted curves according to the monatomic chain model (MCM) for those modes generating from 2L ($N_0$=1), 6L ($N_0$=3) and 10L ($N_0$=5) branches. Another set of modes (shear and LBMs) according to the MCM model are shown by the gray dashed lines in (a) and (b). Taken from Ref. \cite{zhang}.}
 \label{zhang5}
\end{figure}
%The fitting has been done for the shear modes using $\omega_{shear}(N)=\omega_{shear}(2)\sqrt{1 + cos(N_0 \pi/N)}$ and for the LBM modes using $\omega_{LBM}(N)=\omega_{LBM}(2)\sqrt{1 - cos(N_0 \pi/N)}$. Here $ N \geq 2N_0$.

In general, in-order to know the exact trend of softening and stiffening features of low frequency rigid layer modes of layered materials, a reduced monatomic chain model (MCM) has been considered \cite{zhang}. The DCM calculated values showed that the relative displacements between Mo atoms and S atoms are so less ($\sim 0.6 \%$ for 2L) \cite{zhang} and it decreases with increasing N. Therefore, for rigid layer modes, the relative displacements can be ignored and hence, in MCM model the additive mass ($M_{add}=M_{Mo}+2M_{S}$, instead of a reduced mass $\frac{1}{\mu}=\frac{1}{M_{Mo}}+\frac{1}{2M_{S}}$) is taken. All the observed shear modes for ENL-MoS$_2$ (E$^2_{2g}$) and ONL-MoS$_2$ (E$^{'}$) from Figs.~\ref{zhang4}(a) and (b) are plotted together in Fig.~\ref{zhang5}(a) with the DCM and MCM calculated values \cite{zhang} ; and Fig.~\ref{zhang5}(b) for all the LBMs. The interesting feature from Fig.~\ref{zhang5} is that all the experimentally observed low frequency modes originate from  ENL-MoS$_2$ i.e. 2L, 4L, 6L and 10L. The black solid lines and the gray-dashed lines are the fitted curves which, follows $\omega(N)=\omega(2N_0)\sqrt{1 \pm cos(N_0 \pi/N)}$ (+ and - sign for upper and lower branch, respectively ) such that $ N \geq 2N_0$ and $N_0$=1,2,3,..etc. Since, $\omega(2N_0)$ is almost equal to $\omega(2)$ for all rigid layer modes, the fitted frequency obeys $\omega(N)=\omega(2)\sqrt{1 \pm cos(N_0 \pi/N)}$ formula according to MCM. Here, the model (MCM) did not take into account the interactions between one layer of MoS$_2$ and the supported substrate and explains well the experimentally observed frequencies; whereas in contrast, Zeng et al \cite{zeng} included the substrate effect and showed the 1/N behavior of one LBM as shown in Fig.~\ref{zeng1}(c). It was shown that for suspended multilayer graphene, the scaling of the shear mode with thickness follows the trend predicted by MCM calculations \cite{tan}. Recently, suspended 2L-MoS$_2$ shows no substrate effect on the rigid layer phonon frequencies and their FWHMs \cite{zhao}. It would, therefore, be reasonable to assume that the interaction between substrate and MoS$_2$ is not responsible for the observed 1/N scaling with the thickness \cite{zhang}.

\begin{figure}[h!]
 \centering
%\leavevmode
%\includegraphics[trim=0 0 20 25, scale=0.5]{fig_16.pdf}
\includegraphics[trim=0.5cm 18cm 0.5cm 2cm, width=1.0\textwidth]{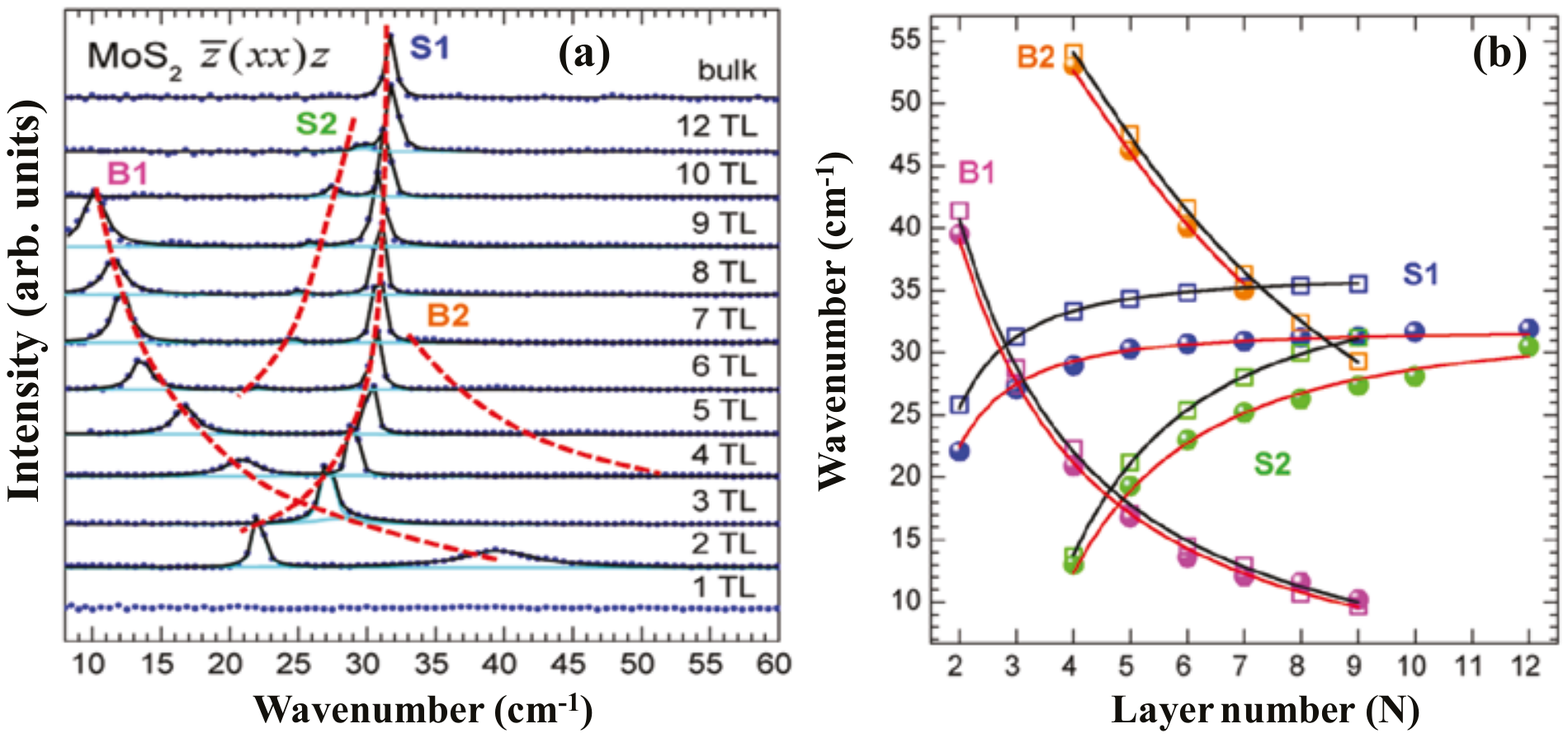}
  \caption{(color online)--(a) Raman spectra of few layers and bulk  MoS$_2$ for the low frequency shear (S) and layer breathing modes (B). Here 1TL means one S-Mo-S unit i.e. one layer (1L). The dashed lines are guide to the eye. (b) The variations of the measured (solid dots) and the calculated frequencies (open squares) from first-principles. Solid lines represent the fitting of the data according to the linear chain model. Taken from Ref. \cite{zhao}.}
 \label{zhao}
\end{figure}

Recently, two new low frequency rigid layer modes are observed \cite{zhao} (Fig.~\ref{zhao}a): one is LBM (B2) and another is a shear mode (S2). Both modes appear from 4L-MoS$_2$ such that B2 mode softens from 54 cm$^{-1}$ (4L) to 35 cm$^{-1}$ (7L) and S2 mode stiffens from 13 cm$^{-1}$ (4L) to 30 cm$^{-1}$ (12L). The frequency variation with the thickness of the sample is shown in Fig.~\ref{zhao}(b). The fitted solid curves are according to linear chain model calculations \cite{zhao}. S2 mode shows softening behavior because, the out of phase displacement between layers increases with N \cite{zhao}. B2 mode stiffens as in-phase motion increases with the thickness N. The appearance of B2 LBM mode for even layers of MoS$_2$ (4L) needs further studies as, according to Zhang et al \cite{zhang} for even layers, there should be no Raman active LBM.

\section{Layer dependence of optical Raman modes: resonance}

\begin{figure}[h!]
 \centering
%\leavevmode
%\includegraphics[trim=0 0 20 25, scale=0.5]{fig_17.pdf}
\includegraphics[trim=0.5cm 16cm 0.5cm 2cm, width=1.0\textwidth]{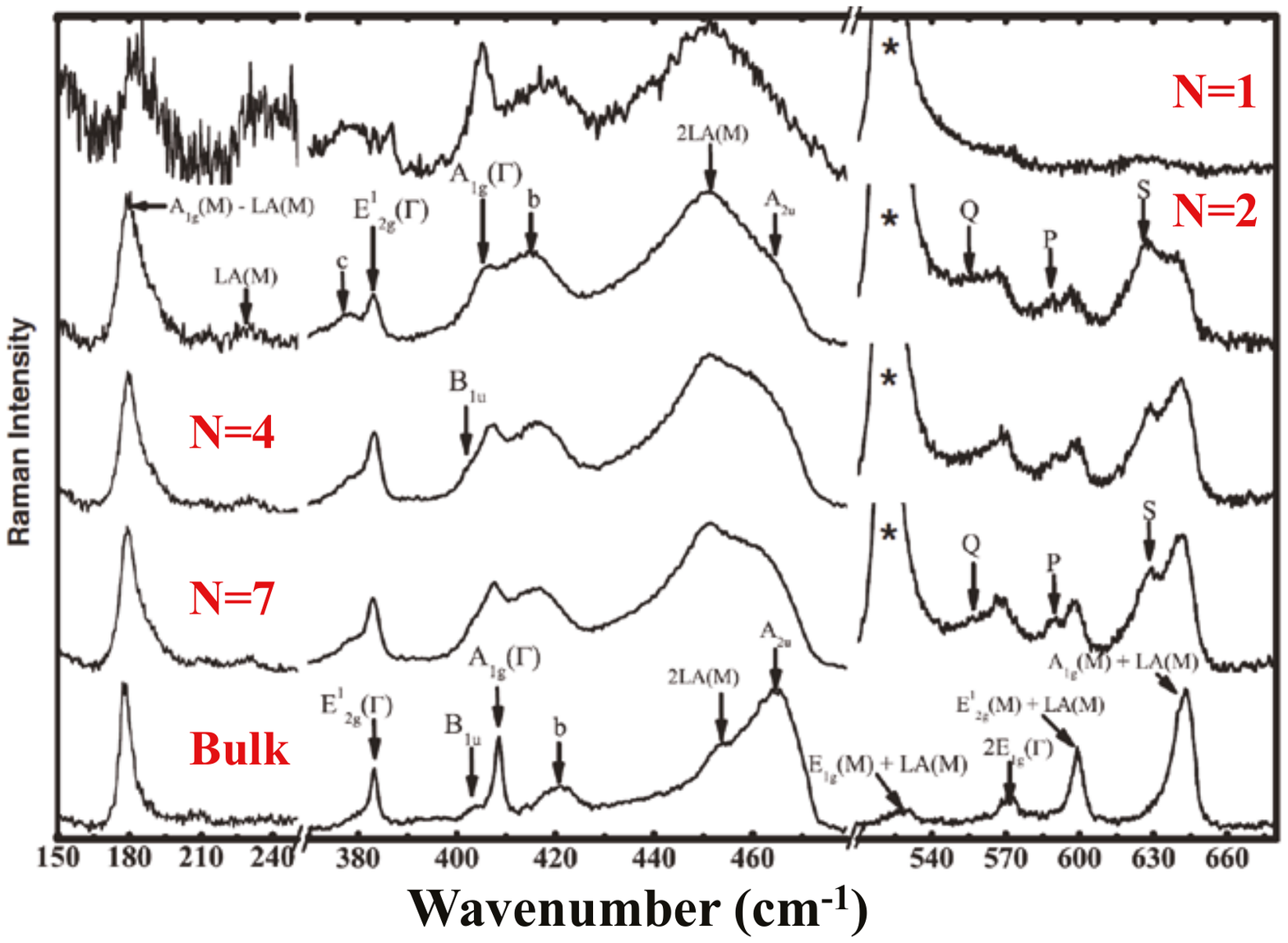}
  \caption{(color online)--Resonant Raman spectra of few layers and bulk MoS$_2$ using 633 nm laser line. The strong mode coming from the SiO$_2$/Si substrate (marked by *) is absent for thick samples (bulk). Here N represents the number of layers. Taken from Ref. \cite{bisu}.}
 \label{bisu1}
\end{figure}

\begin{figure}[h!]
 \centering
%\leavevmode
%\includegraphics[trim=0 0 20 25, scale=0.5]{fig_18.pdf}
\includegraphics[trim=0.5cm 6cm 0.5cm 2cm, width=0.8\textwidth]{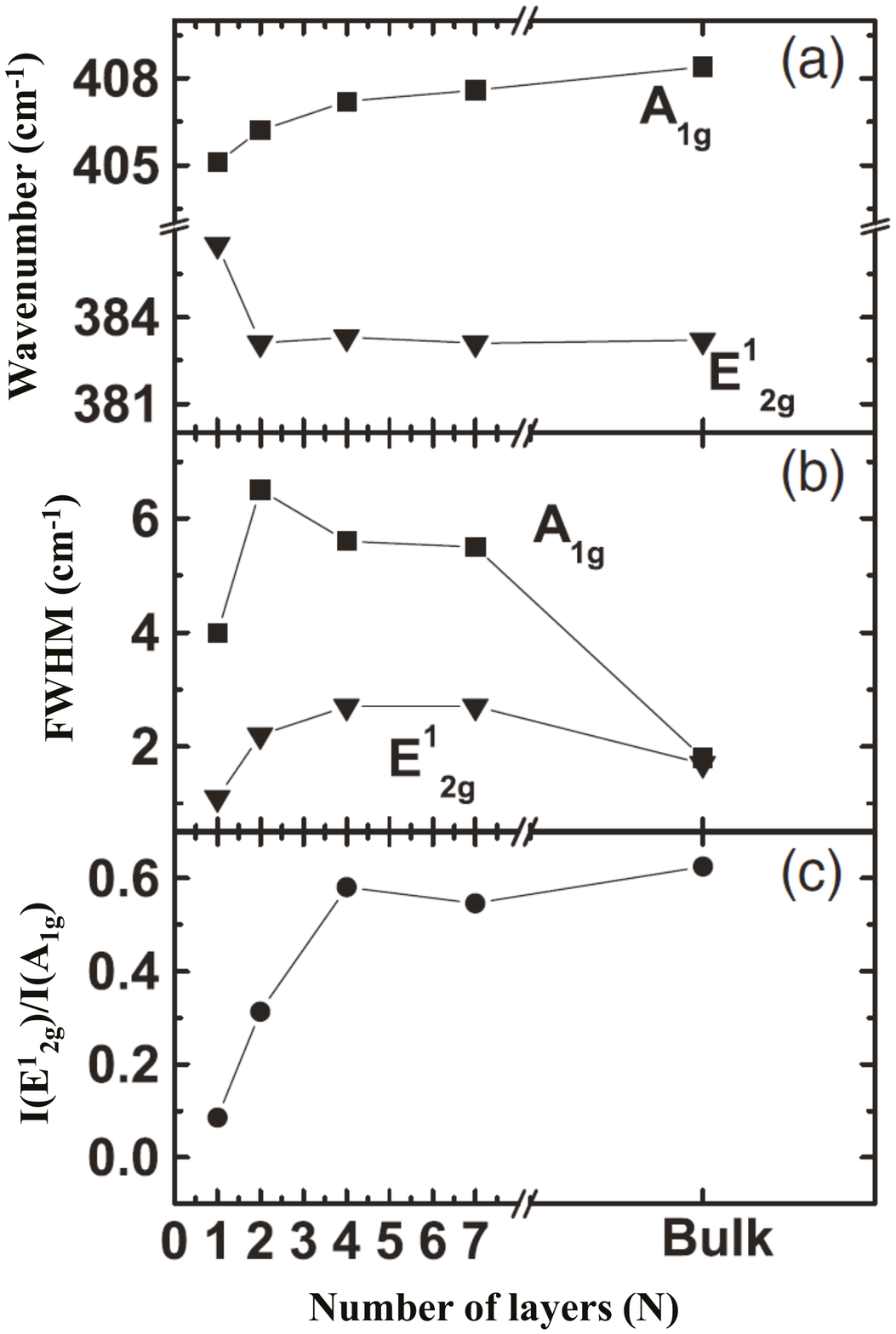}
  \caption{(a) Raman shifts and (b) the line widths (FWHM) of the A$_{1g}$ and the E$^1_{2g}$ modes as a function of the number of layers. (c) The integrated intensity ratio I(A$_{1g}$)/I(E$^1_{2g}$). Taken from Ref. \cite{bisu}.}
 \label{bisu2}
\end{figure}

Resonance Raman scattering (RRS) occurs for a system, when the incident laser excitation energy is close to the electronic band gap. In this section, we will discuss the RRS with the laser energy 1.9 eV (633 nm), which matches with the direct band gap at the K-point of the BZ. Fig.~\ref{bisu1} shows the resonant Raman spectra \cite{bisu} of different layers (N=1L, 2L ,4L and 7L) and bulk MoS$_2$. Since, the RRS enhances the possibility of higher order Raman scattering (multiphonon process) compared to the first order process in normal Raman scattering, Fig.~\ref{bisu1} shows many Raman peaks along with the off-resonant Raman modes. Figs.~\ref{bisu2}(a), (b) and (c) show the variations of frequencies, FWHMs and integrated intensity ratio of two Raman modes A$_{1g}$ and E$^1_{2g}$, respectively for the RRS \cite{bisu}; which show the same trend as for the off-resonant case (see Figs.~\ref{lee2}b, and ~\ref{lee3}). The new Raman bands, which were not observed with the 514 nm and 533 nm laser lines, are understood as follows.

\subsection{New modes in low frequency region of 160 - 230 cm$^{-1}$}
(i) A Raman band $\sim$ 179 cm$^{-1}$ is observed and this mode is assigned as a difference combination mode of A$_{1g}$(M)-LA(M) \cite{chen,stacy}. The frequency of this mode does not show layer dependent feature. The three Raman active (A$_{1g}$, E$_{1g}$ and E$^1_{2g}$) phonons at $\Gamma$ point of BZ are almost dispersionless along the $\Gamma$-M direction; the frequency of rigid layer mode E$^2_{2g}$ increases with q (phonon wave-vector) and reaches to a longitudinal acoustic (LA) mode at M (233 cm$^{-1}$) \cite{stacy}. At M point of BZ, the dispersion of LA mode shows an inverse parabolic nature, i.e. on either side of M-point, the frequency decreases. The asymmetric nature of the peak on the higher side is due to the involvement of the LA(M) mode as a difference combination band \cite{stacy}. (ii) Another band at $\sim$ 233 cm$^{-1}$ is observed for 2L, 4L and 7L and is assigned \cite{souri} as LA(M). The activation of this first order mode implies the relaxation of the q $\sim$ 0 selection rule. The appearance of the zone boundary phonon (LA) for MoS$_2$ nanoparticle is attributed to the defect induced Raman scattering \cite{frey}. In the present case, the observed LA(M) mode for 2L, 4L and 7L MoS$_2$ could be due to the structural defects \cite{bisu}.

\begin{figure}[h!]
 \centering
%\leavevmode
%\includegraphics[trim=0 0 20 25, scale=0.5]{fig_19.pdf}
\includegraphics[trim=0.5cm 4cm 0.5cm 2cm, width=0.8\textwidth]{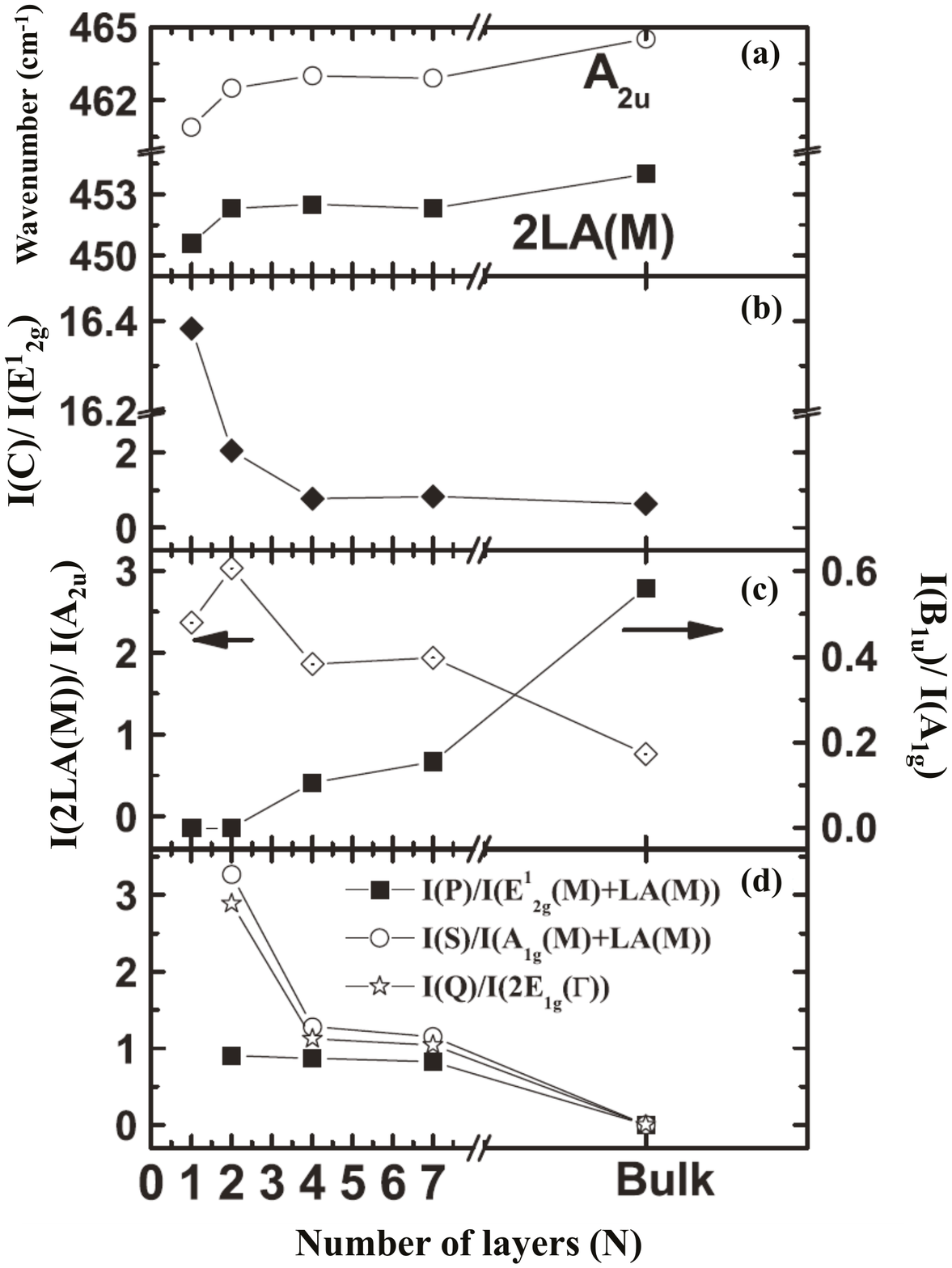}
  \caption{(a) Raman shifts of the 2LA and the A$_{2u}$ modes as a function of the number of layers. (b)--(d) The changes of the integrated intensity ratios for different modes. Taken from Ref. \cite{bisu}.}
 \label{bisu3}
\end{figure}

\subsection{Frequency region of 360 - 430 cm$^{-1}$}

\begin{figure}[h!]
 \centering
%\leavevmode
%\includegraphics[trim=0 0 20 25, scale=0.5]{fig_20.pdf}
\includegraphics[trim=0.5cm 10cm 0.5cm 2cm, width=0.8\textwidth]{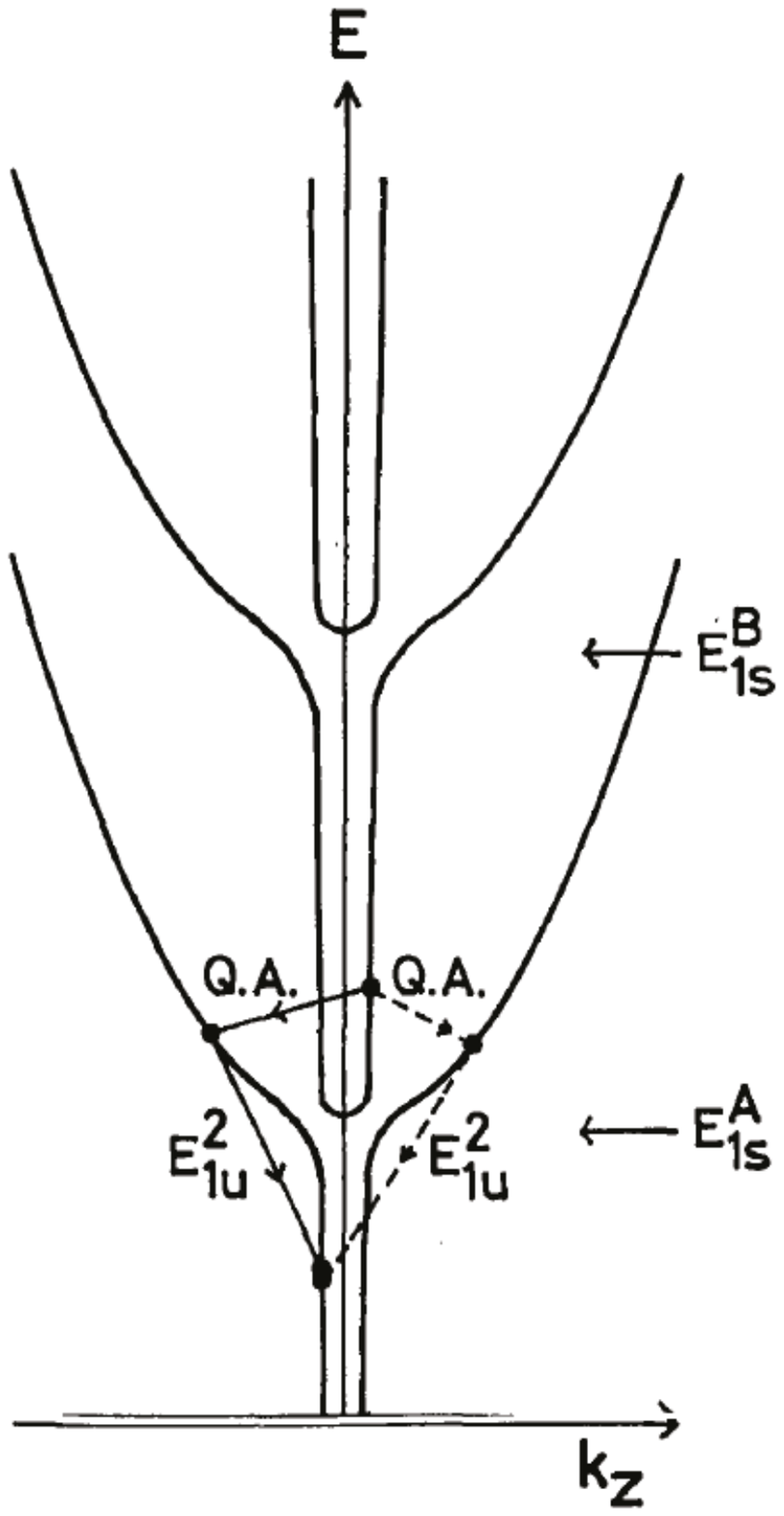}
  \caption{(color online)--Schematic diagram for the dispersion curve of a polariton and the two phonon Raman scattering involving a longitudinal quasi-acoustic (Q.A.) mode and the E$^2_{1u}$ mode. The dashed and the solid lines marked by arrows represent the $\alpha$ and $\beta$ processes, respectively. Here E$^A_{1s}$ and E$^B_{1s}$ represent the energy of 1s level for the A and B excitons, respectively. Taken from Ref.  \cite{sekine}.}
 \label{sekine}
\end{figure}

There are three new Raman active bands in the frequency range of 360 - 430 cm$^{-1}$. (i) a band named as $^{'}$c$^{'}$ $\sim$ 377 cm$^{-1}$ (shown in Fig.~\ref{bisu1}) becomes sharper with increasing the layer from 1L to 7L. Sekine et al \cite{sekine} assigned this mode as E$^2_{1u}$ mode. Since, the experimentally observed E$^2_{1u}$ mode has higher frequency than the E$^1_{2g}$ mode \cite{verbel,ghosh,liv} and these are Davydov pair \cite{sekine}, the observed c mode cannot be assigned as a E$^2_{1u}$ mode \cite{bisu}. The c mode does not show thickness dependent frequency variations. The integrated intensity ratio I(c)/I(E$^1_{2g}$) decreases with the number of layers as shown in Fig.~\ref{bisu3}(b). (ii) One peak $\sim$ 409 cm$^{-1}$ evolves with number of layers and is assigned as B$_{1u}$ mode \cite{liv,ghosh}. The B$_{1u}$ and A$_{1g}$ mode are the small frequency splittings of Davydov pair \cite{sekine} and the appearance of the B$_{1u}$ mode is due to the resonance effect. Fig.~\ref{bisu3}(c) shows that the integrated intensity ratio I(B$_{1u}$)/I(A$_{1g}$) increases with the layers of the sample. (iii) The prominent feature at $\sim$ 420 cm$^{-1}$, marked as $^{'}$b$^{'}$ in Fig.~\ref{bisu1}, is observed under resonant conditions. This mode is assigned as a two-phonon process and the appearance of this mode is related to the A exciton ($\sim$ 1.9 eV) at the K-point \cite{souri,frey,sekine}. The dispersion curve for polariton-exciton is shown in Fig.~\ref{sekine}. Here, the intermediate polariton state involves the two phonons under resonant condition. First, the photon-like state from the high energy upper branch (inner branch) scatters to  exciton-polariton state (outer branch) by scattering a longitudinal optical phonon (this dispersive quasi-acoustic phonon is the silent B$^2_{2g}$ mode at the $\Gamma$ point) and this interaction is due to the deformation potentials; second, the polariton state decays to photon-like state by scattering the transverse optical phonon (with E$^2_{1u}$ symmetry) \cite{sekine}. This mode is a combination band, $\omega_b=\omega_{QA}+\omega_{TO}$ \cite{bisu}.

%FIG-23
%The frequency difference between the two processes ($\alpha$ and $\beta$) is less.

\subsection{Frequency region of 440 - 490 cm$^{-1}$}
(i) The mode $\sim$ 450 cm$^{-1}$ (for bulk) is attributed to second order Raman (2LA(M)) \cite{stacy,frey,sekine}, and (ii) the mode $\sim$ 466 cm$^{-1}$ (for bulk) is assigned as A$_{2u}$ \cite{frey,waka}. The peak positions of the two modes (2LA and A$_{2u}$) increases with the number of layers (N) as shown in Fig.~\ref{bisu3}(a). Fig.~\ref{bisu3}(c) shows the decrease of the integrated intensity ratio I(2LA)/I(A$_{2u}$) with the thickness \cite{frey,bisu}.

\subsection{Frequency region of 510 - 630 cm$^{-1}$}

For the bulk MoS$_2$, (i) the observed mode $\sim$ 526 cm$^{-1}$ is assigned as E$_{1g}(M)+2LA(M)$, (ii) $\sim$ 571 cm$^{-1}$ mode as 2E$_{1g}(\Gamma$), (iii) $\sim$ 599 cm$^{-1}$ as E$^1_{2g}(M)+LA(M)$ and (iv) $\sim$ 642 cm$^{-1}$ as A$_{1g}(M)+LA(M)$ \cite{frey}. In all the multiphonon processes involving LA(M) mode, the sum combination band has asymmetric tail at lower side \cite{stacy}. All the new high frequency bands $\sim$ 554 cm$^{-1}$ (marked as Q in Fig.~\ref{bisu1}), $\sim$ 588 cm$^{-1}$ (marked as P) and $\sim$ 628 cm$^{-1}$ (marked as S) are observed under the resonance condition for 2L, 4L and 7L MoS$_2$ \cite{bisu} and were not observed for bulk MoS$_2$ and nanoparticles of MoS$_2$ \cite{frey}. The peak positions of these modes do not show variations with N. Fig.~\ref{bisu3}(d) shows the variations of the relative integrated intensity of the bands (Q, P and S) with respect to the neighboring band. 

\begin{figure}[h!]
 \centering
%\leavevmode t
%\includegraphics[trim=0 0 20 25, scale=0.5]{fig_21.pdf}
\includegraphics[trim=0.5cm 15cm 0.5cm 2cm, width=0.8\textwidth]{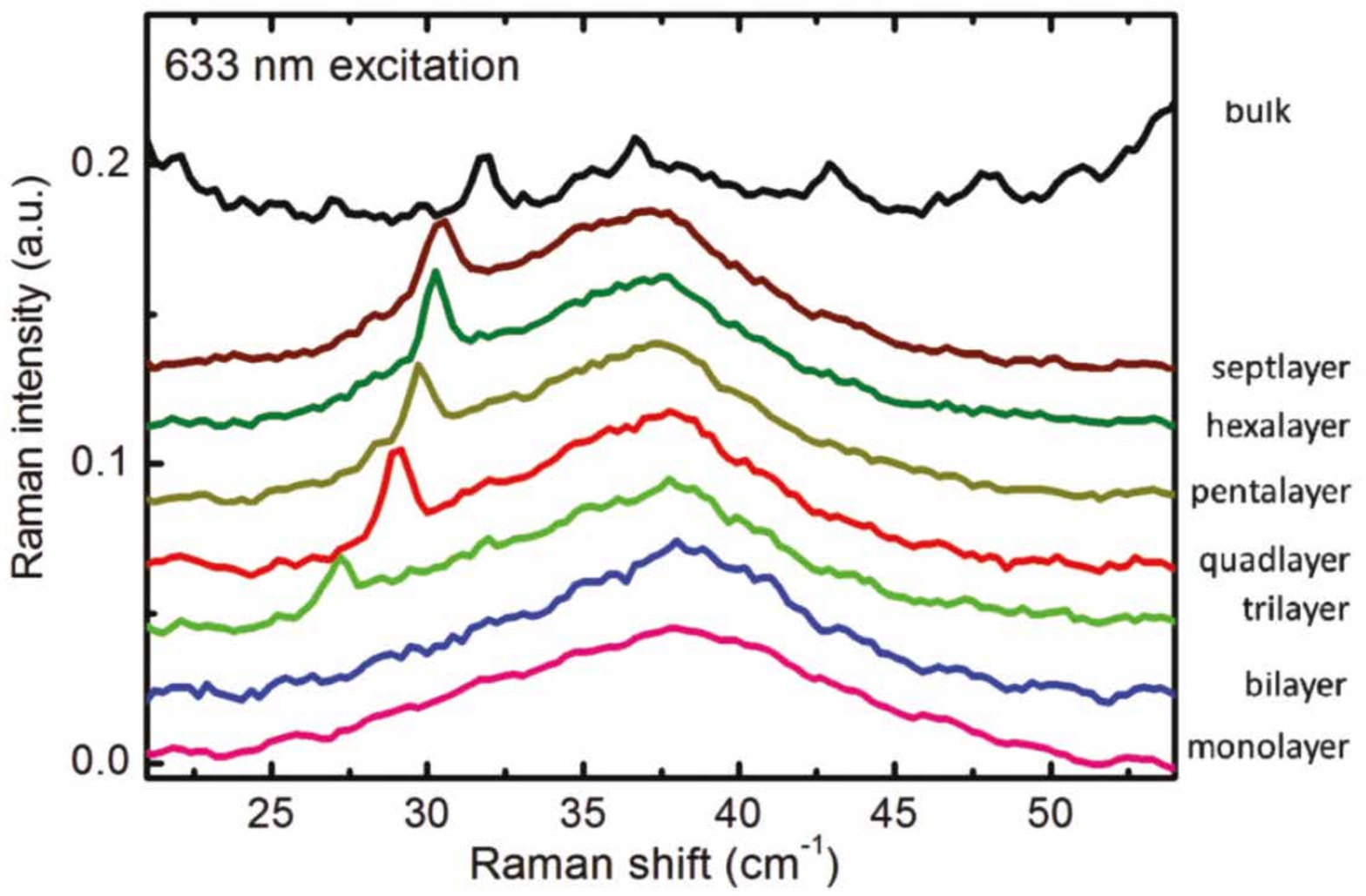}
  \caption{(color online)--Resonant Raman spectra using 633 nm laser line for samples of different thickness. The new broad feature $\sim$ 38 cm$^{-1}$ appears only under resonant condition. Taken from Ref.  \cite{zeng}.}
 \label{zeng3}
\end{figure}

Fig.~\ref{zeng3} shows another new broad feature $\sim$ 38 cm$^{-1}$ (5 meV) \cite{zeng} observed under the resonant condition for the cross polarization geometry. The peak position and the shape of the mode do not change with the layer numbers. Since, the binding energy of the excitons (A and B with energies 42 and 134 meV, respectively \cite{hoorn}) are much greater than the observed peak with 5 meV, the appearance of this mode is not responsible for the exciton mechanisms. Defects and impurity effects are also excluded due to having energy comparable to the excitons in optical measurements \cite{zeng}. The oscillating collective modes (plasmons, charge density waves) are also excluded due to the reluctant behavior of the position and shape of the peak; as the long-range Coulomb interaction would be screened with the layer thickness \cite{zeng}. The spin-orbit coupling is responsible for the splitting near the top of the valence bands at the K-point of BZ, which leads to the origin of A and B excitons. Zeng et al \cite{zeng} argued that the same spin-orbit coupling mechanisms also splits the conduction band at the K-point of $\sim$ 4 meV, which leads to the observed broad feature ($\sim$ 38 cm$^{-1}$) associated with the resonant Raman scattering. The understanding of the appearance of all the new Raman modes under resonant condition requires further theoretical work.

\section{Transport characteristics as a function of temperature and carrier concentration: various aspects towards improvement of mobility}
In technological applications of nanodevices, the two-dimensional (2D) layered-materials have a lot of potential due to easy fabrication. The highest mobility observed for the suspended graphene device is larger than 10$^5$ cm$^2$/ V-sec \cite{bolo}. Due to the lack of an electronic band gap in single layer graphene, the 2D semiconducting materials like MoS$_2$ (on-off ratio is 10$^8$ \cite {radi}) family compounds has potential applications in digital electronics with low power dissipation. In order to replace silicon based logic-devices, a high current on-off ratio $\sim$ 10$^4$ to 10$^7$ \cite{bolo} and the electronic band gap larger than 400 meV \cite{schw} are reasonable. The various scattering mechanisms (intrinsic and extrinsic effects), which are responsible for the momentum relaxations and thereby decreasing the mobility of charge carriers in MoS$_2$, are discussed in the following sections.

\subsection{Temperature dependence of carrier mobility in bulk-MoS$_2$}
Bulk MoS$_2$ is a layered material in which each individual 2D layers are stacked along the z-direction via van der Waals interactions. To capture this anisotropic feature, a potential model was invoked \cite{mooser} such that, total potential V($\textbf{r}$)=V(x,y)+V(z); where V(x,y) and V(z) represent the potential in basal plane and perpendicular to the basal plane, respectively. According to this model, the excess carriers feel a series of parallel potential wells along the perpendicular direction and between each of them, there exist local energy levels for the carriers which depend upon the local width of the wells. This implies that the localization energy within each layer depends on the layer thickness. The variation of the layer thickness associated with thermal phonons may give a friction to the carriers. The energy of the excess carriers (due to decoupled potential) is given by E$(\textbf{p})=\frac{p_x^2+p_y^2}{2m_x}+2I_z(cos\frac{p_z d_z}{\hbar}-1)$ \cite{fiva}; where $m_x$ is the effective mass in basal plane, $d_z$ the inter-layer spacing, $I_z$ small overlap energy between adjacent layers and the second term of E($\textbf{p}$) corresponds to the tight binding model. The corresponding density of states (DOS) is given by $D(E)=\frac{m_x N_z}{2 \pi^2 \hbar^2} cos^{-1}(1-E/2I_z)$ for $0\leq E\leq 4I_z$ and $D(E)=\frac{m_x N_z}{2 \pi^2 \hbar^2}$ for $E\geq 4I_z$ \cite{fiva}; where $N_z$ denotes the number of layers/length along the z-direction. The DOS is almost constant except for the low energy range, which is the characteristic feature for 2-dimensional electron gas.

In most isotropic semiconductors (non-polar), the high temperature mobility is limited by the quasi-elastic scattering with acoustic phonons via deformation potentials \cite{cardona}. The acoustic phonon mode corresponds to local dilatations of the lattice. Since, between two adjacent layers there is a weak  van der Waals force, the change in deformation potentials (CDP) due to the acoustic modes polarized in the basal plane will be more than that by the vertically polarized ones. Therefore, in layered semiconductors, the carriers will be scattered significantly by the horizontally polarized acoustic phonons. The calculated mobility due to the deformation potential scattering by acoustic modes is $\mu_{bulk}^{acoustic} \propto T^{-\gamma^{acoustic}}$ \cite{fiva}; where $\gamma^{acoustic}=1$ which is quite different than the well known value of $\gamma^{acoustic}=3/2$ for isotropic case because of peculiar behavior of DOS.\\

\begin{figure}[h!]
 \centering
%\leavevmode t
%\includegraphics[trim=0 0 20 25, scale=0.5]{fig_22.pdf}
\includegraphics[trim=0.5cm 8cm 0.5cm 2cm, width=1.0\textwidth]{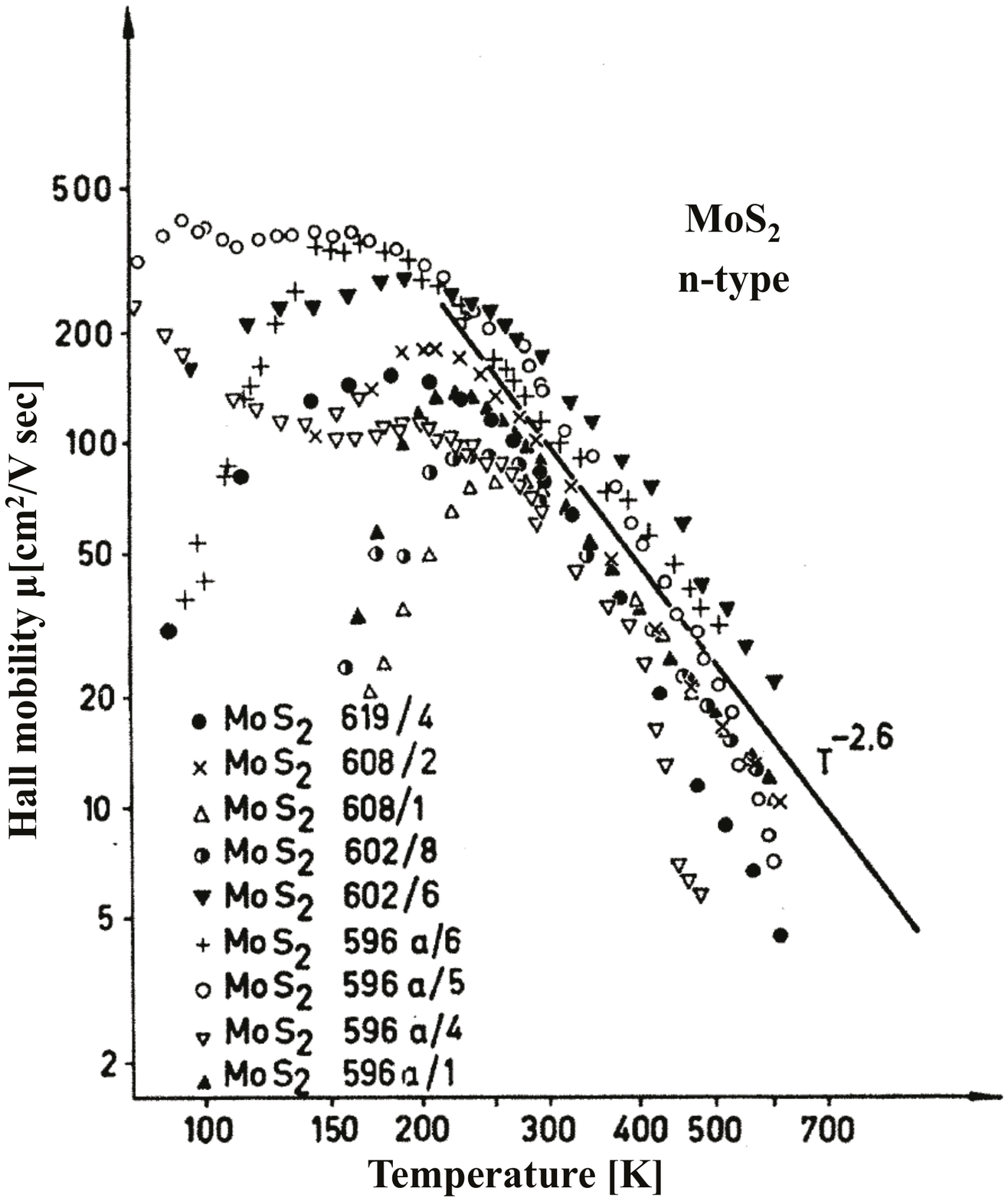}
  \caption{Hall mobility versus temperature for the n-type bulk-MoS$_2$. The solid line represents the fitting to the observed data. At low temperature, the increment of mobility is due to extrinsic properties. Taken from Refs. \cite{mooser,fiva}.}
 \label{fiva1}
\end{figure}

\begin{figure}[h!]
 \centering
%\leavevmode t
%\includegraphics[trim=0 0 20 25, scale=0.5]{fig_23.pdf}
\includegraphics[trim=0.5cm 12cm 0.5cm 2cm, width=1.0\textwidth]{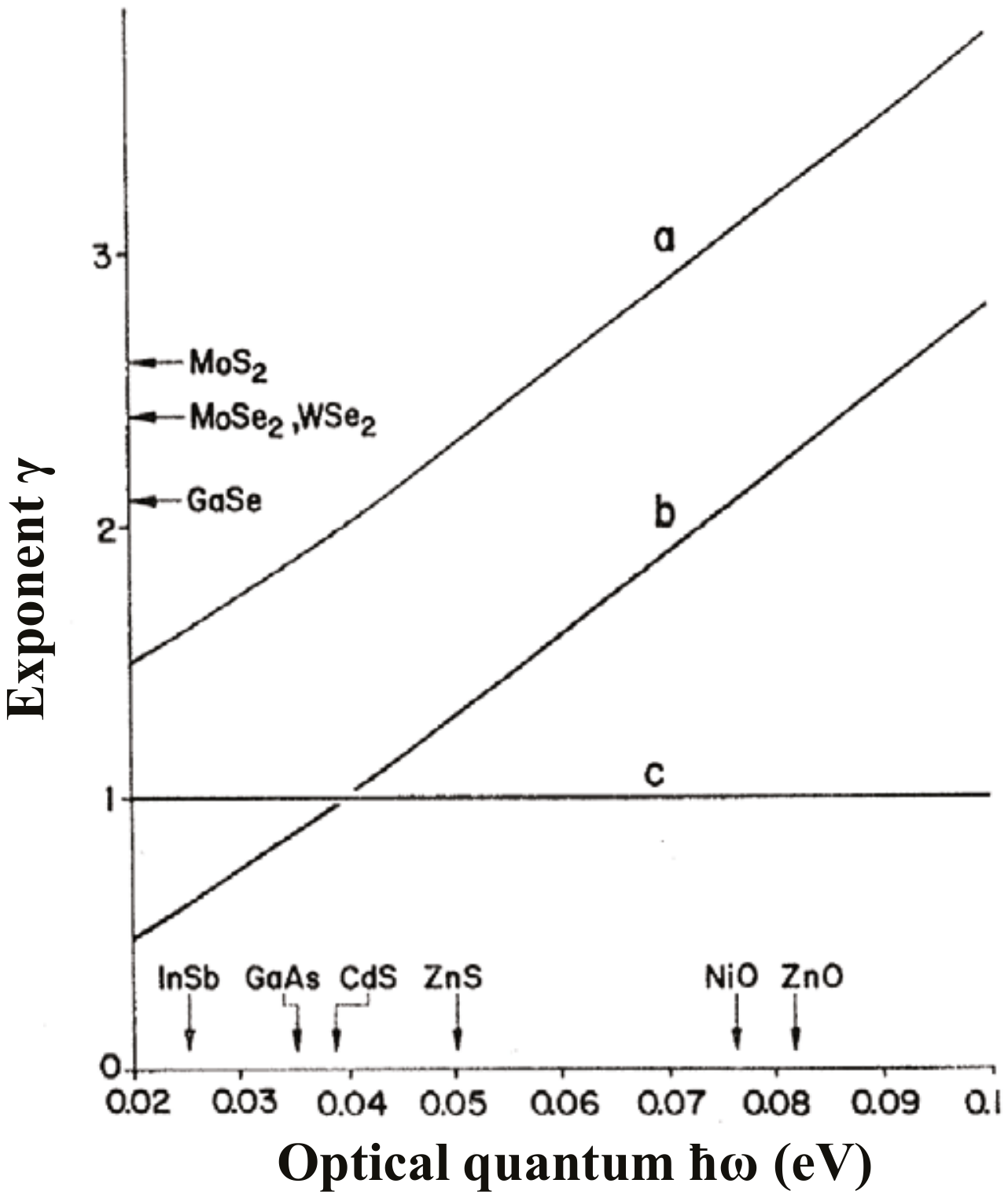}
  \caption{The exponent ($\gamma$) for mobility as a function of phonon energies ($\hbar \omega$). Exponent is given by $\mu_{bulk}=\mu_0 (T/T_0)^{-\gamma}$. The graphs a, b and c correspond to the scattering by homopolar phonon modes, by polar phonon modes and by acoustic phonon modes, respectively. Here, bulk-MoS$_2$ corresponds to $\hbar \omega$=60 meV and $\gamma$=2.6 (indicated by an arrow along the vertical axis). Taken from Ref. \cite{fiva}.}
 \label{fiva2}
\end{figure}

In contrast to the acoustic mode, the optical modes corresponds to the distortion of the unit cell. In isotropic materials (non-polar), the CDP due to optical modes is little and can be neglected. But in layered materials, the deformation potential is highly anisotropic \cite{fiva}. Those optical modes which can modulate the thickness of the individual layer will interact with the charge carriers as discussed above (because of series of parallel potentials along z). These thickness modulating optical modes which do not produce first order dipoles irrespective of having effective charges on each atom are known as \textit{homopolar} modes. The mobility due to the vertically polarized (along z) homopolar optical modes is $\mu_{bulk}^{hpolar}=\mu_0 (T/T_0)^{-\gamma^{hpolar}}$ \cite{fiva}; where $\gamma^{hpolar} > 1$ and  $\mu_0=\mu_{bulk}$ at T=T$_0$. In case of polar materials, where long range Coulomb interaction associated with polar optical modes couples the charge carriers (via Frohlich interaction), it has been shown that $\gamma^{polar} \cong \gamma^{hpolar}- 1$ \cite{fiva}. The experimentally observed Hall mobility of bulk MoS$_2$ is shown in Fig.~\ref{fiva1}. The fitted solid line corresponds to $\mu_{bulk}^e \cong 100 (T/T_0)^{-2.6}$ cm$^2$/V sec, where T$_0$=300K. To get a comparison of different scattering mechanisms of charge carriers by thermal phonons, the exponent ($\gamma$) for the mobility has been plotted with the phonon energy as shown in Fig.~\ref{fiva2}. Since, for bulk MoS$_2$ the only one homopolar mode \cite{fiva} corresponds to 60 meV (A$_{1g} \sim$ 408 $cm^{-1}$ for bulk), it is clear from Fig.~\ref{fiva2} that the vertically polarized non-polar (homopolar mode) mode is responsible for the high-T limited mobility.\\

\subsection{Temperature dependence of carrier mobility in single layer MoS$_2$}

\begin{figure}[h!]
 \centering
%\leavevmode t
%\includegraphics[trim=0 0 20 25, scale=0.5]{fig_24.pdf}
\includegraphics[trim=0.5cm 17cm 0.5cm 2cm, width=1.0\textwidth]{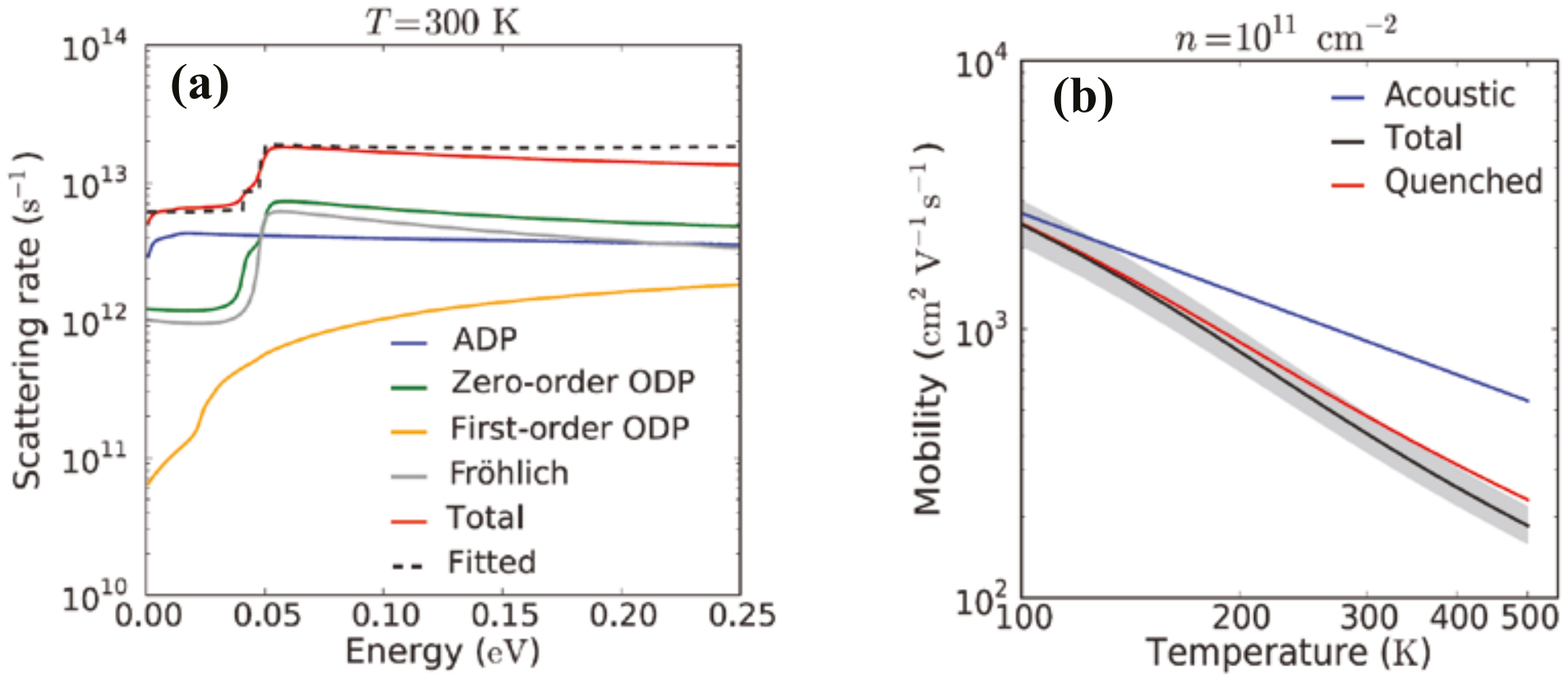}
  \caption{(color online)--(a) The first-principles calculations of different electron-phonon scattering rates (and the total) at room temperature versus the charge carrier energy. Here ADP=A$_{\lambda}$, Zero-order ODP=D$^0_{\lambda}$ and First-order ODP=D$^1_{\lambda}$. The dashed line shows the fitted deformation potential parameters given by $M_{\textbf{q}\lambda}=A_{\lambda}$q for acoustic modes and $M_{\textbf{q}\lambda}=D^0_{\lambda}$q for all the other modes by choosing corresponding deformation potentials.   The kinks in the curves for the optical phonon processes are attributed to the onset of phonon emissions. (b) Calculated mobility for different temperatures. The shaded area denotes the variations of the mobility due to the uncertainty (10\%) in the calculated deformation potentials. For comparison, the mobility because of the presence of the acoustic modes only is plotted, where $\mu$ $\sim$ 1/T. The red solid line represents the improvement of the mobility due to the quenching of the homopolar modes. Taken from Ref.   \cite{kaas}.}
 \label{kaas}
\end{figure}

\textbf{From theoretical point of view:} The electrical properties of single layer MoS$_2$ transistor have been probed extensively in past few years. The different phonon scattering mechanisms with the charge carriers discussed in the previous section is calculated \cite{kaas} and plotted at T=300K as a function of energy for nondegenerate carriers, as shown in Fig.~\ref{kaas}(a). The Brillouin zone for SL- MoS$_2$ is hexagonal and similar to graphene, two valleys exist at K and K$^{'}$ point. Kaasbjerg et al \cite{kaas} calculated both the intravalley and intervalley scattering contributions to the relaxation time of the charge carriers, as these two valleys are occupied by the carriers for n-type SL- MoS$_2$. Here the first order deformation potential is also shown in addition to the other scattering mechanisms. The longitudinal acoustic (LA) and transverse acoustic (TA) modes contribute to quasielastic intravalley scattering process through the acoustic deformation potential (ADP). The contribution from TA is non-zero because of inclusion of Umklapp process and results in a coupling of highly anisotropic nature \cite{kaas}. From Fig.~\ref{kaas}(a), it is clear that at low carrier energy the \textit{total} scattering rate is dominated by ADP. At higher energy the total scattering rate is dominated by zero-order optical deformation potential (ODP) and the scattering via Frohlich interaction; the sudden jump in the optical scattering rates is because of onset of the emission of optical phonons by the charge carriers. At all energy scale, the first-order ODP scattering is one order of magnitude smaller than the other scattering rates. 

Similar to the bulk case, the mobility due to ADP is $\mu_{1L}^{acoustic} \propto T^{-\gamma^{acoustic}}$ \cite{kaas}; where $\gamma^{acoustic}=1$. The calculated temperature dependent mobility for 1L-MoS$_2$ is shown in Fig.~\ref{kaas}(b). At lower temperature ($\sim$ 100K) the value of $\gamma$=1. At higher temperatures the mobility follows according to $\mu_{1L}^{optical} \propto T^{-\gamma}$, where $\gamma = 1.69$ due to zero-order ODP and Frohlich scattering \cite{kaas}. The calculated mobility at room temperature is of $\sim$ 400 cm$^2$/V sec. If we exclude the zero-order ODP (homopolar mode), the resulting mobility is increased by $\sim$ 70 cm$^2$/V sec and the reduced $\gamma$=1.52, as shown in Fig.~\ref{kaas}(b) for the quenched case; although the effect of quenching on the mobility is not much.\\

\begin{figure}[h!]
 \centering
%\leavevmode t
%\includegraphics[trim=0 0 20 25, scale=0.5]{fig_25.pdf}
\includegraphics[trim=0.5cm 15cm 0.5cm 2cm, width=0.8\textwidth]{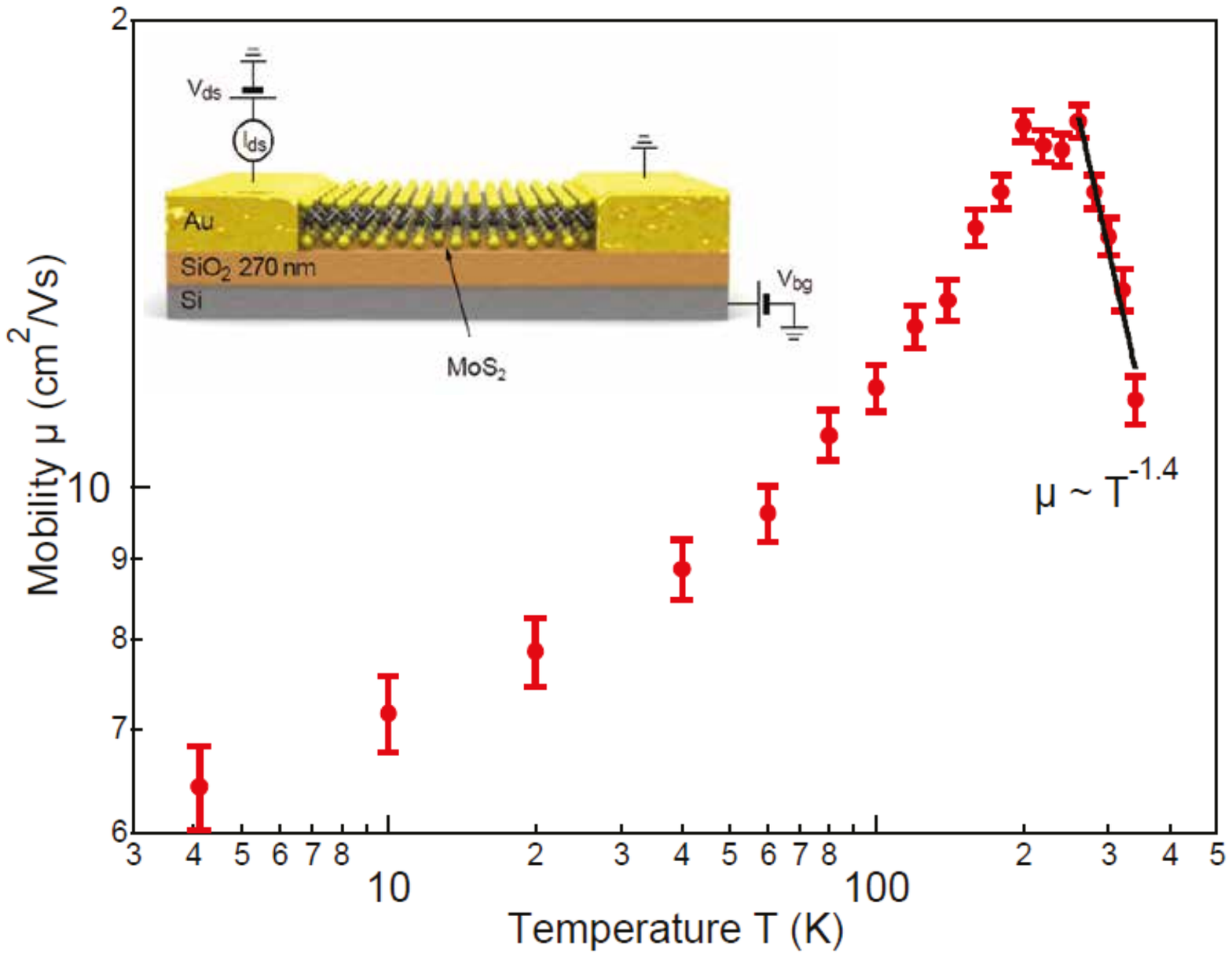}
  \caption{(color online)--The temperature dependent Field effect mobility for single layer MoS$_2$ deposited on SiO$_2$/Si substrate. The back gate voltage (V$_{bg}$) is in the range of 30-40 V to extract the mobility.. In the high-T range, the fitted value of the exponent ($\gamma$) is 1.4. Taken from Ref. \cite{kis}.}
 \label{kis2}
\end{figure}
%FIG-30

\begin{figure}[h!]
 \centering
%\leavevmode t
%\includegraphics[trim=0 0 20 25, scale=0.5]{fig_26.pdf}
\includegraphics[trim=0.5cm 15cm 0.5cm 2cm, width=0.8\textwidth]{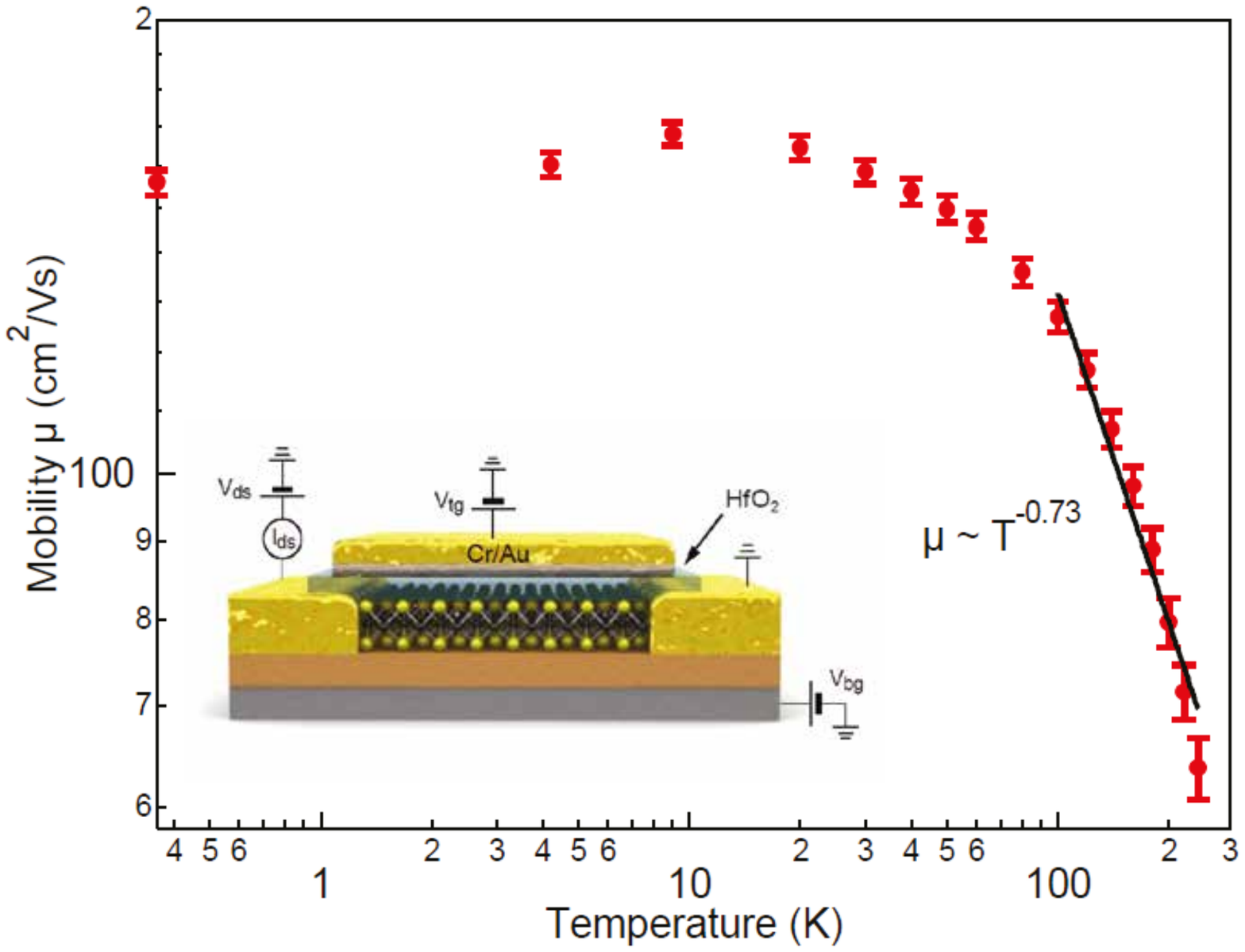}
  \caption{(color online)--The temperature dependent Field effect mobility ($\mu$=[d$\sigma$/dV$_{bg}$][L/WC$_{tg}$) for single layer MoS$_2$ deposited on SiO$_2$/Si substrate. Applied top gate voltage (V$_{tg}$) is in the range of 3-5 V to extract the mobility. In the high-T range, the fitted value of the exponent ($\gamma$) is 0.73. Taken from Ref. \cite{kis}.}
 \label{kis3}
\end{figure}

\textbf{Comparison with the experimentally observed mobility:} The experimentally observed (at room temperature) low mobility in the range of 1-50 cm$^2$/V sec \cite{radi, novo,aya} reported earlier is due to the charge impurity scattering and increased upto 200 cm$^2$/V sec \cite{radi} by depositing a high dielectric material (HfO$_2$) on top of the SL-MoS$_2$. This enhancement of the mobility is partially attributed to the strong damping of Coulombic scattering from charge impurities \cite{jena}. The enhancement of mobility is also achieved by polymer (PEO+LiClO$_4$) top gating from $\sim$ 0.1 to $\sim$ 150 cm$^2$/V sec \cite{ming}. The difference between the calculated value ($\sim$ 410 cm$^2$/V sec) and the experimentally observed ($\sim$ 200 cm$^2$/V sec) is may be due to the defects and surface optical phonon scattering \cite{kaas}. The temperature dependent mobility with the single-gate configuration is shown in Fig.~\ref{kis2}. The field effect mobility is given by $\mu=[d\sigma/dV_{bg}][L/WC_{bg}]$, where $\sigma$ denotes conductance, V$_{bg}$ applied back gate voltage, L channel length, W the width of the channel and  C$_{bg}$ represents back gate capacitance associated with the SiO$_2$/Si layer. At $\sim$ 200K, the mobility reaches a peak value of $\sim$ 18 cm$^2$/V sec. The decrease of mobility with temperature from 200K to 4K is attributed to the well known scattering from charged impurities. Above 200K, the mobility decreases with the increase in temperature. This high-T limited mobility due to phonons is fitted according to $\mu_{SL} \sim T^{-\gamma}$, where $\gamma \approx $1.4. This is in good agreement with the predicted value of $\gamma \approx $1.65 \cite{kaas}. With the dual-gate configuration, the observed mobility is plotted against temperature, as shown in Fig.~\ref{kis3}. The mobility varies from $\sim$ 168 cm$^2$/V sec at 4K to $\sim$ 60 cm$^2$/V sec at 240K. The mobility with dual-gate configuration is quite different than that with the single-gate configuration, where the mobility monotonously decreases with decreasing temperature. This distinct feature with dual-gate device is attributed to the strong damping of charged impurity scattering in presence of dielectric media and to the metallic top gate which also changes the dielectric environment for the SL-MoS$_2$ \cite{kis}. In the high temperature (100-300K), the fitted value of mobility due to the phonon scattering shows $\gamma \approx 0.73$. All the other dual-gate devices of SL-MoS$_2$ show the variations of $\gamma$ from 0.3 to 0.75 \cite{kis}. These observed values are much less than the predicted $\gamma \approx 1.52$ \cite{kaas}. In addition to the quenching of homopolar mode due to the dielectric environment, other phonon scattering mechanisms are also reduced which needs further theoretical work to clarify this discrepancy.\\ 

\subsection{Substrate effect towards improving the mobility of MoS$_2$: suppression of short range and long range interactions}

\begin{figure}[h!]
 \centering
%\leavevmode t
%\includegraphics[trim=0 0 20 25, scale=0.5]{fig_27.pdf}
\includegraphics[trim=0.5cm 12cm 0.5cm 2cm, width=0.8\textwidth]{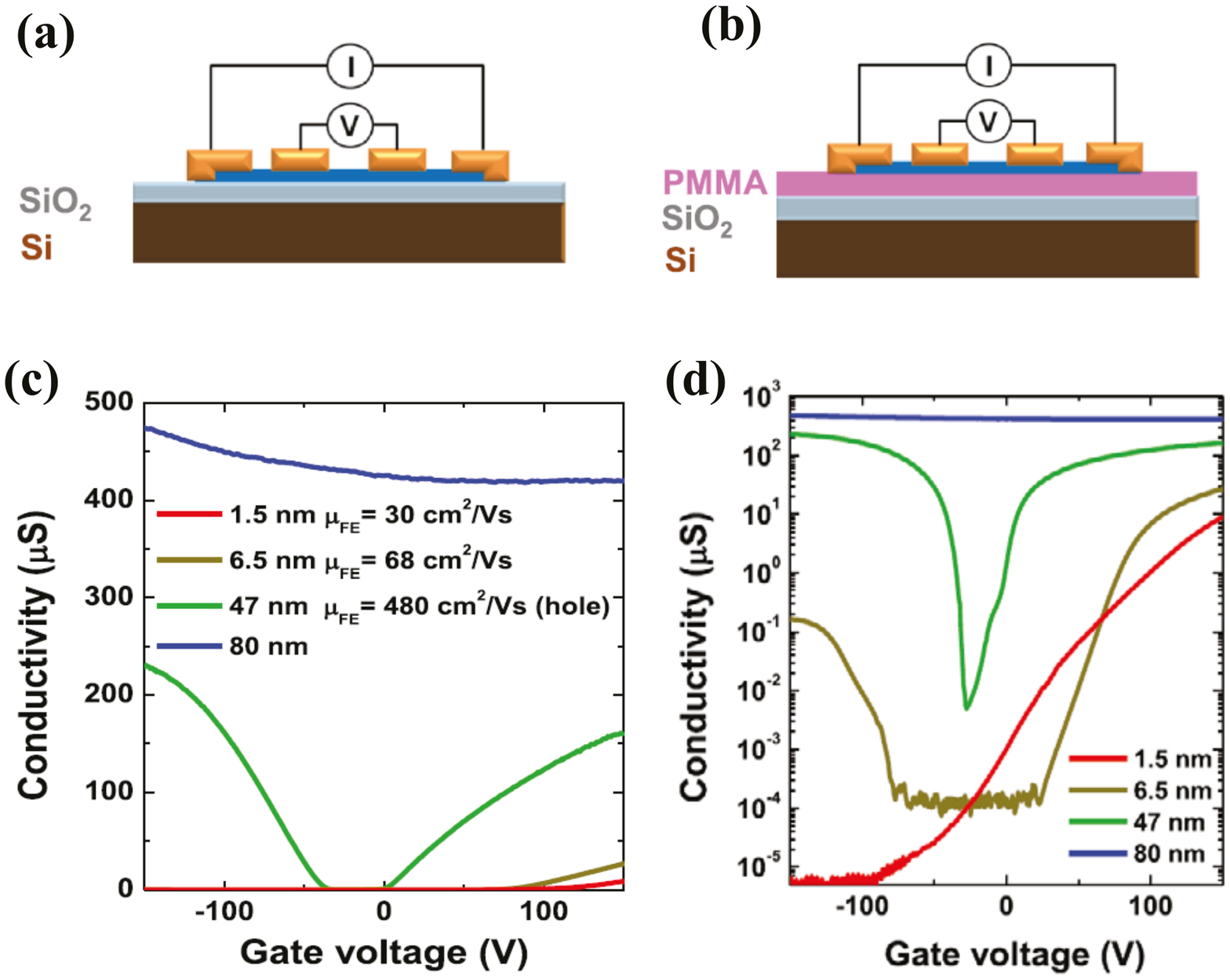}
  \caption{(color online)--Schematic diagrams of the four-probe devices (contacts made of Ti/Al electrodes) (a) without PMMA and (b) with PMMA (300 nm) deposited on 300 nm SiO$_2$/Si substrate. The measured conductivity ($\sigma$) of four MoS$_2$-devices having PMMA with different thicknesses (1.5 nm, 6.5 nm, 47 nm and 80 nm) (c) with linear scale and (d) with the semi-log scale. Taken from Ref. \cite{bao}.}
 \label{bao1}
\end{figure}

It has been established that the presence of the top gate material on MoS$_2$ enhances the mobility of the charge carriers \cite{radi,ming,kis}. Before putting the top-gate material the SL-MoS$_2$ deposited on bare SiO$_2$/Si substrate shows a less amount of mobility in the range of 1-50 cm$^2$/V sec \cite{radi, novo,aya}. In order to investigate the substrate effect on the measured mobility, two different kind of devices are considered for the various thickness (1-80 nm) of MoS$_2$: (i) the MoS$_2$ flakes are deposited on SiO$_2$/Si substrate (shown in Fig.~\ref{bao1}a) and, (ii) those deposited on polymethyl methacrylate (PMMA) (shown in Fig.~\ref{bao1}b). Fig.~\ref{bao1}(c) shows the observed room temperature conductivity ($\sigma$) with the applied back gate voltage (V$_{bg}$) for four different thickness's of sample deposited on 300 nm PMMA \cite{bao}. Fig.~\ref{bao1}(d) shows the semi-logarithmic behavior of the conductivity. While the 6.5 and 47 nm thick MoS$_2$ show the ambipolar behavior, 1.5 nm thick (2L-MoS$_2$) shows unipolar nature with mobility 30 cm$^2$/V sec. The mobilities for the 6.5 nm thick sample are given by $\mu_e(6.5) \sim 68$ cm$^2$/V sec and $\mu_h(6.5) \sim 1$ cm$^2$/V sec for electrons and holes, respectively; for 47 nm thick MoS$_2$, these are given by $\mu_e(47) \sim 270$ cm$^2$/V sec and $\mu_h(47) \sim 480$ cm$^2$/V sec. It is clear from Fig.~\ref{bao1}(d) that the region (off-state), separating electron and hole conductance, decreases with the increasing thickness of the sample; the off-state conductance also increases with the thickness. For 80 nm thick sample, there is no off-state indicating the dominance of conductance from the bulk \cite{bao}.\\

\begin{figure}[h!]
 \centering
%\leavevmode t
%\includegraphics[trim=0 0 20 25, scale=0.5]{fig_28.pdf}
\includegraphics[trim=0.5cm 17cm 0.5cm 2cm, width=0.8\textwidth]{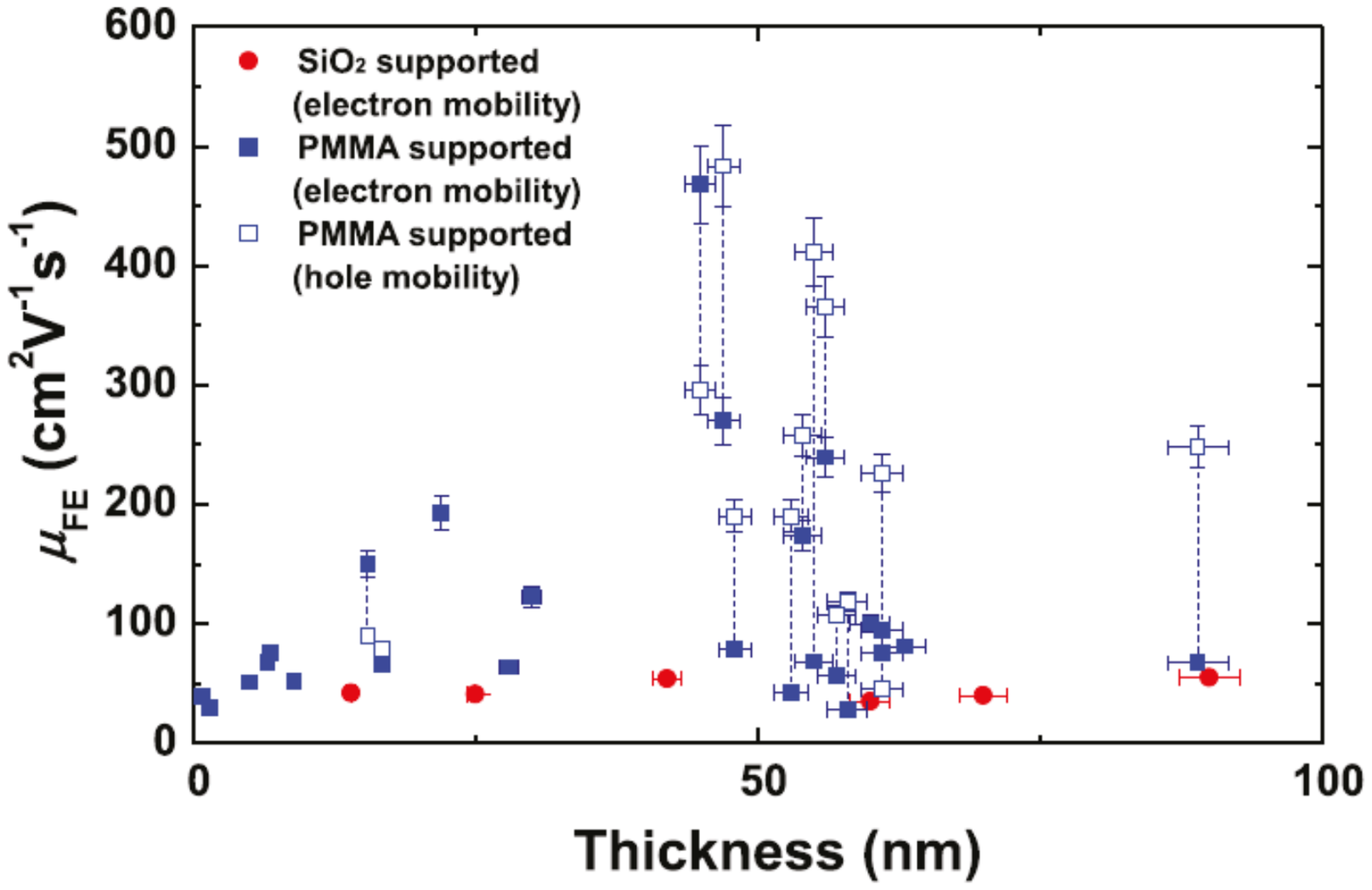}
  \caption{(color online)--The field effect mobility (at 300K) versus thickness of the MoS$_2$ flakes for 25 devices with PMMA and 6 devices without PMMA. For PMMA supported devices (range of V$_{bg}$ $\pm$ 150V), electron (solid squares) and hole (empty squares) mobilities are connected with dashed lines. For the SiO$_2$ supported devices (range of V$_{bg}$ $\pm$ 75V), only electron mobility (red circle) is shown. Taken from Ref. \cite{bao}.}
 \label{bao2}
\end{figure}

The thickness dependent mobility at room temperature for two different kind of device configurations is shown in Fig.~\ref{bao2}. The ambipolar nature has been marked by a dashed line connecting the hole and electron mobility. From Fig.~\ref{bao2}, all the devices can be grouped in two categories: (a) unipolar type--shown by all the devices deposited on SiO$_2$/Si substrate and by all thin samples (1-30 nm) supported by PMMA ; (b) ambipolar type--observed for all the thick sample of width in the range of 40 nm to 70 nm deposited on PMMA \cite{bao}. The observed mobility peaks are for $\sim$ 50 nm thick MoS$_2$ samples of ambipolar type and are given by $\mu_e(50) \sim 470$ and $\mu_h(47) \sim 480$ in cm$^2$/V sec. From Fig.~\ref{bao2} it can be inferred that for most of the devices of ambipolar type, the $\mu_h$ is larger than the $\mu_e$. Hence, the multilayer MoS$_2$ for the p-type operation can be chosen for faster electronics. The thickness independent mobility shown by unipolar-type devices deposited on SiO$_2$/Si substrate is in the range of 30-60 cm$^2$/V sec. The observed higher mobility at room temperature ( $\mu_e(50) \sim 470$ and $\mu_h(47) \sim 480$ in cm$^2$/V sec) for the multilayer MoS$_2$ is larger than the theoretically predicted phonon limited mobility for SL-MoS$_2$ \cite{kaas} and for the thick layer of MoS$_2$ ($\sim$ 100 cm$^2$/V sec) \cite{kim}.\\

\begin{figure}[h!]
 \centering
%\leavevmode t
%\includegraphics[trim=0 0 15 13, scale=0.5]{fig_29.pdf}
\includegraphics[trim=0.5cm 18cm 0.5cm 2cm, width=0.8\textwidth]{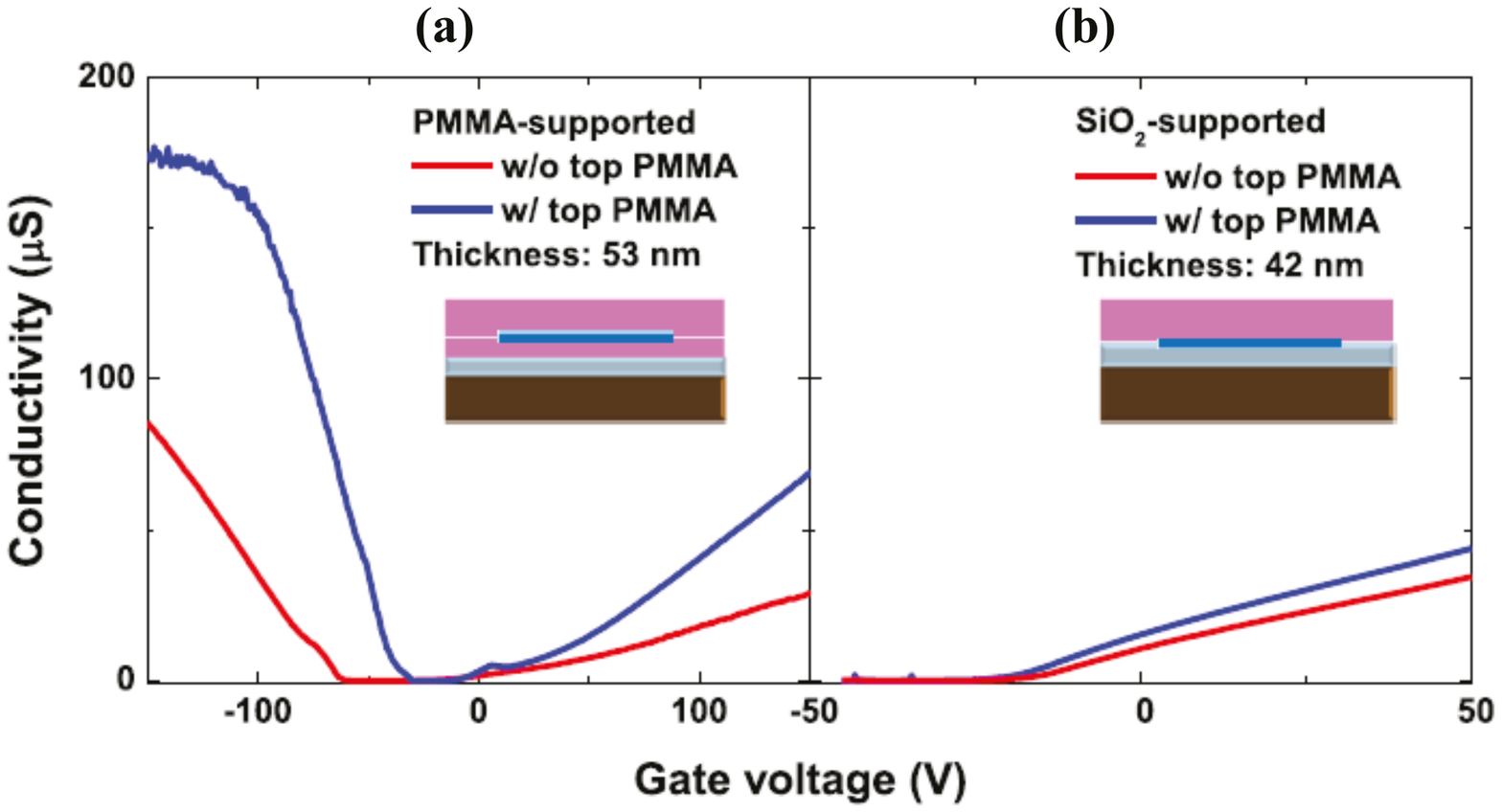}
  \caption{(color online)--The conductivity with back gate voltages (a) for PMMA supported device of 53 nm thick MoS$_2$ and (b) for SiO$_2$ supported device of 42 nm thick MoS$_2$. The red and blue curves show the conductivity before and after PMMA coating (from top) on MoS$_2$ flakes. The insets show the schematics of the devices. Taken from Ref. \cite{bao}.}
 \label{bao3}
\end{figure}

In case of MoS$_2$ devices, the charge carriers are confined close to the dielectric interface (within few nm) used for the back gating \cite{kim}. The thickness dependent enhancement of the carriers mobility can be attributed to the role of the additional MoS$_2$ layers to screen the long range disorder \cite{bao}. To check the screening effect, PMMA coating has been done onto the top of the devices for the two different types of configurations (see Figs.~\ref{bao1}a and b). The resulting measured conductivity is shown in Figs.~\ref{bao3}(a) and (b). While the PMMA coating helps to screen the long range scattering for the PMMA supported devices, but the effect is little for SiO$_2$/Si supported devices. The maximum improvement mobility observed with the PMMA coating is $\sim$ more than 300 $\%$ for 60 nm thick sample \cite{bao}. The observed variation of the mobility on the thickness or to the PMMA coating is attributed to the presence of short range disorder (rough surface) for the SiO$_2$/Si supported devices \cite{bao}.

\subsection{Reduction of Schottky barrier at metal-semiconductor contact to improve the mobility of MoS$_2$ }
\begin{figure}[h!]
 \centering
%\leavevmode t
%\includegraphics[trim=0 0 15 13, scale=0.5]{fig_30.pdf}
\includegraphics[trim=0.5cm 17cm 0.5cm 2cm, width=0.8\textwidth]{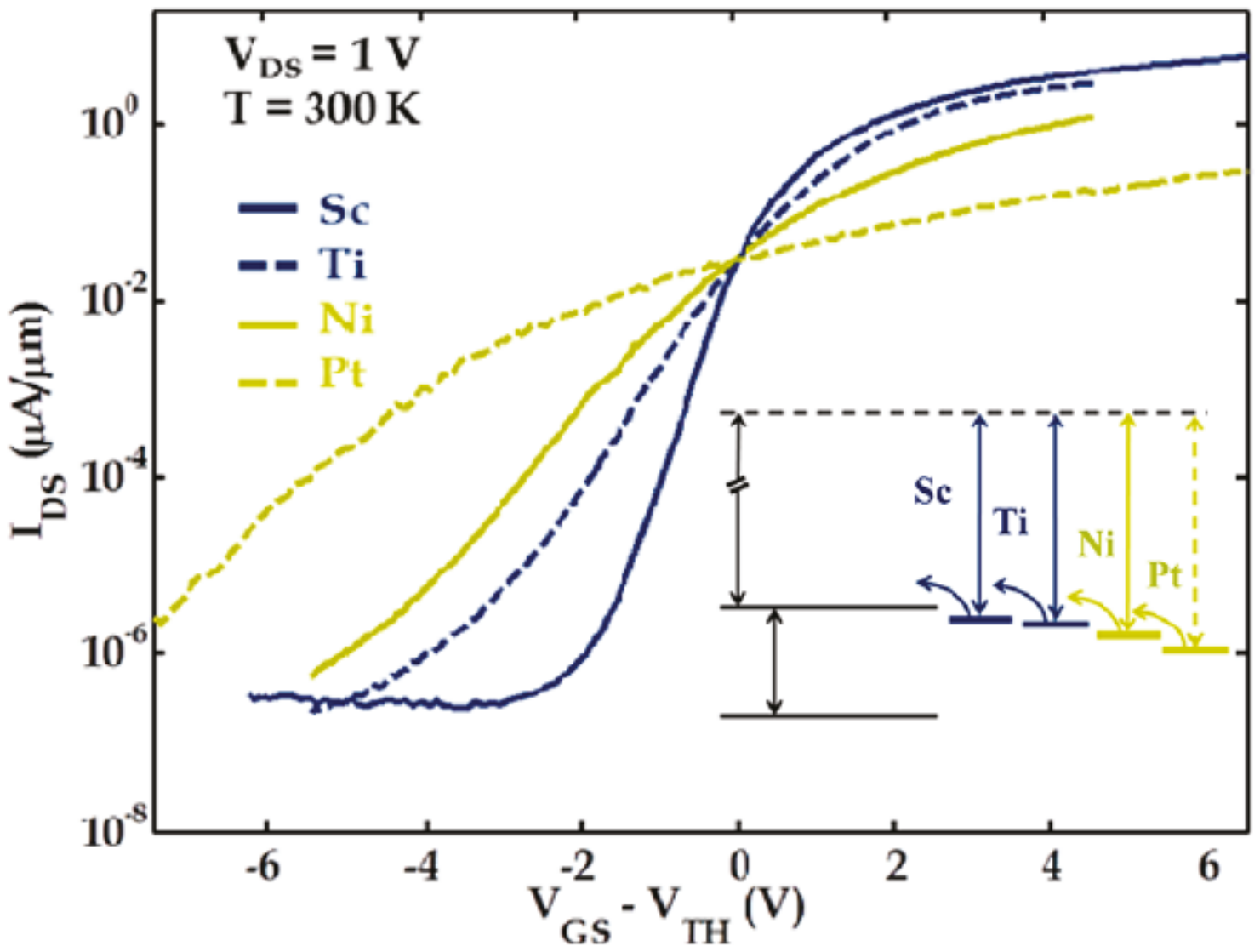}
  \caption{(color online)--The drain-source current versus back gate voltage for 6 nm thick MoS$_2$. The threshold voltage, V$_{Th}$= -6.0, -1.0, 1.5 and 4.0 V for Sc, Ti, Ni and Pt contact devices, respectively. The inset shows the alignment of the Fermi levels of four metals with the conduction band edge (irrespective of different work functions of the metals) according to the observed n-type device characteristics. Taken from Ref. \cite{das}.}
 \label{das2}
\end{figure}
%FIG-35

In all the previous studies, attention has been paid to the interface physics i.e. the interface between gate dielectric and the sample, the interface between the substrate and the sample to improve the mobility and hence device performance. However, the presence of Schottky barrier between metal contacts and semiconductor can affect the mobility significantly. In order to understand the Schottky barrier contacts and their role in improving the mobility, four different metals scandium (work function $\Phi_M$=3.5 eV), titanium ($\Phi_M$=4.3 eV), nickel ($\Phi_M$=5.0 eV) and platinum ($\Phi_M$=5.9 eV) are used to make contacts with the multilayer MoS$_2$ (electron affinity = 4.0 eV \cite{han}) for the devices using a back gate. Fig.~\ref{das2} shows the experimentally measured transfer characteristics for the 6 nm thick MoS$_2$ device \cite{das}. All the four devices show the unipolar n-type operation, irrespective of the different work functions of the four metals. On the positive side of V$_{GS}$-V$_{TH}$, there is a clear decrease in on-current from Sc to Pt contacts, and this is consistent with the formation of Schottky barrier at the metal to channel interface. \\

\begin{figure}[h!]
 \centering
%\leavevmode t
%\includegraphics[trim=0 0 15 13, scale=0.5]{fig_31.pdf}
\includegraphics[trim=0.5cm 14cm 0.5cm 2cm, width=1.0\textwidth]{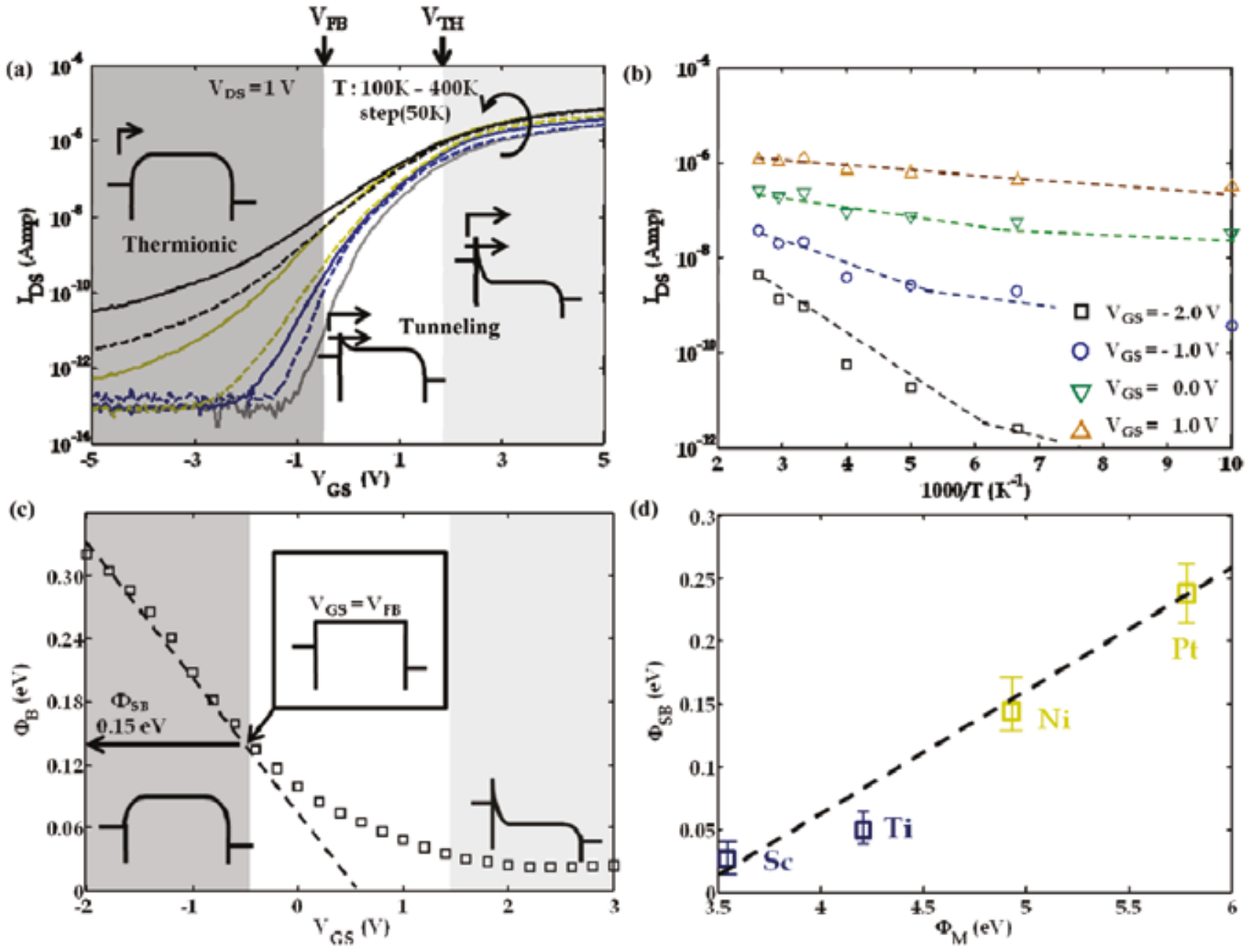}
  \caption{(color online)--(a) The drain-source current versus back gate voltage at different temperatures showing three distinct regions of operations: below flat-band voltage (V$_{FB}$), below threshold voltage and above threshold voltage. The insets show the energy-band diagrams in the corresponding regions. (b) The drain-source current versus the inverse-temperature (1/T) (taken from a). (c) The effective barrier height ($\Phi_{B}$ from b) as a function of gate voltage and the extraction of true Schottky barrier ($\Phi_{SB}$) for Ni-contact device. (d) All $\Phi_{SB}$ versus work functions for all the different metal contact (Sc, Ti, Ni and Pt) devices.  Taken from Ref. \cite{das}.}
 \label{das2}
\end{figure}

In order to estimate the true height of the Schottky barrier ($\Phi_{SB}$) quantitatively, the temperature dependence (100-400K) transfer characteristics is measured for the Ni contact as shown in Fig.~\ref{das2}(a). In the subthreshold region, the number of charge carriers are less and hence diffusive temperature dependent thermionic current (I$_{thermionic}$) and tunnelling current (I$_{tunneling}$) become important. The temperature dependent current (I$_{DS}$) is then plotted for various back gate voltages, as shown in Fig.~\ref{das2}(b). The thermionic emission current is fitted to extract the effective barrier height ($\Phi_B$) with the well know formula $I_{DS}$= I$_{thermionic}$ $=AT^2 exp[\frac{e\Phi_B}{K_BT}][1-exp(\frac{e V_{DS}}{K_BT})]$; where, A is the Richardson's constant, $K_B$ the Boltzmann constant, e the electronic charge. Fig.~\ref{das2}(c) shows the variation of the extracted $\Phi_B$ with the applied V$_{GS}$. The effective barrier height $\Phi_B=\Phi_{SB}-C_{eff}(V_{GS}-V_{FB})$ \cite{das}, for $V_{GS} \leq V_{FB}$ and $\Phi_B=\Phi_{SB}-\Delta$, for $V_{GS} > V_{FB}$. The factor $C_{eff}=1/(1+\frac{C_{it}+C_{ch}}{C_{ox}})$ represents the band movement of the channel with V$_{GS}$, where C$_{it}$ and C$_{ox}$ denotes the capacitance for the interface trap charges and oxide layers, respectively. $\Delta$ is a positive quantity but not constant. V$_{FB}$ is the flat band voltage. The I$_{thermionic}$ is the only current through the channel for the negative V$_{GS} < V_{FB}$ (see Fig.~\ref{das2}a); and hence, the linear relationship between $\Phi_B$ and $\Phi_{SB}$ as shown in Fig.~\ref{das2}(c). When V$_{GS} \geq V_{FB}$, the I$_{tunneling}$ starts to flow through the channel and this current is not included in the fitted equation for I$_{thermionic}$. And because of this fact, the linear relationship between $\Phi_B$ and $\Phi_{SB}$ does not hold and starts to deviate after V$_{FB}$, as shown in Fig.~\ref{das2}(c). The extracted value of $\Phi_{SB}$ is 0.15 eV fo Ni contact. It must be noteworthy that this is the tunneling current which is responsible for getting ohmic contact even for the higher work function metals including gold ($\Phi_M$=5.4 eV) \cite{das}. Fig.~\ref{das2}(d) represents the measured $\Phi_{SB}$ for different metals, which are 30, 50, 150 and 230 meV for Sc, Ti, Ni and Pt, respectively. The fitted dashed line gives the slope d$\Phi_{SB}$/d$\Phi_{M}$ $\sim$ 0.1 (0.27 for Si \cite{sze}) implying that there is a strong pining of Fermi level at the semiconductor interface before the metal contact is made. This pining of Fermi level near neutral level is related to the charge impurities present on the semiconductor-metal interface, which is usually the case with the covalent semiconductors \cite{sze}. This explains clearly why only n-type device characteristics are observed irrespective of high work functions of the metals used for making contacts. 

\begin{figure}[h!]
 \centering
%\leavevmode t
%\includegraphics[trim=0 0 30 35, scale=0.5]{fig_32.pdf}
\includegraphics[trim=0.5cm 19cm 0.5cm 2cm, width=1.0\textwidth]{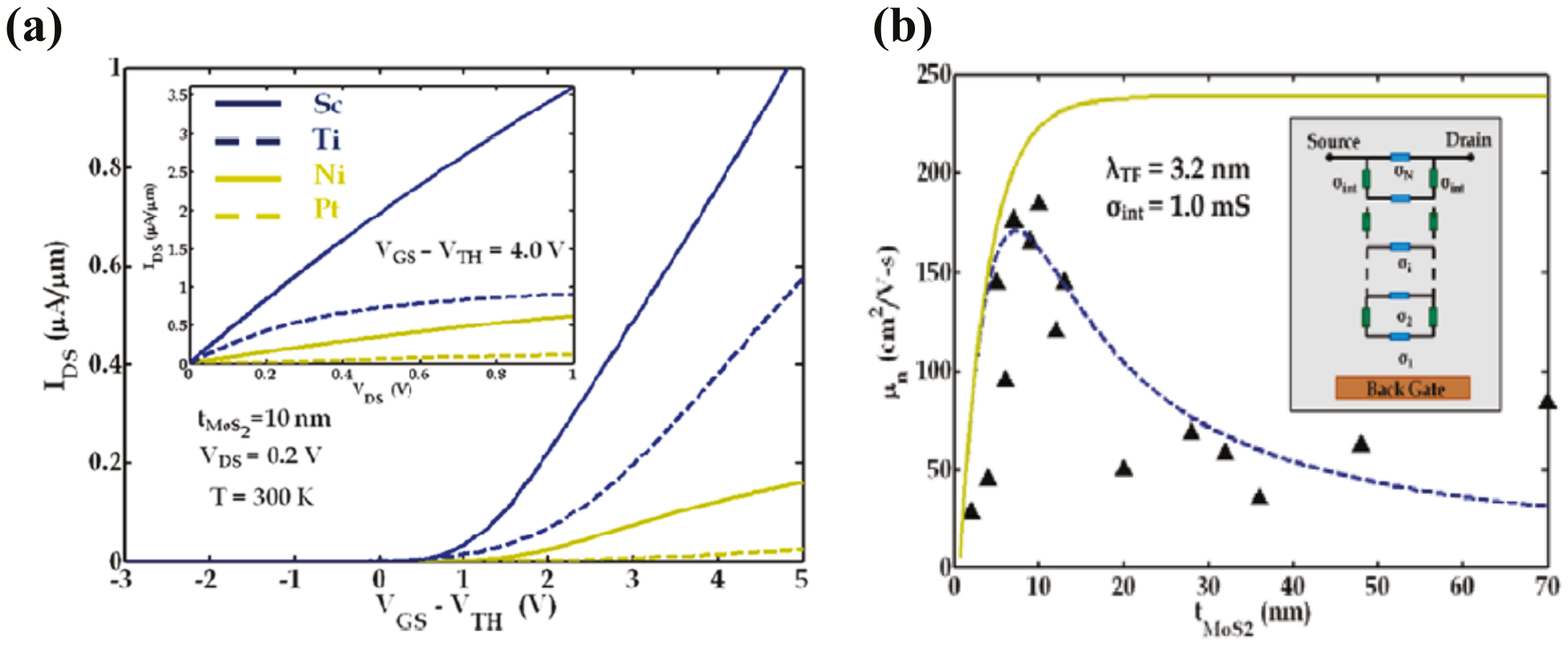}
  \caption{(color online)--(a) The drain-source current versus back gate voltage (T=300K) for Sc, Ti, Ni and Pt-contact devices. Here V$_{GS}$-V$_{TH}$=4 V corresponds to $\sim$ 1.4 $\times$ 10$^{12}$ /cm$^2$. The value of the back gate capacitance is $\sim$ 3 $\times$ 10$^{−4}$ F/m$^{2}$. (b) The extracted field effect mobility for different thicknesses of the sample with Sc-contact devices. The dashed line shows the fitting of the data according to the resistor network model, where two model parameters are Thomas-Fermi screening length ($\lambda_{TH}$ $\sim$ 3 nm) and the interlayer conductivity ($\sigma_{int}$ = 1 mS). The solid line represents the calculated mobility without taking $\sigma_{int}$. The inset shows the schematic of the resistor network model. Taken from Ref. \cite{das}.}
 \label{das3}
\end{figure}

To understand the effect of measured Schottky barrier on mobility, the output and transfer characteristics for 10 nm thin MoS$_2$ are shown in Fig.~\ref{das3}(a). The exponential rise (see the inset) of $I_{DS}$ with $V_{DS}$ for the Pt contact is in accordance with the Schottky barrier of 230 meV. For Ni-contacted devices having $\Phi_{SB}$=150 meV, it shows the linearity. Since at 300K, the temperature assisted tunnelling current is responsible for this observed linear behavior between the current and drain voltage; the issue of getting ohmic contact is irrelevant \cite{das}. The measured field-effect mobilities from transfer characteristics are 21, 36, 125 and 184 in cm$^2$/V sec for the devices having Pt, Ni, Ti and Sc contacts, respectively \cite{das}. The measured low mobility \cite{radi,novo,aya,ming} ($<$ 50 cm$^2$/V sec) using Au or Ti/Au contacts without top gate material is in agreement with the effect of Schottky barrier.\\

\begin{figure}[h!]
 \centering
%\leavevmode t
%\includegraphics[trim=0 0 15 13, scale=0.5]{fig_33.pdf}
\includegraphics[trim=0.5cm 15cm 0.5cm 2cm, width=0.8\textwidth]{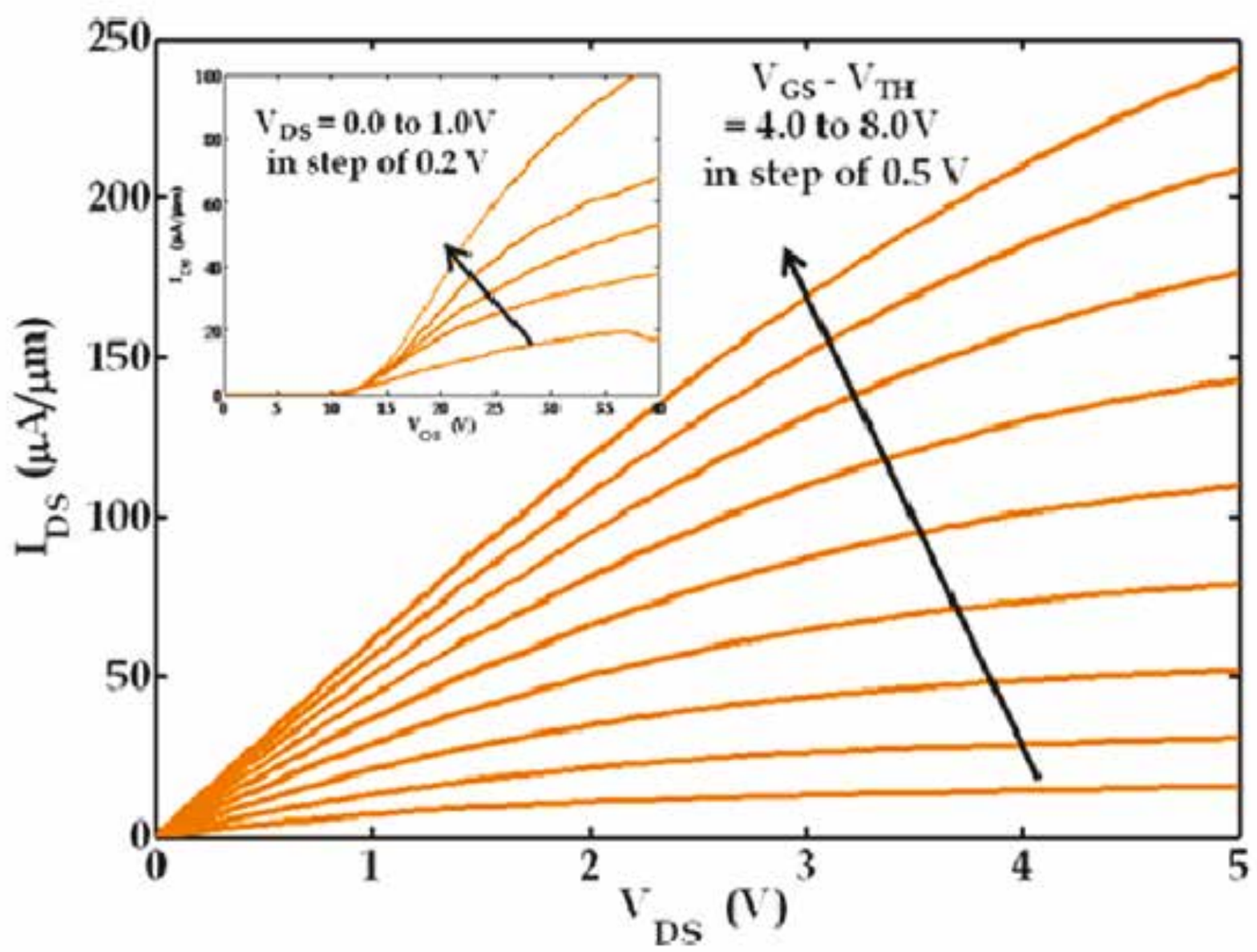}
  \caption{(color online)--The drain-source current versus drain-source voltage (for different top gate voltages) at T=300K for 10 nm thickness of the MoS$_2$ layer (Sc-contact). The improvement of the transport coefficients are achieved by depositing a top-gate material (Al$_2$O$_3$) of 15 nm thickness. The inset shows the corresponding transfer characteristics for different drain-source voltages. Taken from Ref. \cite{das}.}
 \label{das4}
\end{figure}

Fig.~\ref{das3}(b) shows the variations of the mobility with the thickness of the sample in the range of 2-70 nm for the device having Sc contacts. The maximum mobility observed corresponds to the $\sim$ 10 nm thin MoS$_2$ device. The observed mobility is fitted with the resistor network model (see the inset of Fig.~\ref{das3}b). For thicker samples, the access to the lower layer involves the interlayer resistance. For thinner samples, the screening of the substrate is much less than the thicker one. The observed peak is, therefore, related to the competition between the interlayer conductivity ($\sigma_{int}$) and the screening of the substrate \cite{das}. Fig.~\ref{das4} shows the the output and transfer characteristics for 10 nm thin MoS$_2$ with the addition of 15 nm dielectric material (Al$_2$O$_3$) on the top of the sample. The field effect mobility due to the screening of long range scattering is now enhanced to the value of $\sim$ 700 cm$^2$/V sec from 184 cm$^2$/V sec for Sc contacted device. This huge increment is attributed to the (i) use of low Schottky barrier ($\Phi^{Sc}_{SB}$=30 meV), (ii) choosing of optimum value of layer thickness (10 nm) which corresponds to the optimization of suppression of substrate effect as well as the interlayer resistor, and (iii) screening of the long range Coulomb interaction. Finally, a high saturation current density $\sim$ 240 $\mu A/\mu$m at the carrier density of $\sim$ 2.8 $\times$ 10$^{12}$ /cm$^2$ and the higher transconductance $\sim$ 4.7  $\mu S/\mu$m at $V_{DS}$=1.0 V are measured for a 5 $\mu$m channel length. In order to get a higher mobility ($\geq$ 700 cm$^2$/V sec), it is desirable to do experiment on a device having Sc contact deposited on a high dielectric material instead of  SiO$_2$/Si substrate (may be PMMA \cite{bao}) with high dielectric top gating material by varying only the thickness of the sample. 

\section{Phonon-renormalization in a single layer MoS$_2$ transistor: aspects of symmetry}

\begin{figure}[h!]
 \centering
%\leavevmode t
%\includegraphics[trim=0 0 12 18, scale=0.5]{fig_34.pdf}
\includegraphics[trim=0.5cm 16cm 0.5cm 2cm, width=1.0\textwidth]{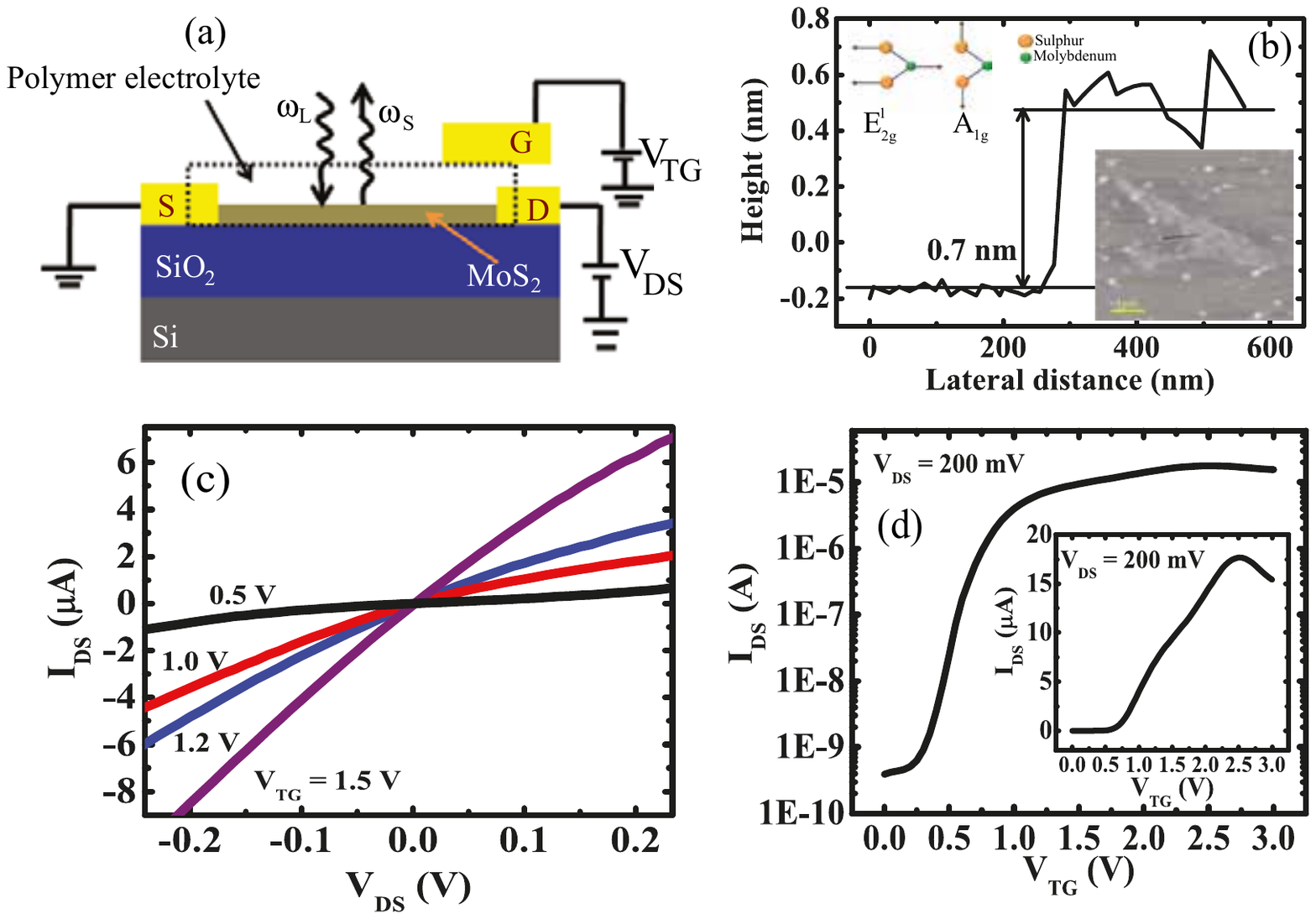}
  \caption{(color online)--(a) Schematic of the single layer MoS$_2$ device (50 nm thick Au contact used). Here $\omega_i$ and $\omega_s$ represent the incoming and scattered photons, respectively. The length of the device is $\sim$ 2.5 $\mu$m and width is $\sim$ 1.5 $\mu$m. (b) The AFM height profile of the single layer. The right inset (AFM image) showing the dotted line on the sample along which the AFM height profile has been taken. (c) The output characteristics of the device at different top gate voltages. (d) The transfer characteristics in semi-log scale (the inset showing in linear scale) at V$_{DS}$=200 mV. It shows the on-off ratio $\sim$ 10$^5$. Taken from Ref. \cite{bera}.}
 \label{bera1}
\end{figure}

It has been established that the high temperature ($\sim$ 200 to 300 K) phonon limited mobility in single and bulk-MoS$_2$ (homopolar mode A$_{1g} \sim$ 408 $cm^{-1}$) is because of dominant scattering of charge carriers by the optical phonons \cite{fiva,kis}. Electron-phonon scattering not only affects carrier momentum relaxation, but also affects the frequency and the full width at half maximum (FWHM) of phonons via renormalization of their self energy; the real part of the self energy is related to the frequency shift and imaginary part to the FWHM \cite{cardona}. The non-destructive Raman spectroscopy is exploited extensively to probe the electron-phonon coupling (EPC) quantitatively for single layer graphene \cite{pisa,yan1,bisu2} and bi-layer graphene \cite{yan2,bisu3,marlad}. To quantify the EPC for single layer MoS$_2$, in-situ gate voltage dependent Raman experiments have been reported \cite{bera}. The schematic diagram for the device is shown in Fig.~\ref{bera1}(a). A mixture of LiClO$_4$ and polyethylene oxide (PEO) in 1:8 weight ratio is used as a top gating material. Fig.~\ref{bera1}(b) shows the atomic force microscopy (AFM) image in contact mode to measure the height (0.7 nm) of the MoS$_2$. The output characteristics of the device is shown in Fig.~\ref{bera1}(c). The observed slight non-linearity between I$_{DS}$ and V$_{DS}$ is due to the Schottky barrier for Au contacts \cite{das} (as discussed earlier). The transfer characteristics in semi-log scale is shown in Fig.~\ref{bera1}(d). The n-type operation was attributed to the Fermi level pinning at the semiconductor interface due to interface charges \cite{das}. An extracted mobility was $\sim$ 50 cm$^2$/V sec and the on-off ratio was 10$^5$ (maximum observed $\sim$ 10$^8$ \cite{radi}).\\

\begin{figure}[h!]
 \centering
%\leavevmode t
%\includegraphics[trim=0 0 15 20, scale=0.5]{fig_35.pdf}
\includegraphics[trim=0.5cm 20cm 0.5cm 2cm, width=1.0\textwidth]{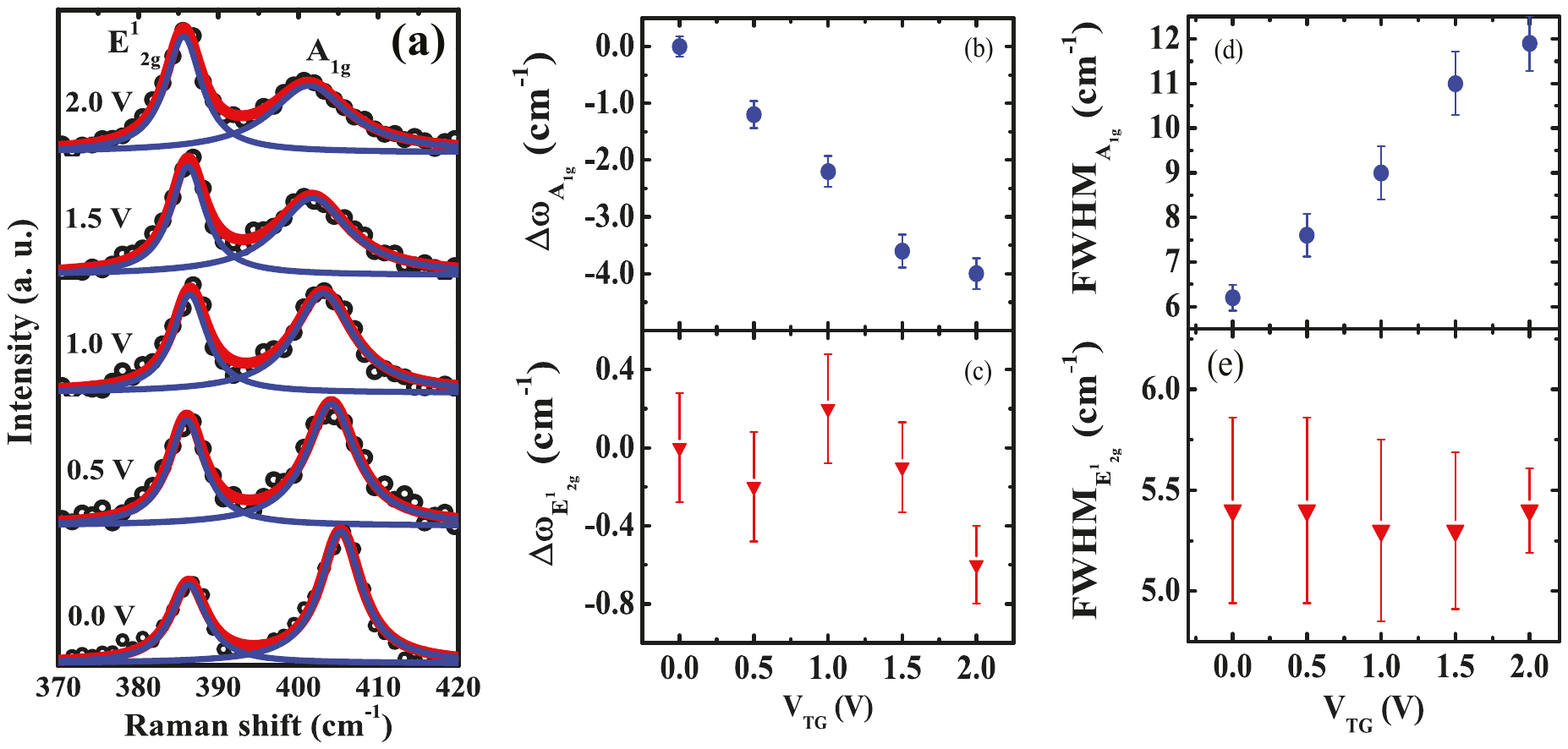}
  \caption{(color online)--(a) Raman spectra along with Lorentzian fits at different top gate voltages for the single layer MoS$_2$. The changes in frequency and FWHM for the A$_{1g}$ (b and d) and E$^1_{2g}$ (c and e) modes as a function of top gate voltages. Taken from Ref. \cite{bera}.}
 \label{bera2}
\end{figure}

The observed Raman spectra with different top gate voltages is shown in Fig.~\ref{bera2}(a). In the spectral range of 350- 450 cm$^{-1}$, two Raman modes are observed. From Table-I, these two mode are A$^{'}_1$ ($\sim$ 402 cm$^{-1}$) and E$^{'}_2$ ($\sim$ 382 cm$^{-1}$) corresponding to the bulk A$_{1g}$ and E$^1_{2g}$ modes, respectively. We will continue to use the notation A$_{1g}$ and E$^1_{2g}$ for the observed two modes. The frequency variations of the two phonon modes with the top gating (V$_{TG}$) are shown in Fig.~\ref{bera2}(b). The A$_{1g}$ mode softens by 4 cm$^{-1}$ and the E$^1_{2g}$ mode shows almost no change ($\sim$ 0.6 cm$^{-1}$) in frequency upto the maximum doping of electron concentration $\sim$ 1.8 $\times$ 10$^{13}$ /cm$^{2}$ (V$_{TG}$ $\sim$ 2 V). The FWHM of A$_{1g}$ mode increases by 6 cm$^{-1}$ (see Fig.~\ref{bera2}c), but the FWHM of E$^1_{2g}$ mode remains constant \cite{bera}. Similar to the detection principle of layer numbers of MoS$_2$ by measuring the $\omega_{A_{1g}}-\omega_{E^1_{2g}}$ \cite{lee}, the $\Delta \omega_{A_{1g}}$ can be used to estimate the doping concentration of the MoS$_2$ flake.\\

\begin{figure}[h!]
 \centering
%\leavevmode t
%\includegraphics[trim=0 0 15 15, scale=0.5]{fig_36.pdf}
\includegraphics[trim=0.5cm 16cm 0.5cm 2cm, width=1.0\textwidth]{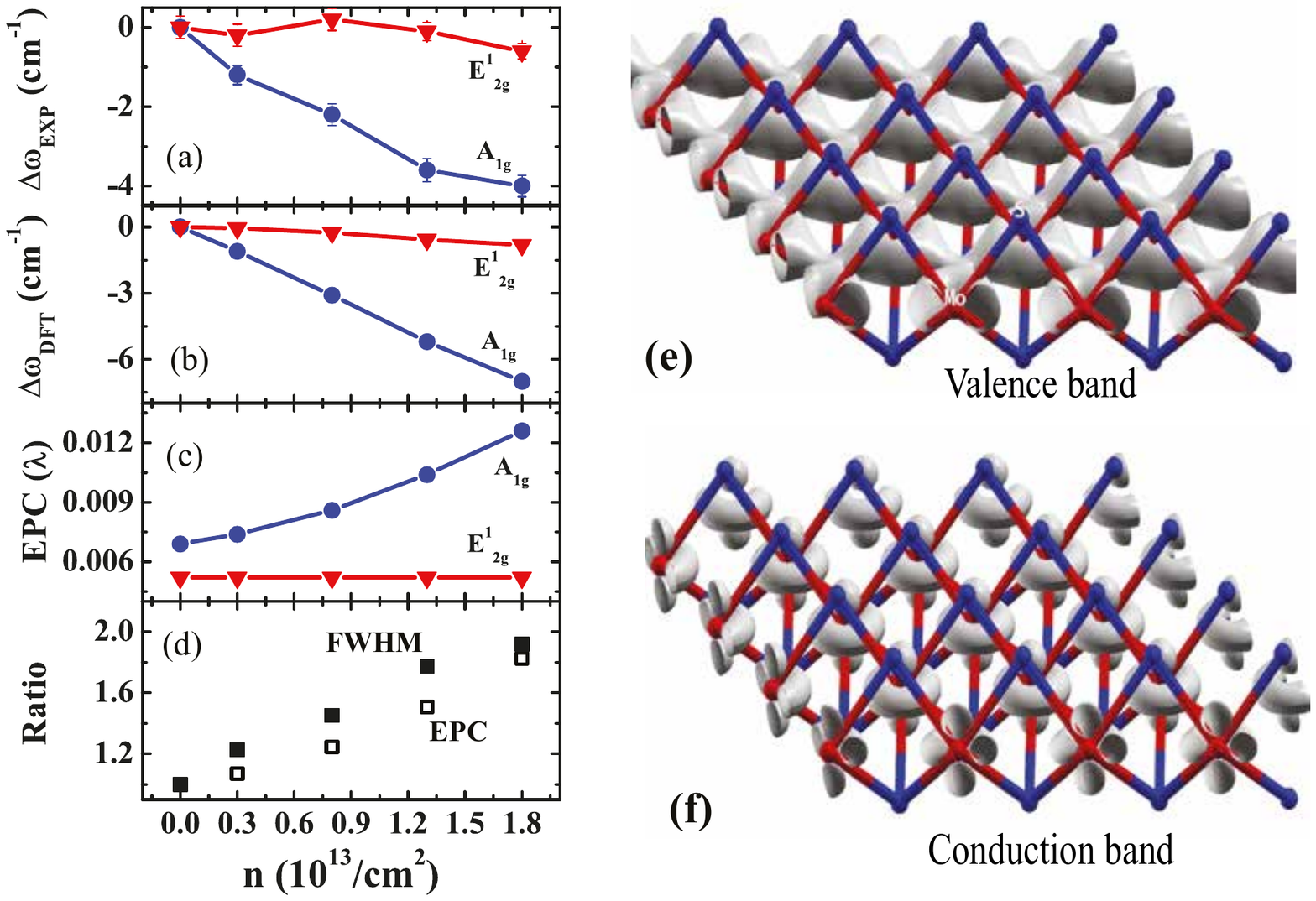}
  \caption{(color online)--Experimentally observed and the DFT calculated changes in frequency of the (a) A$_{1g}$ and (b) E$^1_{2g}$ modes versus electron concentrations (n). (c) Calculated EPC for the A$_{1g}$ and E$^1_{2g}$ modes. (d) Calculated two ratios for EPC [EPC(n $\neq$ 0)/EPC(n=0)] and FWHM [FWHM$^{total}$ (n $\neq$ 0)/FWHM$^{total}$ (n=0)] of the A$_{1g}$ mode. The charge densities ($|\psi(r)|^2$) for the (e) top of the valence band and (b) bottom of the conduction band. Here, The dark gray (red) and black (blue) spheres indicate Mo and sulfur atoms, respectively. Taken from Ref. \cite{bera}.}
 \label{bera3}
\end{figure}

In order to understand the renormalization of the A$_{1g}$ phonon mode, first-principles density functional theoretical (DFT) calculations have been carried out \cite{bera}. Single layer MoS$_2$ has a direct band gap of 1.9 eV at the K-point. The hybridized conduction band edge consists of dominated d$_{z^2}$-type Mo states and the hybridized valence band edge states having d$_{xy}$-type \cite{bera}, as shown in Figs.~\ref{bera3}(e) and (f). The square of the electronic wavefunction ($|\psi(r)|^2$) or the charge density near the valence band top and the conduction band bottom has the same full symmetry as the single layer MoS$_2$ crystal \cite{bera}. The EPC for a phonon mode $\nu$ with momentum \textbf{q} and frequency $\omega_{\textbf{q}\nu}$ is given by \cite{attac}
$\lambda_{\mbox{\scriptsize\boldmath $q$}\nu}=$
$\frac{2}{\hbar\omega_{\mbox{\scriptsize\boldmath q}\nu}N(\epsilon_f)}\sum_{\mbox{\scriptsize\boldmath k}} \sum_{ij}|g_{\mbox{\scriptsize\boldmath k+ q,k}}^{\mbox{\scriptsize\boldmath q}\nu ,ij}|^2\times\delta(\epsilon_{\mbox{\scriptsize\boldmath k+ q},i}-\epsilon_f) {\times}$ 
$\delta(\epsilon_{\mbox{\scriptsize\boldmath k},j}-\epsilon_f),$
such that the EPC matrix is given by
$g_{\mbox{\scriptsize\boldmath k + q,k}}^{\mbox{\scriptsize\boldmath q}\nu ,ij}=\left(\frac {\hbar}{2M\omega_{\mbox{\scriptsize\boldmath q}\nu}}\right)^\frac{1}{2}\langle\psi_{\mbox{\scriptsize\boldmath k + q},i}|\Delta V_{\mbox{\scriptsize\boldmath q}\nu}|\psi_{\mbox{\scriptsize\boldmath k},j}\rangle,$
where N($\epsilon_f$) is the density of states at the Fermi energy for electrons, $\epsilon_{\mbox{\scriptsize\boldmath k},j}$ the electronic energy with momentum $\textbf{k}$ in a band j, $\psi_{\mbox{\scriptsize\boldmath k},j}$ corresponding electronic wave function. Here $\Delta V_{\mbox{\scriptsize\boldmath q}\nu}$ denotes the change in the potential due to the vibrations of the lattice associated with a phonon ($\omega_{\textbf{q}\nu}$). In the A$_{1g}$ phonon mode, the atoms vibrate in such a way that the total symmetry of the single layer MoS$_2$ does not change, and hence it corresponds to the identity representation. For n-type device operations, the electrons are occupying gradually the conduction band edge states and for those electrons $|\psi(r)|^2$ transforms according to the identity representation. Hence, EPC matrix element is non-zero, in which the product of $\Delta V_{A_{1g}}$ and $|\psi(r)|^2$ is involved. The irreducible representation for the E$^1_{2g}$ mode is orthogonal to the A$_{1g}$ mode representation and hence the EPC matrix vanishes leading to no response of this mode to the channel doping \cite{bera}.\\ 

To understand dependence of FWHM on doping, the total FWHM is expressed as FWHM$^{total}$=FWHM$^{EPC}$+FWHM$^{an}$; where FWHM$^{an}$ is due to anharmonic effects. The two ratios FWHM$^{total}$ (n $\neq$ 0)/FWHM$^{total}$ (n=0) and EPC(n $\neq$ 0)/EPC(n=0) have been calculated for the A$_{1g}$ mode and they show the same trend as shown in Figs.~\ref{bera3}(d). This explains that only EPC is responsible to broaden the FWHM of A$_{1g}$ mode with the doping. It is noteworthy that the top gating by an amount of $\sim$ 1.8 $\times$ 10$^{13}$ /cm$^{2}$ leads to hardening of the G phonon mode by $\sim$ 10 cm$^{-1}$ in case of graphene \cite{bisu2}, while the same condition gives rise to softening (by $\sim$ 4 cm$^{-1}$) of A$_{1g}$ mode for single layer MoS$_2$. In case of graphene adiabatic approximation fails, whereas in case of MoS$_2$, EPC is within the adiabatic approximation.

\section{Resonant Raman scattering of bulk MoS$_2$ tuned by temperature and pressure: stability and contributions of A and B excitons to resonance}

\begin{figure}[h!]
 \centering
%\leavevmode t
%\includegraphics[trim=0 0 15 15, scale=0.5]{fig_37.pdf}
\includegraphics[trim=0.5cm 20cm 0.5cm 2cm, width=1.0\textwidth]{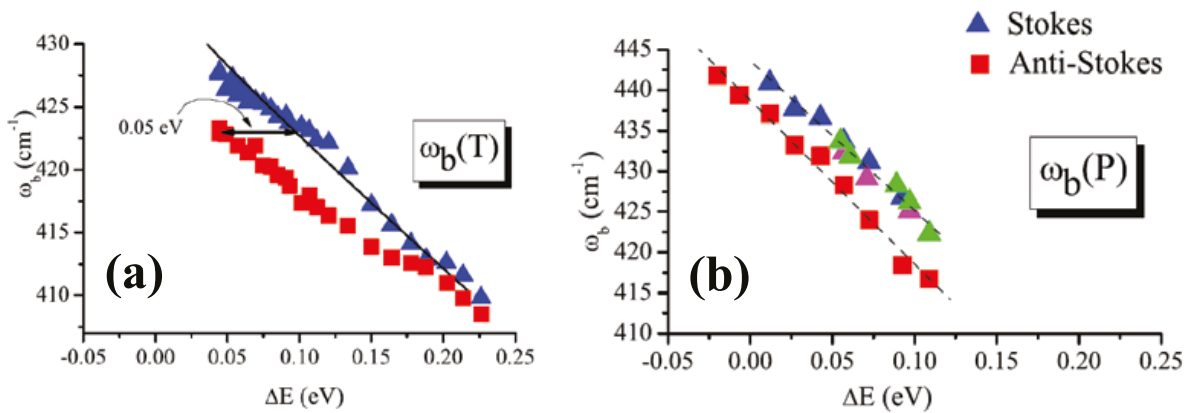}
  \caption{(color online)--(a) Temperature and (b) pressure dependent $\omega^S_b$ and $\omega^{AS}_b$  as a function of the energy difference $\Delta E=E_L-E_{A1}$ . Solid line in (a) indicates the fitting for $\omega^S_b$.  Taken from Ref. \cite{livn}.}
 \label{livn2}
\end{figure}

Resonant Raman scattering at ambient conditions on single and multilayer MoS$_2$ has been discussed \cite{bisu}, where 633 nm red laser line was used for excitation. It has been shown that there exists two excitons corresponding to direct transitions at the K-point \cite{hoorn}. The exciton A having binding energy 42 meV corresponds to K$_4$ $\rightarrow$ K$_5$ optical transition, whereas B exciton of binding energy 134 meV corresponding to K$_1$ $\rightarrow$ K$_5$ optical transition. The energy difference between the A and B excitons is due to the spin-orbit splitting and interlayer interaction \cite{hoorn}. Since, the pressure and temperature variations affect the lifetime of the excitons, we will address their stability and the limits of applied pressure and temperature to which the resonance effects can be observed. The appearance of all the Raman active modes and inactive modes (E$^2_{1u}$ and B$_{1u}$) due to resonance effects is already discussed \cite{bisu} in the section V. Here, our focus will be on the $^{'}$b$^{'}$ mode attributed to the two-phonon Raman process (see Fig.~\ref{sekine}) such that it corresponds to the emission of a dispersive quasi-acoustic (QA) longitudinal phonon followed by emission of a dispersionless transverse optical mode (here c-band) \cite{sekine}. The more interesting feature about the b mode is that the observed frequency for the Stokes ($\omega^S_b$) and anti-Stokes ($\omega^{AS}_b$) spectra is not the same and they differ by $\sim$ 4.7 cm$^{-1}$, as shown in Fig.~\ref{livn2}. \\

Since, longitudinal phonons are strongly coupled to excitons than the transverse ones, the contribution of QA phonons will be more dominant to the b mode \cite{sekine}. The rate at which frequency decreases with temperature ($\partial \omega/\partial T$) for the b mode is $\sim$ 2.66 $\times$ 10$^{-2}$, whereas for all the other modes it is in the range of $\sim$ 1.0--1.6 $\times$ 10$^{-2}$. Fig.~\ref{livn2}(a) and (b) shows the comparison between the shifting of the peak position of the b band for the Stokes and anti-Stokes scattering as a function of temperature and pressure, respectively. Here, the $\omega^{AS}_b$ is redshifted compared to the $\omega^{S}_b$ part.

Since, the intermediate exciton state plays an important role for the observed b band and from the above experimental findings, three major things need to be addressed: (i) identification of tuned excitonic states with respect to temperature and pressure to observe resonance effects for Stokes as well as anti-Stokes processes, (ii) the stability of the A and B excitons and, (iii) the observed redshifted frequency $\omega^S_b$-$\omega^{AS}_b$ $\sim$ 4.7 cm$^{-1}$. As we already noted \cite{bera}, the d$_{z^2}$ ( this orbital points along the c-axis) character of the K$_5$ final state predominantly couples the A$_{1g}$ phonon with the excitons than the E$^1_{2g}$ mode. This means that the A$_{1g}$ phonon mode is much more sensitive to the resonance effect than the E$^1_{2g}$. Hence the resonance effect will be captured from the intensity ratio by comparing $[I_{A_{1g}}/I_{E^1_{2g}}]^{reso}$ with the $[I_{A_{1g}}/I_{E^1_{2g}}]^{non-reso}$ (i.e. with 514 nm laser line). Here, the measured $[I_{A_{1g}}/I_{E^1_{2g}}]^{reso}$ implies the Raman scattering cross-section. The intensity ratio $[I_{A_{1g}}/I_{E^1_{2g}}]^{reso}$ is denoted by R$^S$ for Stokes process and by R$^{AS}$ for anti-Stokes process.

The probability of the Raman scattering of a phonon by an excitonic 1s state is given by \cite{cardona,livn} 
$$ P_{phonon}\approx (2 \pi \hbar) \left|\frac{\bra{0}H_{eR}(\omega_L)\ket{1}\bra{1}H_{e-ion}\ket{1}\bra{1}H_{eR}(\omega_s)\ket{0}}{(E_i-\hbar\omega_L-i\Gamma_1)(E_i-\hbar\omega_s-i\Gamma_1)}\right|^2.$$ Here, $H_{eR}$ and $H_{e-ion}$ represents the electron-photon and electron-ion interaction Hamiltonian, respectively. The states indicated by 0 and 1 correspond to the ground and the exciton states (A$_1$ and B$_1$), respectively. The incident and scattered laser frequencies are given by $\omega_L$ and $\omega_s$, respectively. E$_i =\hbar\omega_L$ or =$\hbar\omega_s$ depending on incoming or outgoing resonance, respectively. It has been shown experimentally that the Raman scattering cross section for the b-band shows an peak at E$^A_{1s}+\hbar\omega_b$ (for the A exciton), but does not correspond to energy of the exciton 1s state i.e. at E$^A_{1s}$ \cite{sekine}. Here, the outgoing probability for the Stokes and anti-Stokes Raman scattering of A$_{1g}$ mode is given by $P^{S}_{A_{1g}}$ and $P^{AS}_{A_{1g}}$ corresponding to $E_i+\hbar\omega_{A_{1g}}$ and $E_i-\hbar\omega_{A_{1g}}$ energy levels, respectively. Ho et al \cite{ho} have measured the temperature dependence of the exciton energies by fitting a formula given by $E_{i}(T)=E_{iPo}-a_{iP}[1+\frac{2}{exp(\Theta_{iP}/T)-1}]$; where, $a_{iP}$ and $\Theta_{iP}$ represent the exciton-phonon interaction strength and the average phonon temperature, respectively. The measured values of E$_{iPo}$, $a_{iP}$ and $\Theta_{iP}$ are for A$_1$ are 1.976 eV, 46 meV and 220 K, respectively; and for B$_1$ those are 2.179 eV, 42 meV and 200 K, respectively \cite{ho} for T $\leq$ 300K. The broadening parameter has also been measured by fitting the equation $\Gamma(T)=\Gamma_0+\frac{\Gamma_{LO}}{exp(\Theta_{LO}/T)-1}$; where $\Gamma_{LO}$ = 75 meV, represents the exciton-LO phonon interaction strength and $\Theta_{LO}$=560 K. The $\Gamma_0$ for A$_1$ and B$_1$ excitons are 18.0 and 37.4 meV, respectively. Hence, it is clear that the broadening is because of exciton-optical phonon interactions \cite{ho}.\\

\begin{figure}[h!]
 \centering
%\leavevmode t
%\includegraphics[trim=0 0 15 15, scale=0.5]{fig_38.pdf}
\includegraphics[trim=0.5cm 15cm 0.5cm 2cm, width=1.0\textwidth]{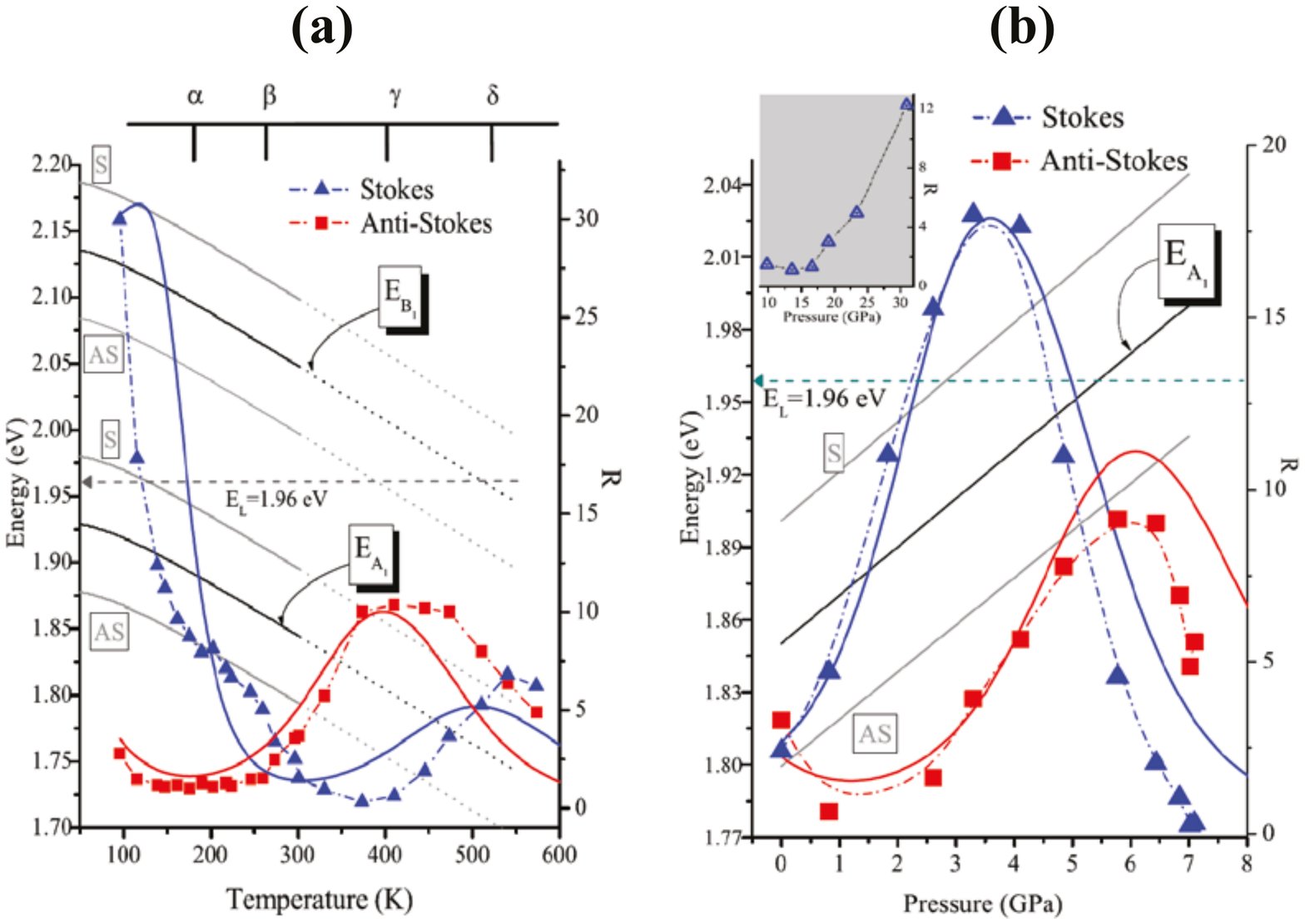}
  \caption{(color online)--(a) (corresponding to the left scale) Temperature (measured for T $\leq$ 300K and then extended to 550K) dependent binding energies (E$_{A_1}$ and E$_{B_1}$) of the A$_1$ and B$_1$ excitons. (b) Pressure dependent binding energy for the A$_1$ exciton is plotted (corresponding to the left scale). Here AS and S represent the outgoing resonance energies $E_i+\hbar\omega_{A_{1g}}$ and $E_i-\hbar\omega_{A_{1g}}$ for anti-Stokes and Stokes Raman scattering, respectively. The measured intensity ratios R$^S$ for Stokes process, R$^{AS}$ for anti-Stokes process and the corresponding calculated probabilities ($P^{S}$ and $P^{AS}$) are plotted together (corresponding to the right scale) in (a) and (b). The inset in (b) shows the R$^S$ for pressures above 10 GPa. The upper scale in (a) showing $\alpha$-$\delta$ represent different temperature regions of operations discussed in the text. Taken from Ref. \cite{livn}.}
 \label{livn4}
\end{figure}

The temperature dependent binding energies of the A$_1$ and B$_1$ excitons are shown in Fig.~\ref{livn4}(a). The measured intensity ratio R$^{S}$ and R$^{AS}$, and the calculated probabilities $P^{S}_{A_{1g}}$ and $P^{AS}_{A_{1g}}$ are also plotted (blue and red solid lines). The outgoing resonance energies for the $A_{1g}$ mode are also shown for Stokes and anti-Stokes scattering processes of A$_1$ and B$_1$ excitons. The interactions of excitons with the acoustic and optical phonons are responsible for the observed temperature dependent shift; and also for the variations of the lattice parameters. Since the exciton-optical phonon interaction holds for the high temperature, the fitted parameters are extended to 550 K. The agreement between the theory and experiment is reasonable. From Fig.~\ref{livn4}(a), it is clear that at T=0K the excited laser line (E$_L$=1.96 eV) lies between $E_{A1}$ and $E_{A1}+\hbar\omega_{A1g}$. As we increase the temperature the $E_{A1}+\hbar\omega_{A1g}$ (Stokes-process) approaches to E$_L$ and around $\approx$ 95 K, outgoing resonance shows an peak for the R$^{S}$ i.e. for the Raman cross section. Now with the further increase of temperature for T $>$ 175 K (marked as $\alpha$ on the upper scale) $E_{A1}+\hbar\omega_{A1g}$ goes away from the E$_L$ and $E_{B1}-\hbar\omega_{A1g}$ approaches E$_L$ and continues upto $\sim$ 260 K (marked as $\beta$). In the boundary of $175K \leq T\leq260K$, the resonance effects are of mixed type: due to the proximity effects of E$_L$ to the Stokes/anti-Stokes shifted outgoing resonance for the A$_1$ and B$_1$ excitons. With increase of T further, R$^{S}$ goes to a minimum value around 350K, whereas R$^{AS}$ reaches a maximum value at $\sim$ 400K (marked as $\gamma$). For T $\geq$ 520 K (marked as $\delta$), R$^{S}$ and R$^{AS}$ follows the non-resonant temperature dependence behavior of $exp(-\hbar\omega_{A1g}/kT)$ \cite{livn}.

Connell et al has shown experimentally that the binding energy of the A$_1$ exciton decreases with pressure at the rate of $\sim$ 7 meV/GPa at room temperature \cite{conel}. The decreasing of the exciton binding energy is associated with the pressure induced free charge carriers and hence more screening. Therefore, at $\sim$ 6 GPa and higher values, the A$_1$ exciton cease to exist. The pressure variations of the binding energy ($\sim$ 0.02 eV/GPa \cite{livn}), Stokes and anti-Stokes shifted energies of the A$_1$ exciton are plotted in Fig.~\ref{livn4}(b) along with the measured (R$^{S}$ and R$^{AS}$) and calculated ($P^{S}_{A_{1g}}$ and $P^{AS}_{A_{1g}}$) resonance Raman scattering cross sections. The pressure dependent broadening parameter $\Gamma(P) \approx 35+3P$ (in meV) has been taken to calculate the $P^{S}_{A_{1g}}$ and $P^{AS}_{A_{1g}}$ values. According to calculations, the peak positions for the $P^{S}_{A_{1g}}$ and $P^{AS}_{A_{1g}}$ are $\sim$ 3.8 GPa and $\sim$ 6.3 GPa, respectively \cite{livn}. Since, $E_{A1}+\hbar\omega_{A1g}$ is closer to E$_L$ at low pressure regime, R$^{S}$ increases and peaks at $\sim$ 3.8 GPa in accordance with the theoretical value. At higher pressure, $E_{B1}-\hbar\omega_{A1g}$ approaches to E$_L$ (not shown in the figure) and hence, R$^{AS}$ increases to the peak value at $\sim$ 6 GPa. The deviation of $P^{AS}_{A_{1g}}$ from R$^{AS}$ for the pressure $\geq$ 6 GPa is because of the dissociation of the A$_1$ exciton into free electron-hole pairs \cite{livn}. Although the A$_1$ exciton is unstable in the high pressure range ($\geq$ 6 GPa), the measured R$^{S}$ values with the increased pressure (P$\geq$ 10 GPa) is shown in the inset of Fig.~\ref{livn4}(b). Livneh et al attributed this observed high pressure regime of R$^{S}$ to the pressure induced change of electronic band structure, and by that the enhancement of the resonant sensitive A$_{1g}$ mode couplings. This point needs to be cleared further from theoretical point of view.

To address the stability of the B exciton, Livneh et al \cite{livn} compared the observed intensity ratio I$^{AS}$/I$^S$ and the calculated ratio $P^{AS}_{A1g}$/ $P^{S}_{A1g}$ (normalized by $n(\omega_{ph})/[n(\omega_{ph})+1]$, where $n(\omega_{ph})=1/[exp(\omega_{ph})-1]$ is the Bose-Einstein factor) of the A$_{1g}$ and E$^1_{2g}$ modes for anti-Stokes and Stokes Raman scattering as a function of 1/T and proposed that as long as the ratio I$^{AS}$/I$^S$ of the A$_{1g}$ mode follows the resonant feature, B$_1$ exciton is stable; and hence it is unstable for T $\geq$ 520 K. 

It has been observed that the b-band shifts to lower frequency with the increase of the laser energy and reaches to the frequency of A$_{1g}$ mode \cite{sekine}. This dispersive mode was assigned as the two phonon process having dominant contribution from the dispersive QA mode. In order to address the observed $\omega^S_b$-$\omega^{AS}_b$ $\sim$ 4.7 cm$^{-1}$ feature, the temperature and pressure dependent $\omega^S_b$ and $\omega^{AS}_b$ are plotted with the energy difference $\Delta E=E_L-E_{A1}$ in Figs.~\ref{livn2}(a) and (b), respectively. The temperature dependent $\omega^S_b$ through $\Delta E$ data is fitted (see Fig.~\ref{livn2}a) according to the equation $\hbar\omega_b(q_z,T)=\Delta E (T)-\hbar^2\left(|q_z|\pm \omega_L \frac{n_0}{c}\right)^2/2M_{II}^{A1}+\hbar\omega_{TO}(T).$ Here, n$_0$ represents the ordinary refractive index. M$_{II}^{A1}$ represents the effective mass of the $A_1$ exciton along the c-axis. The $\pm$ sign refers to the $\alpha$ and $\beta$ processes (see Fig.~\ref{sekine}). The q$_z$ and k$_z$ represent wavevector of the QA mode and the exciton pointing along the c-axis, respectively. The value of $\hbar\omega_b(0,300)=52.4$ cm$^{-1}$ is used as a single fitted parameter. The extracted value of the frequency for the QA mode $\sim$ 56 $cm^{-1}$ is consistent with the earlier observations from neutron scattering \cite{waka}. The calculated temperature coefficient of the QA mode is given by $\left(\frac{\partial\omega_{QA}}{\partial T}\right)_{P=0}\approx -2 \times 10^{-3} $ in cm$^{-1}$/K for $\Delta E \leq$ 110 meV \cite{livn}; where, $\omega_{QA}=\omega_{b}-\omega_{c}$. The observed $\left(\frac{\partial\omega_{b}}{\partial T}\right)_{P=0}\approx -2.66 \times 10^{-2} $ in cm$^{-1}$/K is much larger than the $\left(\frac{\partial\omega_{QA}}{\partial T}\right)_{P=0}$. Hence, the observed high temperature coefficient of the b-band compared to all the observed mode is attributed to the tuning of the exciton energy with respect to the temperature \cite{livn}. The largest shift between the Stokes and anti-Stokes scattering of the b-band for which $\omega^S_b$=$\omega^{AS}_b$ (see Fig.~\ref{livn2}a) is given by 50 meV for $\Delta E \leq$ 110 meV, which is close to the energy of c-band (TO mode $\sim$ 377 cm$^{-1}$) \cite{bisu,livn}. Finally, $\omega^S_b$-$\omega^{AS}_b$ is attributed to the reverse order processes for the Stokes and ant-Stokes scattering: the inner photon-like state scatters to the outer exciton-polariton state by absorbing a TO phonon and then scatters back to inner high-energy photon-like state by absorbing a QA mode, i.e. for anti-Stokes process it is TO+QA order, whereas for Stokes scattering it is of QA+TO order. So, the redshift between $\omega^S_b$ and $\omega^{AS}_b$ is due to the order of the involvement of the dominant quasi-acoustic (QA) phonon in resonant two-phonon scattering process.   

\section{Photoluminescence of MoS$_2$: control of valley polarization by optical helicity and observation of trions}

\subsection{Layer Dependence} 
\begin{figure}[h!]
 \centering
%\leavevmode t
%\includegraphics[trim=0 0 15 15, scale=0.5]{fig_39.pdf}
\includegraphics[trim=0.5cm 23cm 0.5cm 1cm, width=1.0\textwidth]{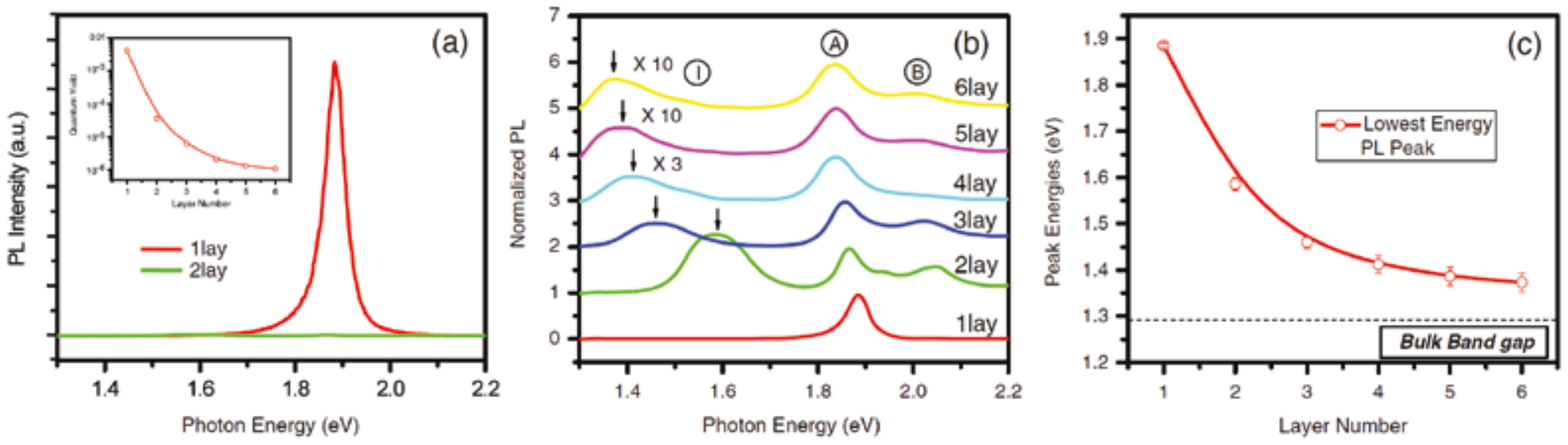}
  \caption{(color online)--(a) Room temperature PL spectra for suspended monolayer and bilayer samples. The inset shows the changes in QY with different layers (N=1-6). (b) The normalized PL spectra (with respect to the A-peak intensity) for all the layers of MoS$_2$. Here, $\times$ 3 and $\times$ 10 indicate the magnification of the I-peak in the corresponding spectra. (c) The A-peak energy for N=1 and the I-peak energy for N=2-6 plotted together showing the band-gap variations with the number of layers. The dashed line represents the indirect band gap of the bulk-MoS$_2$. Taken from Ref. \cite{mak}.}
 \label{makn1}
\end{figure}

The phenomena of radiative recombination of optically excited electron (e) and hole (h) pairs of a band-gap semiconductor is known as photoluminescence (PL). PL study has been employed extensively to probe excitons and the band gap of a semiconductor. As discussed that the bulk MoS$_2$ undergoes a transition from an indirect band gap to a direct one in the monolayer limit \cite{mak}, PL experiments are useful to probe the transition with the number of layers. Fig.~\ref{makn1}(a) shows the comparison of the PL spectra for the monolayer and bilayer MoS$_2$. The striking difference with the addition of another layer is observed for the PL spectra, which is characterized by the photoluminescence quantum yield (QY). The PL spectra with the layer N=1 to 6 is shown in Fig.~\ref{makn1}(b). The observed peaks are marked as 'A', 'B' and 'I'. The extracted QY for the bright PL of the monolayer is $\sim$ 4 $\times$ 10$^{-3}$; whereas for N=2-6, it is of $\sim$ 10$^{-5}$ to 10$^{-6}$ \cite{mak}. The inset of Fig.~\ref{makn1}(a) shows the variations of the QY with the layer numbers. The observed peak A centered at $\sim$ 1.90 eV corresponds to the direct exciton transition at the K-point of the BZ (K$_4$ $\rightarrow$ K$_5$ optical transition); while the B peak centered at $\sim$ 2.05 eV corresponds to the direct exciton transition K$_1$ $\rightarrow$ K$_5$. The B peak is also observed for the monolayer MoS$_2$ deposited on SiO$_2$/Si substrate \cite{splend,eda}. The A peak having width $\sim$ 50 meV (for N=1) redshifts very little and broadens with the thickness of the sample. The peak marked as I starts to appear with the bilayer and shifts to lower energy with the N. The broad peak I is attributed to phonon assisted PL peak of the indirect band gap MoS$_2$. Since, the states near conduction band edge at the K-point of the BZ have d$_{z^2}$ character and Mo atoms reside in between two planes of S atoms, these states are less affected for the interlayer couplings; whereas at the $\Gamma$ point, both valence and conduction bands are linear combination of d-states (Mo) and p-states (S) and hence are more affected with the layer stacking. Therefore, indirect band gap is increased due to the confinement effects, direct band gap at the K-point remains almost constant. The peak energy of the A peak for monolayer and the peak energy of the I peak for all the other layers are plotted together in Fig.~\ref{makn1}(c), which shows that the band gap decreases and approaches to the bulk value with the layers.\\

The direct band gap at K-point does not change with the layers, but the direct excitonic transition at the K-point gives different QY for the monolayer and the thick layer/bulk. In order to address this difference, the quantum efficiency for the MoS$_2$ is given by $\eta \approx \tau_{rad}/(\tau_{rad}+\tau_{defect}+\tau_{relax})$ \cite{splend}; where, $\tau_{rad}$ represents the radiative recombination rate, $\tau_{defect}$ the rate of defect induced scattering, and $\tau_{relax}$ the rate of intraband relaxation of charge carriers. The phonon-assisted $\tau_{relax}$ is very large for the indirect band gap semiconductor. Because of the presence of the direct band gap of the monolayer MoS$_2$, the decay rate via phonons decreases enormously i.e. $\tau_{relax} \approx 0$. And $\tau_{rad}$ is nearly constant for both the bulk and monolayer. Hence, the enhanced QY limited by $\tau_{defect}$ for monolayer is due to the quenched channel of $\tau_{relax}$ with respect to the bulk value.\\

\begin{figure}[h!]
 \centering
%\leavevmode t
%\includegraphics[trim=0 0 15 10, scale=0.5]{fig_40.pdf}
\includegraphics[trim=0.5cm 15cm 0.5cm 1cm, width=1.0\textwidth]{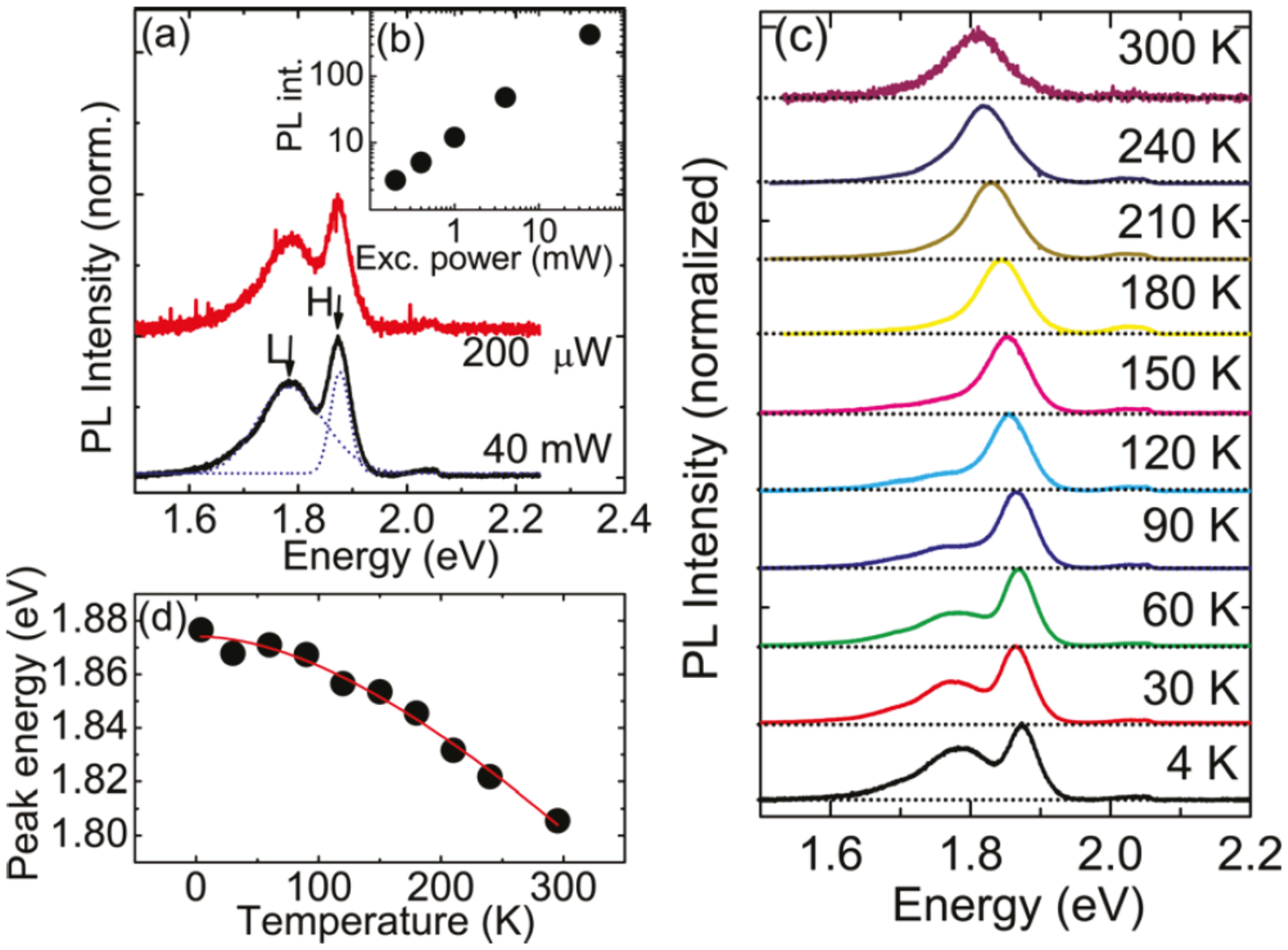}
  \caption{(color online)--(a) The normalized PL spectra measured at 4K on monolayer MoS$_2$ deposited on SiO$_2$/Si substrate for two different laser (532 nm) powers. The two peaks are fitted with Gaussian functions. (b) The total PL intensity (integrated area of the two peaks) as a function of laser power. (c) The normalized PL spectra for different temperatures. (d) The energy of the H-peak versus temperature. Solid line represents the fitting of the Varshni formula. Taken from Ref. \cite{korn}.}
 \label{korn1}
\end{figure}

\begin{figure}[h!]
 \centering
%\leavevmode t
%\includegraphics[trim=0 0 15 11, scale=0.5]{fig_41.pdf}
\includegraphics[trim=0.5cm 16cm 0.5cm 1cm, width=1.0\textwidth]{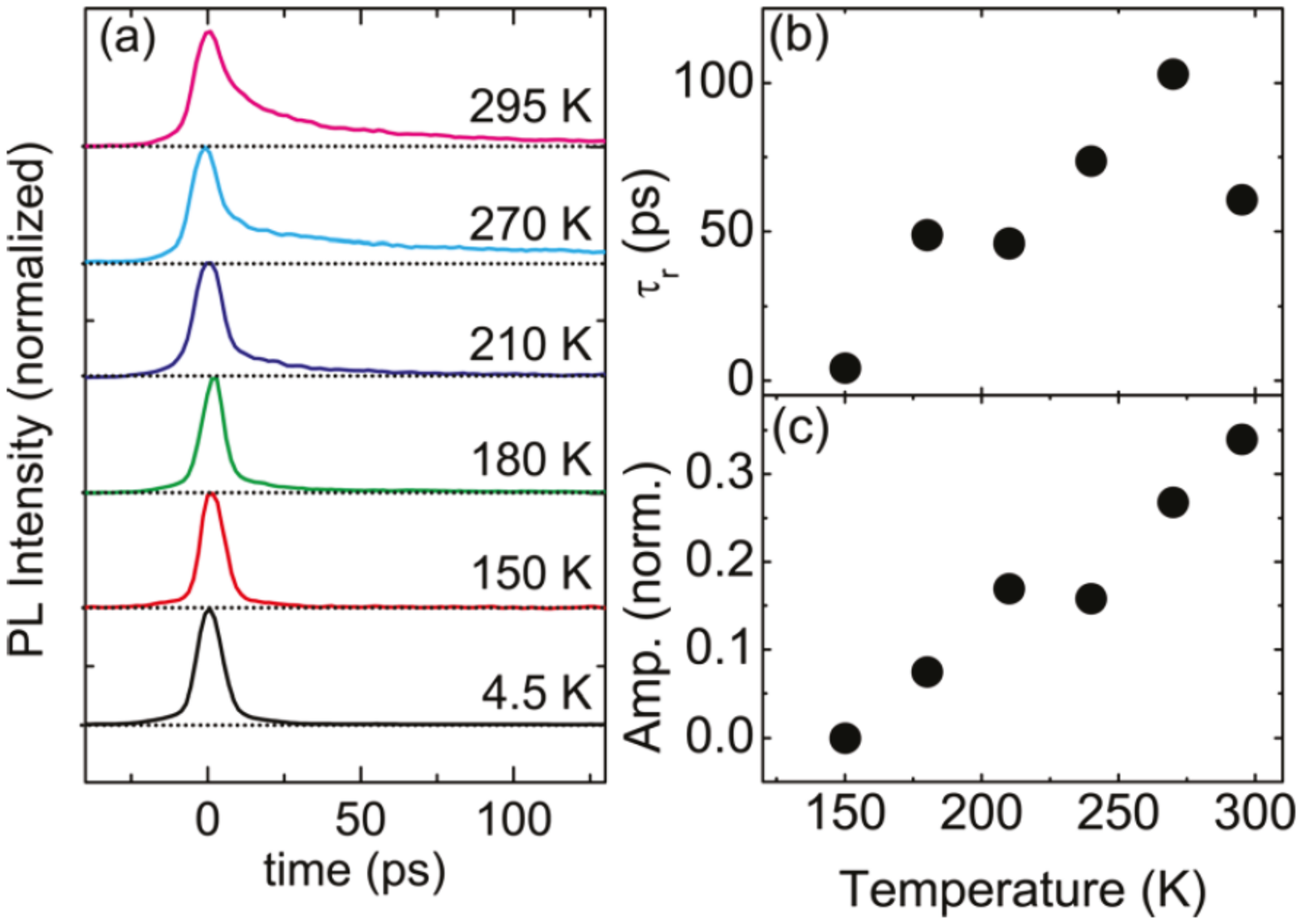}
  \caption{(color online)--(a) Time-resolved PL for different temperatures using 402 nm laser line. (b) The extracted decay rate and (c) the amplitude of the LLC. Taken from Ref. \cite{korn}.}
 \label{korn2}
\end{figure}

\subsection{Temperature Dependence} At the lowest temperature ($\sim$ 4 K), the observed PL spectra of monolayer MoS$_2$ for the two different laser powers (200 $\mu$W and 40 mW) are shown in Fig.~\ref{korn1}(a). The lowest broad peak is marked as 'L' and the higher narrow peak is marked as 'H'. The L peak is separated from H by $\sim$ 90 meV ($\Delta E_s$). With the increasing power of the laser line by more than two orders, the shape of the spectra does not change. Fig.~\ref{korn1}(b) shows the variation of the total integrated intensity of the two peaks with the laser power, which shows linear relationship. It does not show the onset of the PL emission due to the presence of the non-radiative recombination channel of charge carriers to the defects at the low power. The constant relative intensity of the two peaks with the laser power indicates that the two radiative recombination channel do not interfere each other with the increasing population of the charge carriers \cite{korn}. PL spectra at various temperatures (T) are shown in Fig.~\ref{korn1}(c). The two peaks redshift and broaden with the increased T. The intensity of the L peak decreases and almost vanishes above 120K. The $\Delta E_s$ remains constant  before the vanishing of L peak. The redshift of H peak is plotted with T as shown in Fig.~\ref{korn1}(d). The observed data is fitted with the Varshni formula, $E_g(T)=E_g(0)-[\alpha T^2/(T+\beta)]$; where E$_g(0)$=1.874 eV, $\alpha$=5.9 $\times$ 10$^{-4}$ eV/K and $\beta$=430K. The H peak is attributed (tentatively) to the direct free excitonic direct transition at the K-point. The L peak is assigned due to the bound excitons with the defects. Due to the attraction of the neutral impurities/defects the binding energy of the bound-exciton decreases \cite{cardona}. Because of the distribution of various defects with different energies in the sample, the observed L peak is broadened. The non-radiative recombination process increases with T and it is the responsible for the disassociation of the bound-exciton above 120K rather than the thermal activation processes (the binding energy is larger than the 26 meV).

Fig.~\ref{korn2}(a) shows the time resolved PL spectra at different temperatures. It is clear that the PL decays in 5 ps from 4K to 150K. This fast decay is in accordance with the observation of relative intensity dependence on laser power at low T i.e. there is no non-radiative relaxation process of the optically generated e-h pairs down to the minimum energy. We can see from Fig.~\ref{korn2}(a) that above 150K, there is a an appearance of another long-lived component (LLC). The extracted decay rate and the amplitude of the LLC are plotted in Figs.~\ref{korn2}(b) and (c), respectively. The decay rate of LLC increase from 50 ps at 150K to 100 ps at 270K and then decreases to 70 ps st 300K. The amplitude of LLC also increases with T, shown in Figs.~\ref{korn2}(c). The appearance of LLC and its increased decay rate are attributed to the onset of non-radiative phonon assisted scattering channel of exciton-polaritons \cite{korn}.\\

\begin{figure}[h!]
 \centering
%\leavevmode t
%\includegraphics[trim=0 0 15 11, scale=0.5]{fig_42.pdf}
\includegraphics[trim=0.5cm 14cm 0.5cm 1cm, width=1.0\textwidth]{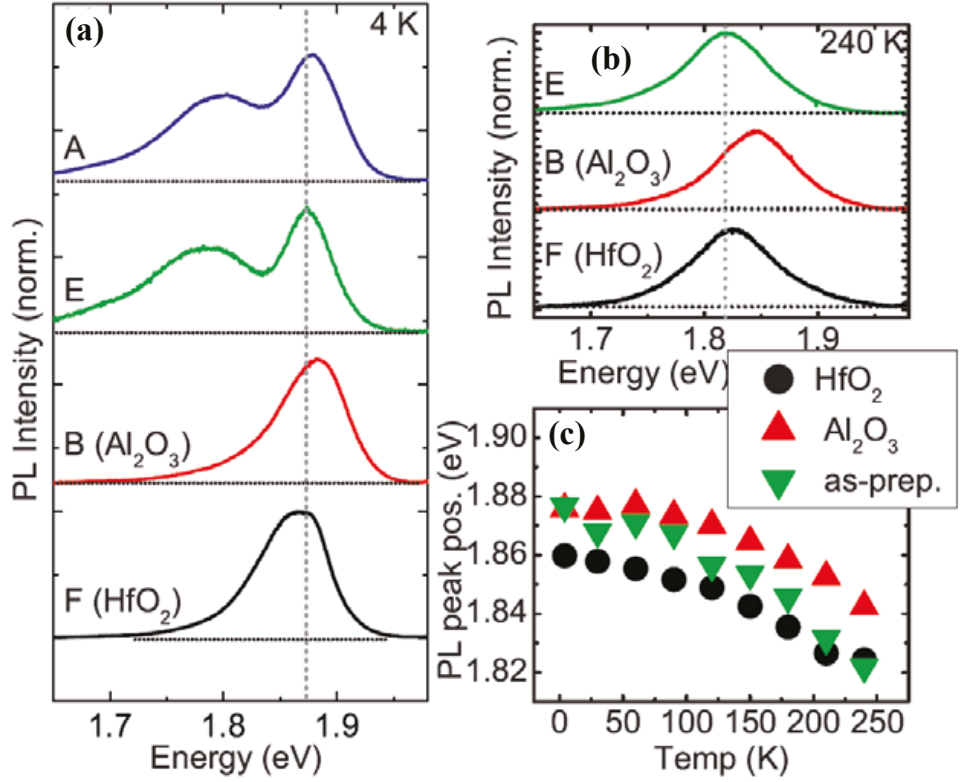}
  \caption{(color online)--PL spectra of (a) four samples (measured at 4K) and (b) three sample (measured at 240K). The A and E samples (in a) are not covered with oxide layers, whereas B and F are covered. (c) The high-energy peaks for three samples (shown in b) as a function of temperature. Taken from Ref. \cite{plech}.}
 \label{plech}
\end{figure}

Plechinger et al \cite{plech} made two kind of samples of monolayer of MoS$_2$ to record the PL spectra: first kind (A and E) is of by mechanical exfoliation method on SiO$_2$/Si substrate and the second type of sample (B and F) is covered with the oxide layer (HfO$_2$ or Al$_2$O$_3$) of 15 nm thickness. The observed PL spectra at 4K, shown in Fig.~\ref{plech}(a), present two kind of features: (1) A and E samples not covered with the oxide layer show two prominent peaks, and (2) B and F covered with the oxide layers show only one prominent peak. Since, the top oxide layers and the high temperatures used to make them remove most of the impurities from the surface, the low energy broad peak is attributed to the bound-excitons (to the surface impurities). The vertical dashed line in Fig.~\ref{plech}(a) indicates the redshift of F sample covered with HfO$_2$ ($\epsilon_r$ $\approx$ 20) and blueshift of B sample covered with Al$_2$O$_3$ ($\epsilon_r$ $\approx$ 10) compared to the without oxide layer samples (A and E). The screening of the long range Coulomb interaction does not play the role for the shifting of the B and F samples, otherwise for F sample the blueshift could have been be more than the B sample. Fig.~\ref{plech}(c) shows the temperature variations (4 to 240K) of the high energy peak for the two kind of samples. The largest shift ($\approx$ 55 meV) occurs for the A and E type samples, whereas for the B sample it is of $\approx$ 34 meV and for F type it is of $\approx$ 36 meV. Around 240K, the oxide covered samples show the blueshited peak compared to the without covered samples, shown in Fig.~\ref{plech}(b). Since, the atomic layer deposition process of making oxide layers involves temperature variations from 520K to below room temperature and the thermal expansion coefficient of MoS$_2$ is larger than the oxide layers, the sandwiched monolayer MoS$_2$ remain strained and hence the larger peak shifts are observed for the as-prepared samples than the oxide coated \cite{plech}.

\subsection{Effect of circular polarization of the incident light on photoluminescence: valley selectivity} 

\begin{figure}[h!]
 \centering
%\leavevmode t
%\includegraphics[trim=0 0 15 11, scale=0.5]{fig_43.pdf}
\includegraphics[trim=0.5cm 5cm 0.5cm 2cm, width=0.7\textwidth]{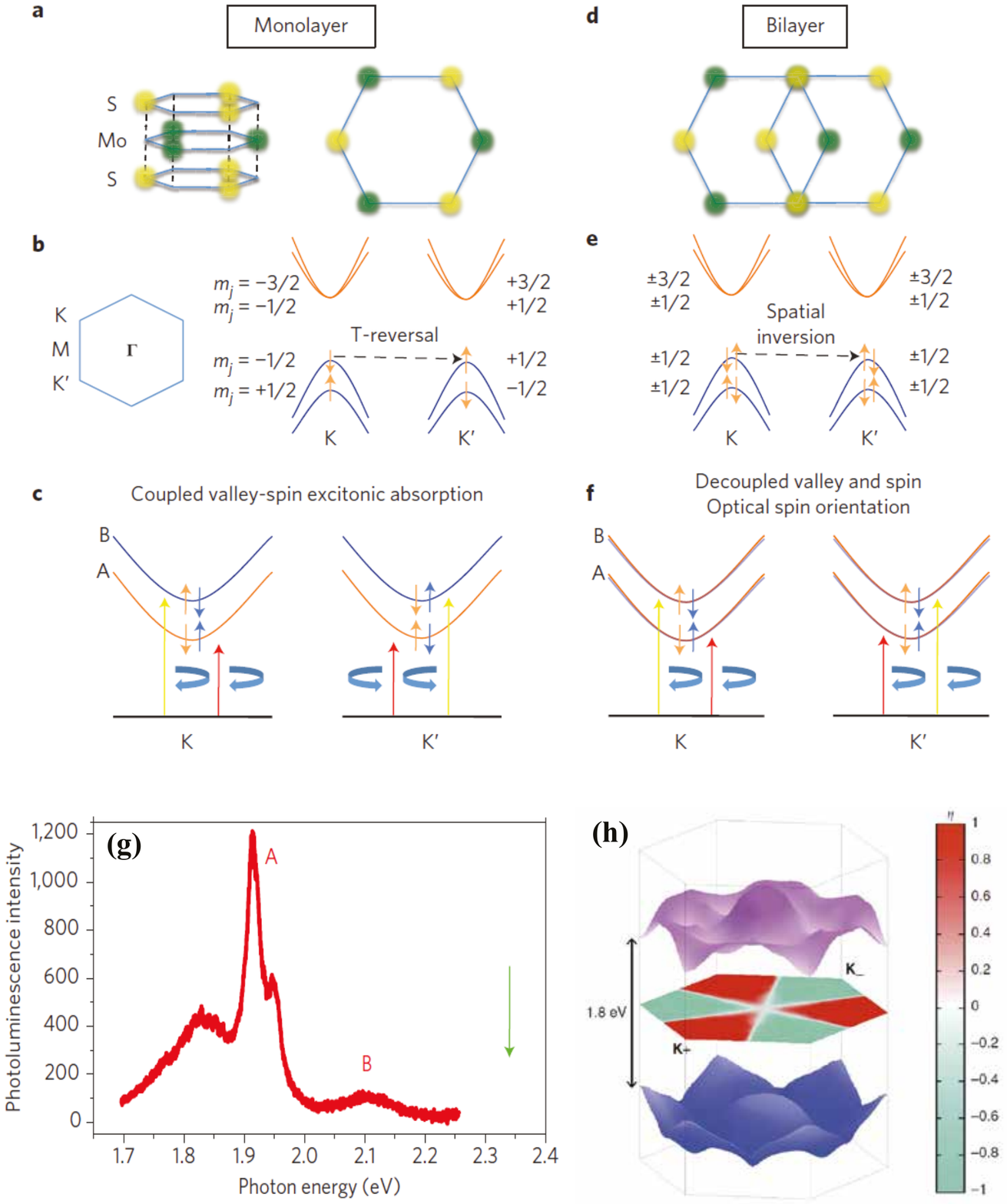}
  \caption{(color online)--(a) The hexagonal lattice structure for single layer MoS$_2$ and (d) the Bernal stacking for bi-layer MoS$_2$ . The valence and the conduction bands (marked by the corresponding total angular momentum) across the energy gap at the K and K$^{'}$ points for (b) single layer and (e) for bilayer. Optical selection rules with the applied left-handed and right-handed circularly polarized lights (c) for single layer MoS$_2$ at the K and K$^{'}$ points, respectively; whereas for left-handed circularly polarized lights (f) in case of bilayer. (g) The unpolarized PL spectrum observed at 14K using 2.33 eV laser line for single layer MoS$_2$. Taken from Ref. \cite{mak2}. (h) The top and the bottom figures represent the conduction (pink) and the valence (blue) bands; and the middle one represents the calculated circular polarization $\eta (\textbf{k},\omega_{cv})$ at different points of the Brillouin zone. The right scale bar represents the corresponding colors of the calculated $\eta (\textbf{k},\omega_{cv})$. Taken from Ref. \cite{cao}.}
 \label{makn2}
\end{figure}

Even layers of MoS$_2$ has inversion symmetry, whereas odd layers and hence monolayer MoS$_2$ (D$^1_{3h}$) has no inversion symmetry. Because of no inversion symmetry, spin-orbit coupling splits the valence band of monolayer MoS$_2$ by $\sim$ 160 meV at the K and K$^{'}$ points \cite{xiao,rama} (see Fig.~\ref{makn2}b). At the band edge of K points, the symmetry adapted hybridized d-states on Mo are given by $\ket{\phi_c}=\ket{d_{z^2}}$ (l=0) and $\ket{\phi^v_\tau}=\frac{1}{\sqrt{2}}(\ket{d_{x^2-y^2}}+i\tau\ket{d_{xy}})$ (l=$\pm$ 2). The hybridized p-states on S atom are $\ket{\phi^v_\tau}=\frac{1}{\sqrt{2}}(\ket{p_{x}}+i\tau\ket{p_{y}})$ (l=$\pm$ 1) ; where $c$ and $v$ represent the conduction and valence band, respectively. Here, $\tau=\pm 1$ indicates the valley index for K and K$^{'}$ points connected by time reversal symmetry (TRS). Those hybridized states interact with each other to form symmetry adapted linearly combined (SALC) states for the valence and conduction bands of monolayer MoS$_2$. Now, the effect of the threefold rotational symmetry (C$_3$) on those states are given by $C_3 \ket{v(K,K^{'})}=\ket{v(K,K^{'})}$ and $C_3\ket{c(K,K^{'})}=e^{\mp i 2\pi/3} \ket{c(K,K^{'})}$ \cite{cao}. Therefore, the SALC states near the top of the conduction band dominated by ${d_{z^2}}$ states on Mo carry an overall azimuthal quantum number $m_{\pm}=\pm 1$ at K and K$^{'}$, whereas for the valence band SALC states it is given by $m_{\pm}=0$. The optical selection rule for the direct transition at the K and K$^{'}$ becomes $\Delta m_{\pm}=\pm 1$. Since, the carriers have well defined angular momentum associated with the K and K$^{'}$ states, we need to have photons with particular helicity (i.e circularly polarized light) to excite them. This phenomena of absorbing left-handed and right-handed circularly polarized lights by the two valleys is known as circular dichrosim (CD), see Fig.~\ref{makn2}(c). The TRS requires that $E_{\downarrow}(\textbf{k})=E_{\uparrow}(\textbf{-k})$. Hence, we have the inherently coupled valley and the spins of the carriers because of the spin-filtered valence band splitting together with the TRS, as shown in Fig.~\ref{makn2}(b). Fig.~\ref{makn2}(g) shows the PL spectrum (not polarization resolved) observed at T=14K using 2.33 eV laser excitation. The prominent feature around 1.9 eV is due to the A exciton complexes (AEC) consisting of two peaks: at higher energy it is due to the neutral exciton, and the lower component redshifted by 40 meV is attributed to the charged exciton (discussed later in details). The PL peak $\sim$ 2.1 eV is due to the B exciton and $\sim$ 1.8 eV arises because of excitons bound to the defects, as discussed earlier.\\

\begin{figure}[h]
 \centering
%\leavevmode t
%\includegraphics[trim=0 0 15 11, scale=0.5]{fig_44.pdf}
\includegraphics[trim=0.5cm 17cm 0.5cm 1cm, width=1.0\textwidth]{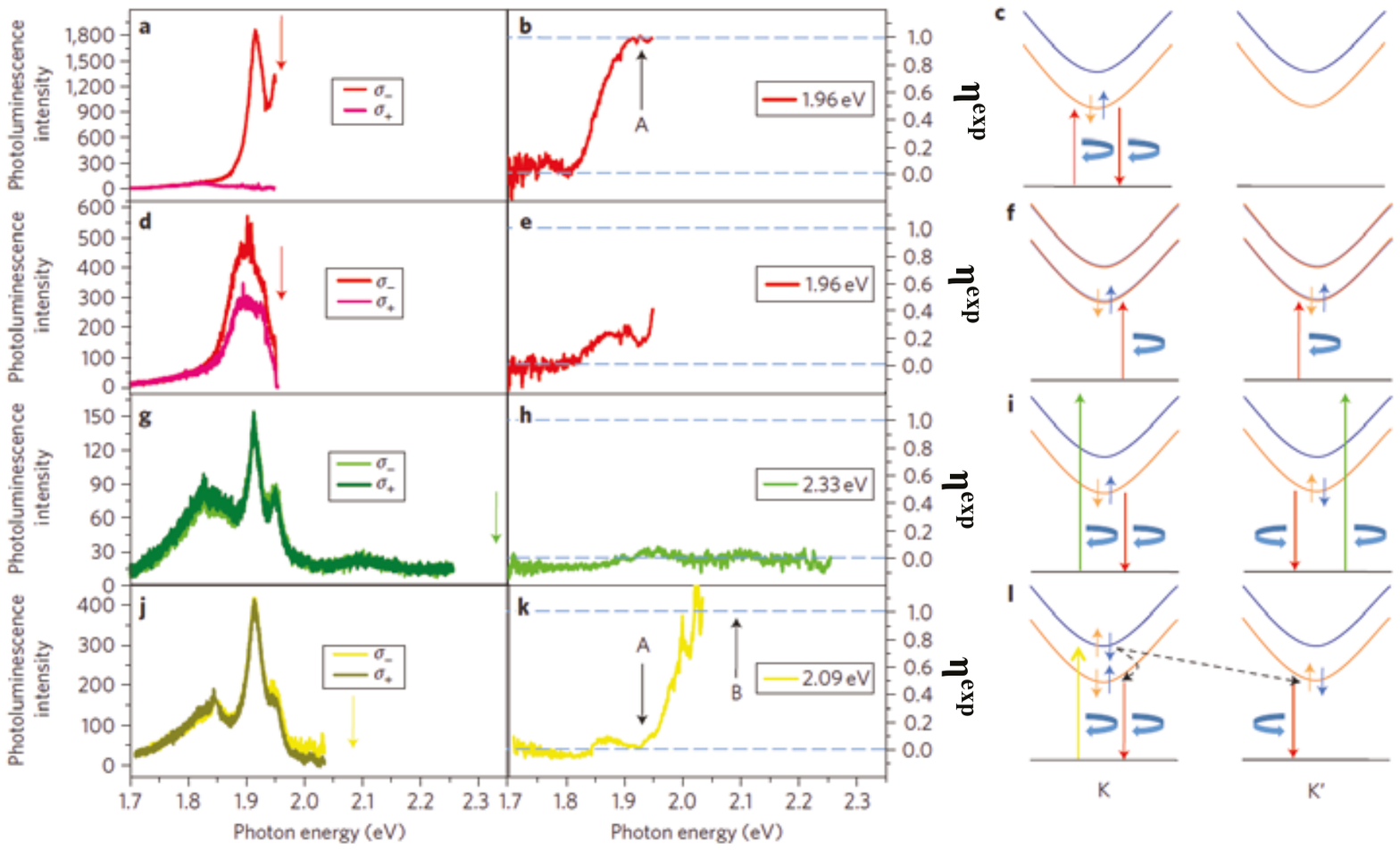}
  \caption{(color online)--Left column represents the PL spectra for (a,g and j) single layer and (d) for bilayer  MoS$_2$ (at 14K). Here the arrows indicate the laser energy used in the experiments. The left-handed circularly polarized light ($\sigma_{-}$) is used for all the excitations. The middle column represent the corresponding observed helicity in PL emissions. The right column represents the schematics of the corresponding absorption and emission of lights. Taken from Ref. \cite{mak2}.}
 \label{makn3}
\end{figure}

The degree of circular polarization for the PL emission is given by $\eta (\textbf{k},\omega_{cv})=\frac{|P^{cv}_+(\textbf{k})|^2-|P^{cv}_{-}(\textbf{k})|^2}{|P^{cv}_+(\textbf{k})|^2+|P^{cv}_{-}(\textbf{k})|^2}$. $P^{cv}_{\pm}(\textbf{k})$ representing the coupling strength with the circularly polarized lights ($\sigma_{\pm}$) is given by, $P^{cv}_{\pm}(\textbf{k})=\frac{1}{\sqrt{2}}\{P^{cv}_{x}(\textbf{k}) \pm P^{cv}_{y}(\textbf{k})\}$, where $P^{cv}_{\alpha}(\textbf{k})=\bra{\Psi_{c\textbf{k}}}\mathbf{P^{cv}_{\alpha}}\ket{\Psi_{v\textbf{k}}}$. The calculated $\eta (\textbf{k},\omega_{cv})$ is shown in Fig.~\ref{makn2}(h) indicating the exact value of $\pm$1 at the K and K$^{'}$ points \cite{cao}. The polarization resolved ($\sigma_{+}$ and $\sigma_{-}$) PL spectrum observed at T=14K using the left-handed circularly polarized ($\sigma_{-}$) lights (1.96 eV) is shown in Fig.~\ref{makn3}(a). The observed spectrum consists of only two peaks: The AEC peak is $\sigma_{-}$ polarized and the peak due to bound-exciton is almost unpolarized. Mak et al \cite{mak2} used a parameter to quantify the experimentally observed PL emission polarization, which is given by $\eta^{exp}=\frac{I(\sigma_{-})-I(\sigma_{+})}{I(\sigma_{-})+I(\sigma_{+})}$; where I($\sigma_{\pm}$) denotes the polarization resolved measured PL intensity. The observed value is given by $\eta^{exp}$=1.00$\pm$0.05 and drops to $\sim$ 0.05 below the photon energy 1.8 eV, as shown in Fig.~\ref{makn3}(b). With the excitation of right-handed circularly polarized light ($\sigma_{+}$), Mak et al observed the $\eta^{exp}$=-1.00$\pm$0.05 (not shown). In general, the quantity $\eta^{exp}$ depends on the two timescales: (i) exciton lifetime given by $\tau^{-1}=\tau^{-1}_{r}+\tau^{-1}_{nr}$, where r and nr represent the radiative and non-radiative recombination processes, respectively, and (ii) valley lifetime or hole-spin lifetime, $\tau_v$ \cite{cao}. In terms of these two timescales, the helicity for the AEC is given by $\eta^{exp}_{A}=\frac{1}{1+2\tau_A/ \tau_{Av}}$ for A-neutral exciton and $\eta^{exp}_{A^{-}}=\frac{\eta^{exp}_{A}}{1+2\tau_{A^{-}}/ \tau_{A^{-}v}}$ for A-charged exciton \cite{mak2}. Mak et al estimated that the observed helicity $\eta^{exp}$=1.00$\pm$0.05 is in accordance with the exciton lifetime $\tau$ $>$ 50 ps from the observed high QY on BN substrate and the valley lifetime $\tau_v$ $>$ 1 ns in monolayer MoS$_2$ \cite{mak2}.\\ 

To compare with the monolayer MoS$_2$, the polarization resolved PL measurement has also been done on the Bernal-stacked (see Fig.~\ref{makn2}d) bilayer MoS$_2$. Schematic diagram of the valleys at the K and K$^{'}$ points with the corresponding spin configurations of the carriers, and the diagram for the optical transitions are shown in Figs.~\ref{makn2}(e) and (f), respectively. Fig.~\ref{makn3}(d) and (e) show the observed PL spectra and the measured helicity for the bilayer, respectively. The observed low helicity $\eta^{exp}(bilayer)$=0.25$\pm$0.05 for the A exciton is in accordance with the hole-spin lifetime ($\tau_v$) of a few hundreds of femtoseconds (from the observed 20 times less QY) \cite{mak2}. The hole-spin lifetime for the bilayer, therefore, is more than three orders of magnitude less compared to the monolayer MoS$_2$, say by $\tau^{diff}_v$. The estimated $\tau^{diff}_v$ is attributed to the inversion symmetry of the bilayer MoS$_2$. The $\sigma_{-}$ polarization couples charge carriers only from one valley K (see Fig.~\ref{makn3}c), and excites an exciton with an electron having spin down and a hole of spin up at the K point. Now the intra-valley scattering involves the spin flip of hole which is forbidden by the spin-filtered energy gap of $\sim$ 160 meV in absence of magnetic impurity scatterings. The inter-valley scattering (from K to K$^{'}$) involves large momentum transfer with the requirement of the spin-flip as well, and hence this channel is also forbidden. Because of the absence of the inter as well as intra-valley scattering, the high valley lifetime is observed for monolayer MoS$_2$ \cite{mak2}. Bilayer MoS$_2$ has inversion symmetry which ensures that the $E_{\uparrow}(\textbf{k})=E_{\uparrow}(\textbf{-k})$. The both TRS and the inversion symmetry implies that at each $\textbf{k}$ point the spin degeneracy remains irrespective of valley index i.e. $E_{\downarrow}(\textbf{k})=E_{\uparrow}(\textbf{k})$. Hence, for the bilayer the spin and the valley index are not coupled and with the $\sigma_{-}$ excitation populations of both the valleys (K and K$^{'}$) having electron of down-spin and hole of up-spin (see Fig.~\ref{makn3}f) are achieved, which allows intravalley hole-spin relaxation process via Elliot-Yafet processes \cite{mak2} leading to the low observed helicity. \\     

\begin{figure}[h]
 \centering
%\leavevmode t
%\includegraphics[trim=0 0 15 11, scale=0.5]{fig_45.pdf}
\includegraphics[trim=0.5cm 19cm 0.5cm 1cm, width=1.0\textwidth]{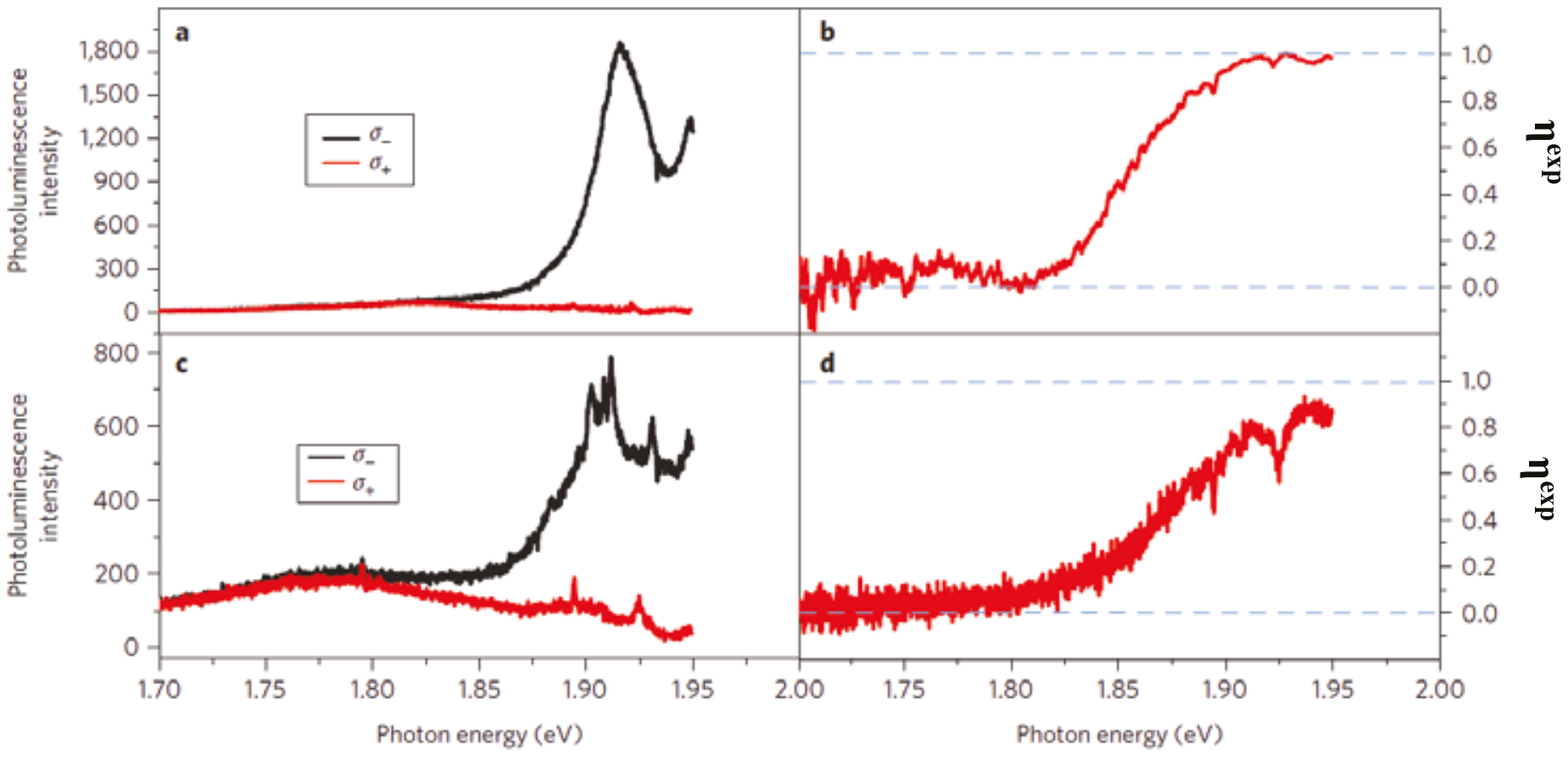}
  \caption{(color online)--PL spectra observed for single layer MoS$_2$ (a) on h-BN substrate and (c) on SiO$_2$/Si substrate; and the corresponding observed helicity in (b) and (d), respectively.  Taken from Ref. \cite{mak2}.}
 \label{makn4}
\end{figure}

The zero helicity is observed from monolayer MoS$_2$ with the $\sigma_{-}$ polarization under the laser excitation of 2.33 eV (532 nm), as shown in Figs.~\ref{makn3}(g) and (h). Since, the probing has been done with the higher energy than the resonance energy of A and B excitons, it populates the charge carriers at both the valleys, as shown in  Fig.~\ref{makn3}(i). In this case, no particular spin-valley selection is occurred and hence the observed zero spin-valley polarization. Figs.~\ref{makn3}(j) and (k) show the PL spectrum and the helicity of monolayer MoS$_2$ with the laser of 2.09 eV energy (594 nm), respectively. Because of the resonance with the B excitons, the observed helicity is $\sim$ 1, whereas the helicity for the A exciton is zero. In this case, the B-exciton relaxes through non-radiative channel to the A-exciton states (phonon-assisted process) and populates again both the valleys, as shown in Fig.~\ref{makn3}(l). The polarization resolved PL spectra for the monolayer MoS$_2$ deposited on BN and SiO$_2$/Si substrates are shown in Figs.~\ref{makn4}(a) and (c) and the corresponding measured helicity in Figs.~\ref{makn4}(b) and (d), respectively. The intensity for BN substrate is $\sim$ 10 times larger than that on SiO$_2$/Si, whereas the helicity is almost same for both the substrates. The robustness of the observed valley selective CD indicates how good the charge carriers preserve the information of the valley index and the spin index with them in their lifetimes i.e. before recombination.\\

\begin{figure}[h]
 \centering
%\leavevmode t
%\includegraphics[trim=0 0 15 11, scale=0.5]{fig_46.pdf}
\includegraphics[trim=0.5cm 15cm 0.5cm 2cm, width=0.8\textwidth]{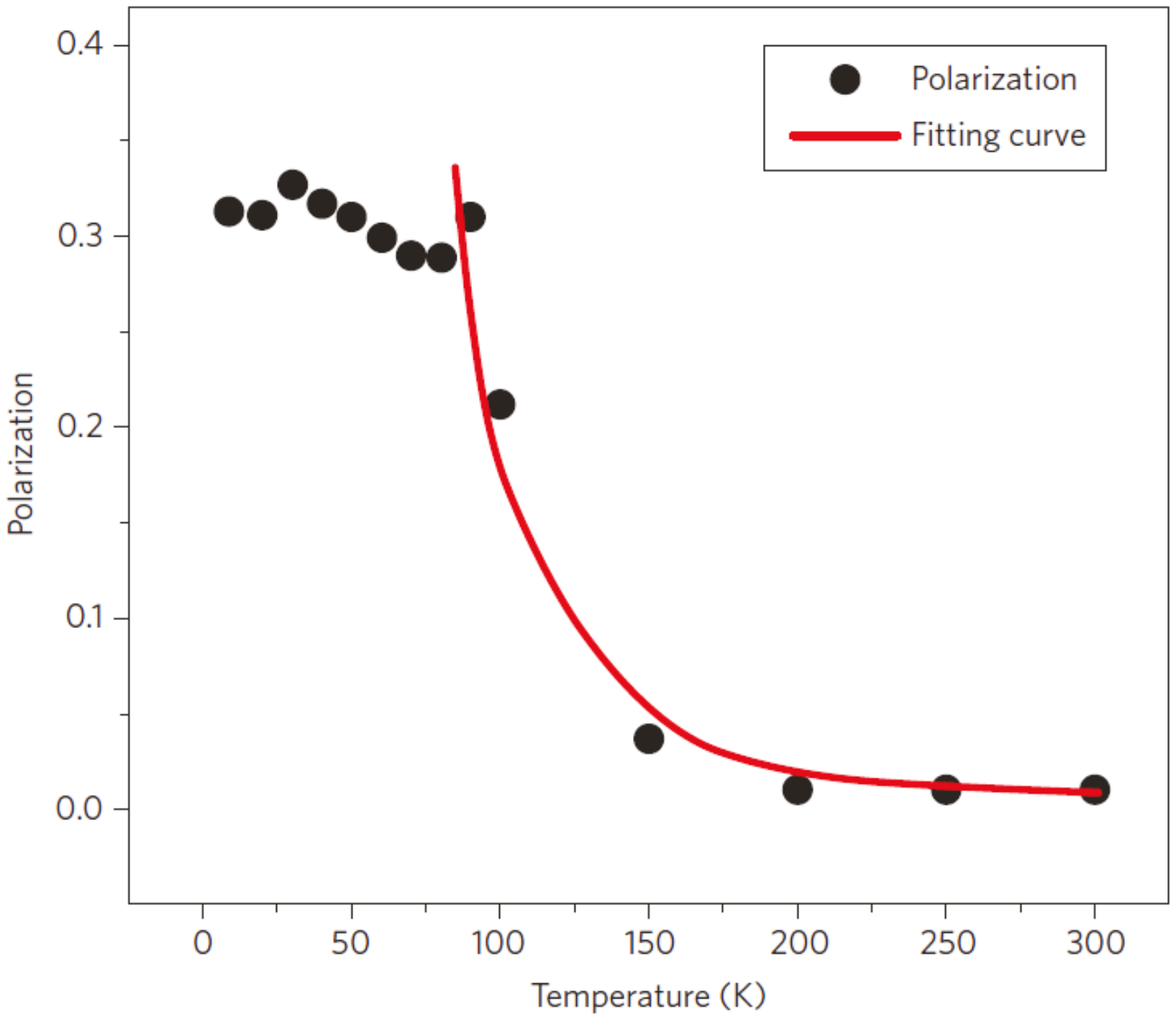}
  \caption{(color online)--Temperature dependent variations of the circular polarizations observed in PL spectra for single layer MoS$_2$. The experimental data have been fitted by considering the inter-valley scattering due to the acoustic phonons. Taken from Ref. \cite{zeng2}.}
 \label{zengn1}
\end{figure}

The temperature dependent helicity of the PL emission spectra for monolayer MoS$_2$ is shown in Fig.~\ref{zengn1}. The circular polarization remains almost constant in the range of 0 to 90K. With the increasing temperature, the observed helicity decreases. As the $\eta^{exp}\propto \tau_v$, with the increase of temperature the phonon populations and hence the scattering via them also increases by providing the required large momentum transfer from K to  K$^{'}$ point in BZ. The inter-valley scattering rate is  $\tau^{-1}_v \propto e^{-E_{ph}/K_BT}$, where $E_{ph}$ is the phonon energy at the K-point. The extracted value of the phonon frequency by fitting the observed PL helicity is given by $E_{ph}$ $\approx$ 240 cm$^{-1}$ in accordance with the acoustic phonon near the K-point of the monolayer MoS$_2$ \cite{waka}. Although, the fitting is in agreement with the observed helicity above 90K, more experimental as well as theoretical studies are needed to explore the full understanding of the underlying physics.\\

\subsection{Gate-voltage dependent absorption and photoluminescence of the A exciton complexes: observation of tightly bound trions}

\begin{figure}[h!]
 \centering
%\leavevmode t
%\includegraphics[trim=0 0 15 11, scale=0.5]{fig_47.pdf}
\includegraphics[trim=0.5cm 15cm 0.5cm 2cm, width=1.0\textwidth]{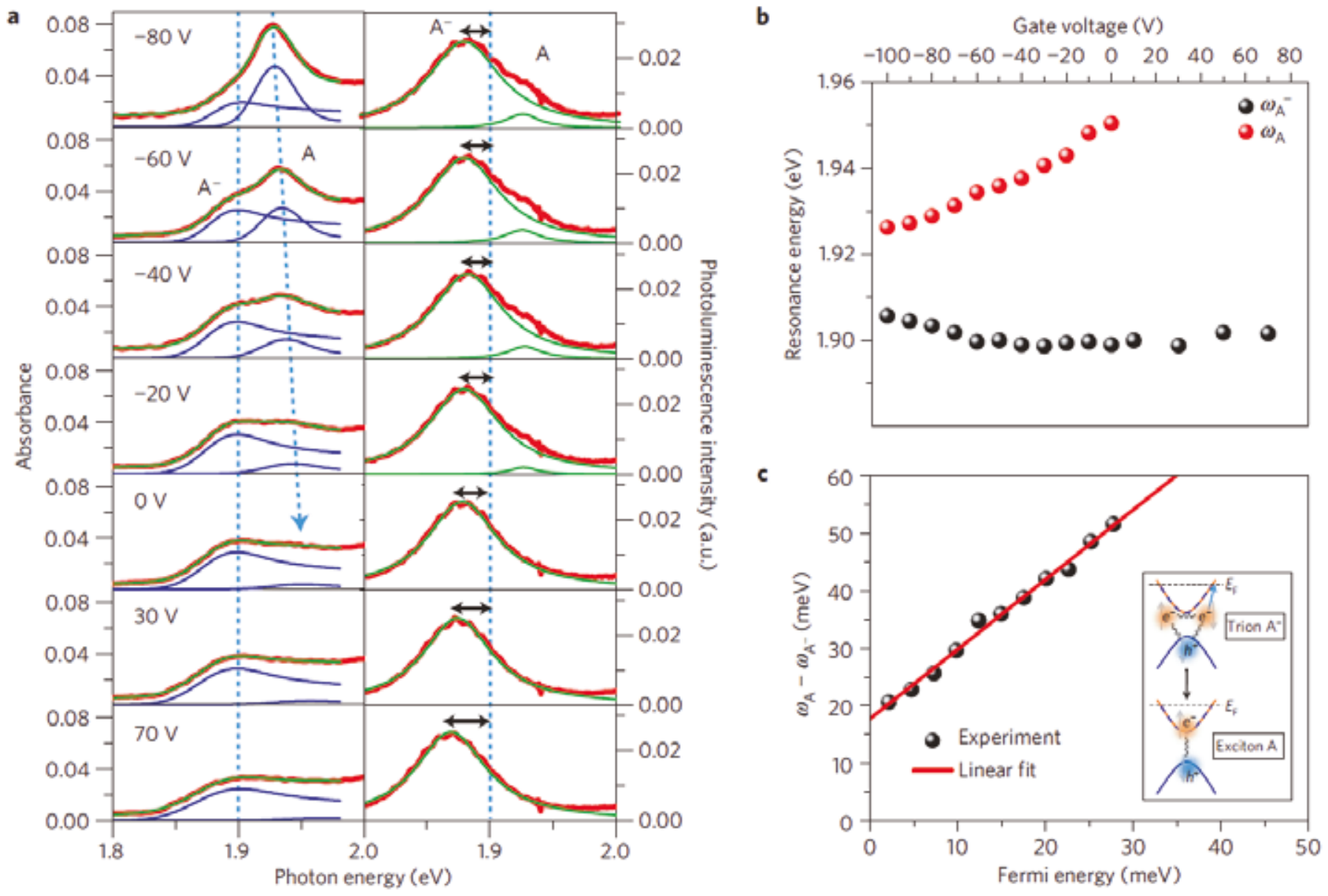}
  \caption{(color online)--(a) Back-gate voltage dependent absorption spectra (left panel) and the PL spectra (right panel) measured at 10K. The power-law has been used to fit (blue solid lines) the data for the absorption spectra and Lorentzian fits has been done for PL spectra. Here the dashed blue lines are guide to the eye. The arrows in PL spectra indicate the shift of the trion with the gating. (b) The extracted (from absorption spectra) peak positions of trion ($\omega_{A^{-}}$) and the neutral exciton ($\omega_{A}$) as a function of gate voltages (upper scale) and the corresponding Fermi energy (bottom scale); and the difference $\omega_{A}-\omega_{A^{-}}$ in (c). The E$_F$ dependence linear fit has been done to the data. The inset in (c) shows the schematics of the breaking of the trion into a neutral exciton and an electron at the Fermi level. Here the offset gate voltage =-107$\pm$6 V for E$_{F}$=0. Taken from Ref. \cite{mak3}.}
 \label{makn5}
\end{figure}

A few absorption spectrum for the monolayer MoS$_2$ at different back gate voltages in the range of -100 to +80 V using 532 nm laser line are shown in Fig.~\ref{makn5}(a). As we discussed, the monolayer MoS$_2$ device shows n-type transfer characteristics at ambient conditions because of the Fermi level pining. Therefore, the device operated at -100V is having almost the undoped channel. Here we discuss the absorption and PL spectra for the AEC only. The absorption spectrum at -80V is dominated by an intense neutral A-exciton and an little shoulder due to the negatively charged A-exciton. This charged exciton consisting of one hole of up-spin and two electrons of opposite spins is known as trion ($A^{-}$) \cite{mak3}. As the negative gate voltage decreases, the $A^{-}$ evolves and around +70V, the spectrum is dominated completely by $A^{-}$ exciton. The neutral A-exciton peak gradually diminishes with the decrease of the negative gate voltages, and after a certain voltage (0V) it is hard to track the peak. The peak positions of these two peaks are plotted versus back gate voltages (Fermi energy in lower axis) in Fig.~\ref{makn5}(b), and their difference ($\omega_{A}-\omega_{A^{-}}$) versus Fermi energy (gate voltages in upper axis) is plotted in Fig.~\ref{makn5}(c). The overall reduction of the intensity for the absorption spectra is due to the Pauli blocking associated with the increasing electrons in the conduction band \cite{mak3}. In the right panel of Fig.~\ref{makn5}(a), the PL spectra at different gate voltages is shown using 532 nm laser line. The neutral A-exciton behaves similarly as in absorption spectra i.e. after a positive gate voltage it is turned off. The feature of the blueshifts of the neutral A-exciton and the almost constant feature of the $A^{-}$-exciton with gate voltages are attributed to the combined effect of the many-body interactions and the Pauli blocking \cite{mak3}. The neutral A-exciton can be thought of an ionized trion such that the splitting between the peaks is given by $\omega_{A}-\omega_{A^{-}}=E_{A^{-}}+E_F$; where, $E_{A^{-}}$ is the binding energy of the trion $A^{-}$. At $E_F$=0, $E_{A^{-}}$ is the separation between the two peaks and the extracted value by fitting the linear relation (see Fig.~\ref{makn5}c) is $\sim$ 18.0$\pm$1.5 meV. The separation $\omega_{A}-\omega_{A^{-}}$ behaves linearly with the Fermi energy ($E_F$), and is in accordance with the earlier observations of trions for the quantum wells \cite{huard}. \\

\begin{figure}[h!]
 \centering
%\leavevmode ture
%\includegraphics[trim=0 0 15 15, scale=0.5]{fig_48.pdf}
\includegraphics[trim=0.5cm 20cm 0.5cm 2cm, width=1.0\textwidth]{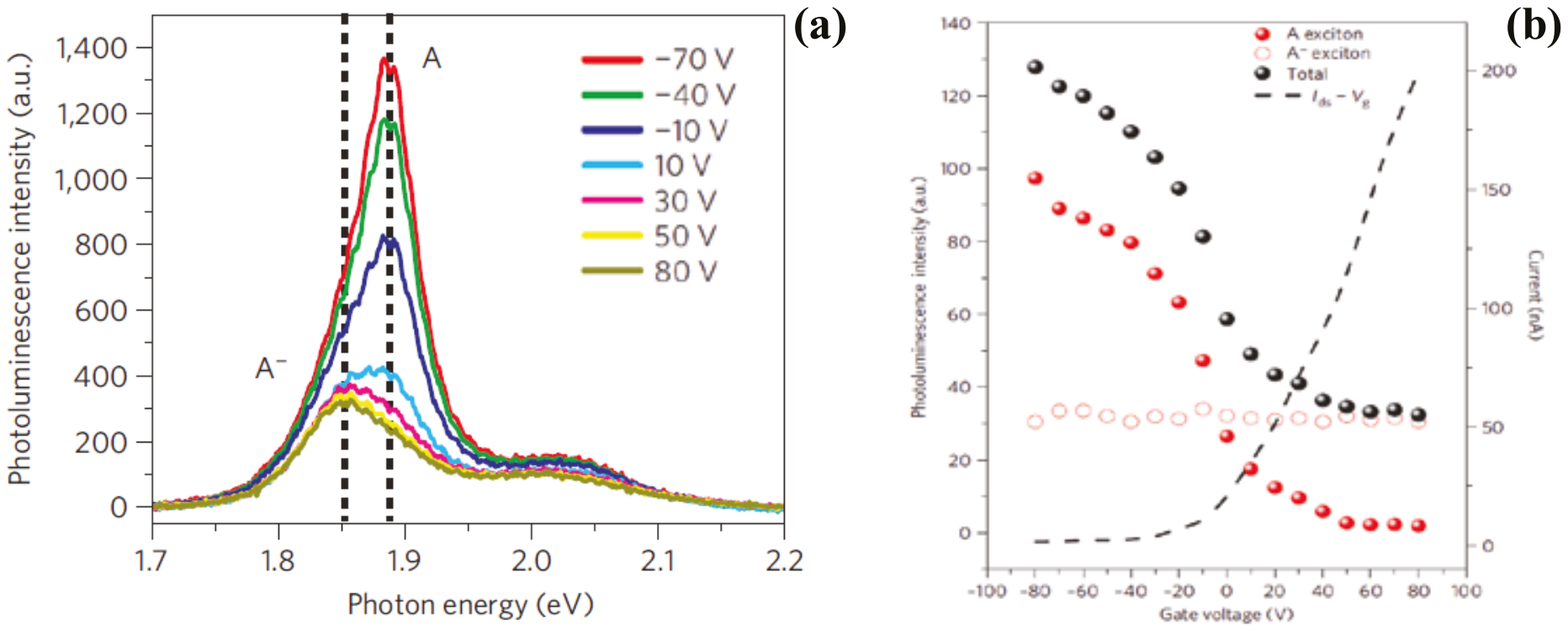}
  \caption{(color online)--(a) The gating voltage dependent PL spectra measured at 300K. (b) The integrated intensity of both the neutral exciton and the trion and their total intensity (left scale). Right scale represents the transfer characteristics. Taken from Ref. \cite{mak3}.}
 \label{makn6}
\end{figure}

Figs.~\ref{makn6}(a) and (b) show the room temperature PL spectra (with broadened features) and the corresponding gate voltage dependent measured PL intensity, respectively. The PL intensity for the neutral A-exciton decreases with gating by almost two orders of magnitude, whereas for the trion it remains constant in the range of -100 to +80 V. The decreased PL intensity of the neutral A-exciton is attributed to the spectral weight reduction and the transfer of it to the trion with the doping as for the absorption spectra at 10K (see Fig.~\ref{makn5}a). This high tunability of the PL intensity, which is not observed at 10K, is attributed to the thermally activated carrier populations from the trion state.\\

\begin{figure}[h!]
 \centering
%\leavevmode ture
%\includegraphics[trim=0 0 15 15, scale=0.5]{fig_49.pdf}
\includegraphics[trim=0.5cm 20cm 0.5cm 1cm, width=1.0\textwidth]{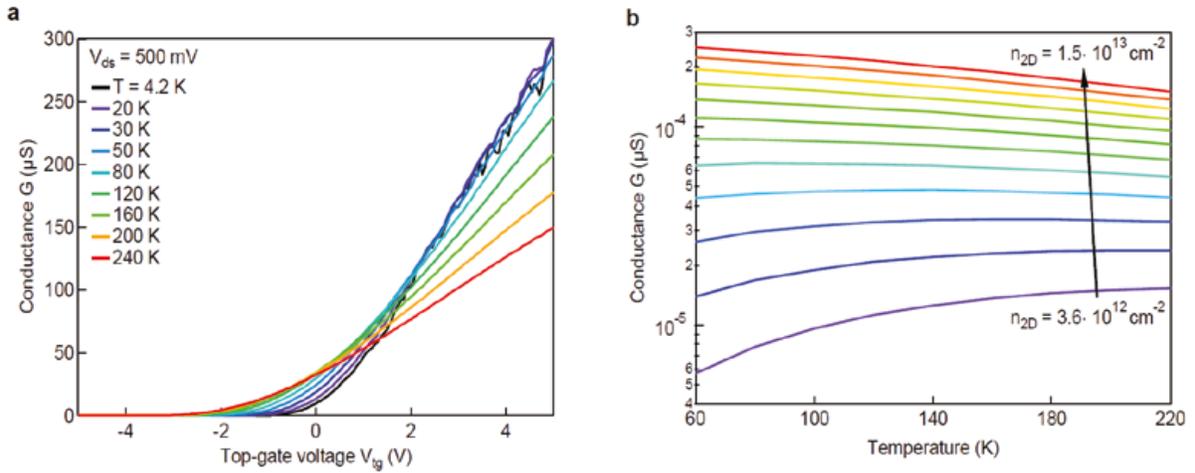}
  \caption{(color online)--(a) The top gate (V$_{tg}$) dependent conductance (G) of single layer MoS$_2$ for various temperatures. The G shows the thermally activated behavior for low V$_{tg}$ and it decreases with decreasing temperature. Single layer MoS$_2$ shows a transition from an insulating regime to a metallic regime above V$_{tg}$ of $\sim$ 1-2 for the decreased temperatures such that the G increases with decreasing temperature and this cross over is clear for the lower temperature curves. (b) The temperature dependent G of the devices for different values of the electron density (n$_{2D}$). Taken from Ref. \cite{kis}.}
 \label{kisn1}
\end{figure}

The gating voltage dependent optical response might be associated with the metal-insulator transition (MIT) as observed for the two-dimensional electron gas (2DEGs) \cite{fink}. At $\sim$ -100V, the channel is undoped (T=10K) and hence the optically excited excitons dominate the absorption spectra. Now as we increase the doping, for positive gate voltages the conduction band is populated with the free carriers and the neutral A-exciton combines with them to form trion. Hence, at higher positive gate voltages, the absorption spectra is dominated completely by the trions (see Fig.~\ref{makn5}a). As we depletes the channel more and more, the decrease in the number of carriers leads to the much less screening of the long range Coulomb interactions due to the presence of the extrinsic disorders (ionized) in the sample. This implies an increase in potential fluctuations in the sample which in turn localize the electrons. This feedback mechanisms results in non-linear screening and hence, can reduce the MIT region to a smaller gate voltage. Recently, it has been shown that the monolayer MoS$_2$ indeed goes to MIT at the top gate voltage of $\sim$ 2.2V (the carrier concentration $\sim$ 1$\times$ 10$^{13}$ cm$^{-2}$) below 80K \cite{kis}, as shown in Fig.~\ref{kisn1}. If 2DEG was uniformly distributed instead of localized, we should not observe the improvement of the spectral weight of the neutral A-exciton peak in absorption spectra. Since localized electrons do not play an effective role to the screening, the more localization means more prominent neutral A-exciton peak. Now, if the charge carriers would form the patches of inhomogeneous carrier density, the associated tuning of the PL intensity should not show any on-set. In contrast, experimental results for the PL intensity show an on-state (threshold) with the gate voltage (see Fig.~\ref{makn6}b), which implies the possibility of the localization of electrons (Wigner crystallization) instead of forming small patches \cite{fink}. If the localized electrons were effective in the screening, then we should not observed the enhancement of the A-exciton in absorption spectra; and hence, we can say that this MIT may be due to the Mott-Hubbard type instead of Anderson-type. More experimental works are required to explore the physics in the insulating region and the corresponding theoretical supports to explain the observed MIT in the monolayer MoS$_2$. \\   

\section{Conclusions}
In conclusion, single and a few layer MoS$_2$ devices show reasonably high carrier mobility ($\sim$ 700 cm$^2$/V-sec) and on-off ratio $\sim$ 10$^8$. The increment of the high temperature (100-300 K) mobility has been engineered by quenching the homopolar and other phonon scattering processes with the help of dielectric coating. The observed robustness of the valley selective circular dichrosim suggest that we should explore experimentally valley Hall effect and magnetism in single-layer MoS$_2$ without applying magnetic field. The observation of negatively charged excitons or trions and the associated metal-insulator-transition at low temperature can be an indication of Wigner-crystal formation in 2-D MoS$_2$ material, which needs to be further explored. Along with that, the measured dimensionless parameter (ratio of Coulomb potential energy to kinetic energy) r$_s$ $\approx$ 60 make the MoS$_2$ as an ideal system for further exploration of many-body phenomena. In addition, other dichalcogenide materials like WS$_2$, MoSe$_2$ and WSe$_2$ \cite{ton} are exciting to explore the novel physics in coming years.

AKS thanks Department of Science and Technology, India for funding under the Nanomission Project. AB thanks CSIR for a research fellowship.

\end{document}